# Event Generators for Bhabha Scattering

*Conveners:* S. Jadach and O. Nicrosini

*Working Group:* H. Anlauf, A. Arbuzov, M. Bigi, H. Burkhardt, M. Cacciari, M. Caffo, H. Czyż, M. Dallavalle, J. Field, F. Filthaut, F. Jegerlehner, E. Kuraev, G. Montagna, T. Ohl, B. Pietrzyk, F. Piccinini, W. Płaczek, E. Remiddi, M. Skrzypek, L. Trentadue, B. F. L. Ward, Z. Wąs,

## Contents

# 1 Introduction

The main goals of the Bhabha working group are to make an inventory of all the available Monte Carlo (MC) event generators for small-angle (SABH) and large-angle (LABH) Bhabha processes at LEP1 and LEP2, and to improve our understanding of their theoretical uncertainties through systematic comparisons of the MC event generators (developed independently) among themselves and with other non-MC programs. The presented activity is of course an obvious continuation of the previous workshops on LEP1 physics [1, 2]. In the beginning of the present workshop the theoretical uncertainty at LEP1 for the SABH process was typically estimated as 0.16%, and for the LABH process was estimated at 0.2% level at Z peak and 1% on the wings of the Z resonance. There were no estimates specific to LEP2.

We shall concentrate on the comparison of all the presently available theoretical calculations (published and unpublished). This will be done for several kinds of event selection (ES), defined as a set of experimental cuts and apparatus acceptances, starting from ES's unrealistic, but useful for some studies oriented towards the QED matrix element, and ending on ES's very close to the experimental ones.

Let us add a few comments to clarify our priorities and to set the proper perspective for our work. In spite of the considerable effort of several theoretical groups, at present the theoretical error on the small-angle Bhabha cross section dominates the luminosity error at LEP1. This inhibits from taking full advantage of the high experimental precision of the final LEP1 data for precision tests of the Standard Model. As a consequence, the reduction of the theoretical error in the SABH process at LEP1 is the biggest challenge, and was the main objective of our working group. The precision requirements of LEP2 are lower than those of LEP1. The total cross section of W pair production will be measured with 1.0% to 0.5% precision at best, so it is sufficient to keep the theoretical uncertainty of the SABH process at the 0.25% level. Furthermore, at LEP2 the detectors and experimental techniques for measuring the SABH process are almost the same as for LEP1[1]. Radiative corrections to the SABH cross section depend on the center of mass energy, but smoothly; moreover, in the small-angle regime the center of mass energy is not so important from the point of view of the physics involved: we are always faced with a $t$-channel photon-exchange dominated process; hence, improving the small-angle Bhabha generators for LEP1 is generally a sufficient condition for improving them also for LEP2. The only subtle point concerns the error estimate: a 0.1% error at LEP1 does not guarantee such a small error also at LEP2, so that an additional analysis has to be performed. For the LABH process, the final LEP1 data analysis requires a theoretical uncertainty of the codes used to be at the 0.5% level. The LABH process at LEP2 is not of major interest, and we think that a precision of the order of 2% is enough. Nevertheless, the physics of the LABH process at LEP2 is significantly different from LEP1 (different Feynman diagrams rise to importance), so performing additional study for the LABH process at LEP2 is a new nontrivial

[1]Actually, the main difference is that, due to machine background radiation, the internal part of luminosity detectors may be obscured by special masks. We shall discuss the impact of such modification on the theoretical errors. This aspect was brought to our attention by B. Bloch-Devaux during our WG meeting in January 95.

work[2].

In view of the above, our strategy was to do all the work for the SABH and the LABH processes first for LEP1 experimental conditions, and to supplement it with all necessary work/discussion which would assure control of the precision at the level sufficient for LEP2 experiments. This practically means that all the numerical comparisons were done for LEP1 and repeated for LEP2, or, in rare cases, a convincing argument was given that it is not necessary (sometimes numerical results for LEP2 were obtained, but are not shown in full form because they were trivially identical to those for LEP1).

We include in our report two main parts: one part on the SABH process and a second one on the LABH process, with the cases of LEP1 and LEP2 discussed in parallel. These two processes are governed by different physics (i.e. dominated by different Feynman diagrams). Also, the theoretical precision requirements in calculating SABH and LABH cross sections are different by a factor of five-ten. These two parts are followed by a section including short descriptions of all the involved Monte Carlo (MC) event generators or other codes, and a final section on conclusions and outlook.

# 2 Small-angle Bhabha scattering

Small-angle Bhabha (SABH) scattering is used at LEP1 and LEP2 to measure the accelerator luminosity. The LEP1 experiments have reached in 1993-1994 a systematic uncertainty of better than 0.10% in selecting luminosity Bhabha events, see Ref. [3] and Refs. [4,5].

On the theory side, QED calculations have still an uncertainty larger than 0.16% [6] in determining the Bhabha cross section in the detector acceptance, which is caused mostly by the non-existence of a Monte Carlo program including complete $\mathcal{O}(\alpha^2)$ next-to-leading terms. Actually, there exist $\mathcal{O}(\alpha^2)$ calculations with complete next-to-leading contributions [7,8] which claim a precision of the order of 0.1%, but they can not be used in a straightforward way, because they are not implemented in the Monte Carlo event generators. The size of the $\mathcal{O}(\alpha^2)$ contributions depends not only on the angular range covered by the detector and on the electron energy cut-off, but also on crucial experimental aspects, such as the sensitivity to soft photons or such as the electron cluster size. This means that the main interest is in the theoretical predictions for the Bhabha process, including as many higher order radiative corrections at it is necessary to reach a precision of 0.05%, in a form of a Monte Carlo *event generator*.

Monte Carlo event generators are very powerful tools because they are able to provide a theoretical prediction, cross sections and any kind of distribution, for arbitrary ES's. However, event generators are difficult to construct and, what is even more serious, they are very difficult to test – one has to have at least two of them to compare with one another for a wide range of

[2]The radiative LABH process is an important background to other processes, like $\tau$-pair production, $W^+W^- -> ee\nu\nu$, "new physics" like SUSY processes and so on, but a detailed analysis of these items goes beyond the aims of the present study.

ES's.

For the SABH process, the task of comparing various Monte Carlo event generators was the *main goal* of the Bhabha Working Group. There were only a few comparisons of independently developed Monte Carlo event generators for the SABH process in the past. A few examples can be found in Ref. [2]. However, we shall include in the comparisons results from non-Monte Carlo calculations, as well. They are usually limited to certain special (primitive) ES's. Nevertheless they provide additional valuable cross-checks.

What shall we learn from these comparisons? The calculations from various Monte Carlo event generators will of course differ. The differences have to be understood. In a certain class of the comparisons, the underlying QED matrix element will be the same and in that case the differences will be only due to numerical effects. The results from two or more computer programs will differ due to rounding errors, programming bugs, numerical approximations. The difference measures uncertainties of this kind, and we say that we are determining the *technical precision* of the tested programs. One has to remember that the technical precision is dependent on the ES, and it is therefore absolutely necessary to use several at least semi-realistic, quite different, examples of ES's. In other cases, we shall compare Monte Carlo event generators which are based on different QED matrix elements. In this case, the difference between results will tell us typically about higher order effects which are *not included* in some of these event generators, or which are *approximated* differently in these programs. In this situation we shall talk about exploring the *physical precision* of the tested Monte Carlo event generators. Needless to say, the physical precision is the main goal, but one has to remember that without a technical precision of at least a factor of two better than the physical precision it is pointless to discuss the physical precision at all!

Before we come to the actual comparisons of the programs, let us characterize various contributions/corrections to the SABH process. We shall also characterize briefly the various Monte Carlo event generators and non-Monte-Carlo calculations involved in the comparisons.

If we remember that the SABH process was chosen for the luminosity measurement because it is calculable from first principles within quantum electrodynamics (QED), then it is natural to group corrections to the SABH process into *pure-QED* and *non-QED* corrections. The latter ones are due to s-channel Z-exchange, and the corrections induced by low energy strong interactions (QCD) through vacuum polarization and light quark pair production. Among the pure QED corrections, we may distinguish *photonic* (bremsstrahlung) corrections, related to multiple photon emission, and *non-photonic* corrections – for instance lepton pairs, leptonic vacuum polarization, multiperipheral diagrams. Numerically, the biggest ones are the photonic corrections and the vacuum polarization correction. They also contribute the most to the physical precision. Photonic corrections dominate completely the technical precision, due to the MC integration over the complicated multi-body phase space. The QED non-photonic corrections are small, but are difficult to calculate and quite uncertain (technical precision).

For all the comparisons of the event generators it is crucial (especially for SABH) to understand the experimental ES. In the main comparisons we shall compare *all* the available event generators for four types of ES's. However, the problem of the variation of the parameters in the ES is so important that we include also a separate subsection on this subject, in which, for a limited number of three event generators, we perform a detailed study of the dependence of the higher order corrections on all possible cut-offs involved in the real experiment. This will allow us to see all our work in the proper perspective from the point of view of the experimental analysis, and will also give us clear hints on the dependence of the higher order corrections on the fine details of the ES. This study will be limited to the SABH process.

## 2.1 Sensitivity of LEP1 observables to luminosity

The importance of the improvement of the theoretical luminosity error on the LEP1 results is shown in Table 1. The results of the lineshape parameter fits made with the theoretical luminosity error of 0.16% and 0.11% are given [9], corresponding to the reduction of error achieved during this workshop. A projection concerning a further reduction of the theoretical luminosity error to 0.06% is also given. The results of the four LEP1 experiments used as input to the fits, as well as the fitting procedure, are described in Ref. [10]. From the five parameter fit, only $\sigma_h^0$ is sensitive to the luminosity error. The decreased error in this variable causes a reduction of the errors of the derived parameters shown in the lower part of Table 1. As we see, the above improvement in the theoretical luminosity error influences significantly not only quantities like the "number of light neutrino's" $N_\nu$, but also other LEP1 observables used routinely in the tests of the Standard Model.

| | theoretical luminosity error | | |
|---|---|---|---|
| | 0.16% | 0.11% | 0.06% |
| $m_Z$ [GeV] | $91.1884 \pm 0.0022$ | $91.1884 \pm 0.0022$ | $91.1884 \pm 0.0022$ |
| $\Gamma_Z$ [GeV] | $2.4962 \pm 0.0032$ | $2.4962 \pm 0.0032$ | $2.4961 \pm 0.0032$ |
| $\sigma_h^0$ [nb] | $41.487 \pm 0.075$ | $41.487 \pm 0.057$ | $41.487 \pm 0.044$ |
| $R_l$ | $20.788 \pm 0.032$ | $20.787 \pm 0.032$ | $20.786 \pm 0.032$ |
| $A_{FB}^{0,l}$ | $0.0173 \pm 0.0012$ | $0.0173 \pm 0.0012$ | $0.0173 \pm 0012$ |
| $\Gamma_{had}$ [GeV] | $1.7447 \pm 0.0030$ | $1.7447 \pm 0.0028$ | $1.7446 \pm 0.0027$ |
| $\Gamma_{ll}$ [MeV] | $83.93 \pm 0.13$ | $83.93 \pm 0.13$ | $83.93 \pm 0.12$ |
| $\sigma_{ll}^0$ [nb] | $1.9957 \pm 0.0044$ | $1.9958 \pm 0.0038$ | $1.9959 \pm 0.0034$ |
| $\Gamma_{had}/\Gamma_Z$ [%] | $69.90 \pm 0.089$ | $69.90 \pm 0.079$ | $69.89 \pm 0.072$ |
| $\Gamma_{ll}/\Gamma_Z$ [%] | $3.362 \pm 0.0037$ | $3.362 \pm 0.0032$ | $3.362 \pm 0.0028$ |
| $\Gamma_{inv}$ [MeV] | $499.9 \pm 2.4$ | $499.9 \pm 2.1$ | $499.9 \pm 1.9$ |
| $\Gamma_{inv}/\Gamma_{ll}$ [%] | $5.956 \pm 0.030$ | $5.956 \pm 0.024$ | $5.956 \pm 0.020$ |
| $N_\nu$ | $2.990 \pm 0.015$ | $2.990 \pm 0.013$ | $2.990 \pm 0.011$ |

Table 1: Line shape and asymmetry parameters from 5-parameter fits to the data of the four LEP1 experiments, made with a theoretical luminosity error of 0.16%, 0.11% and 0.06% [9]. In the lower part of the Table also derived parameters are listed.

At LEP2, the normalization of the total cross section for the WW production process enters in a nontrivial way into tests of the W boson coupling constants. The precision requirements for the total cross section is limited by statistics of the WW process, and a luminosity error at the 0.25% level is sufficient (see the chapter "WW cross-sections and distributions", these proceedings).

## 2.2 Higher order photonic corrections at LEP1 and LEP2

| Canonical coefficients | | | | | |
|---|---|---|---|---|---|
| | | $\theta_{min}=30$ mrad | | $\theta_{min}=60$ mrad | |
| | | LEP1 | LEP2 | LEP1 | LEP2 |
| $\mathcal{O}(\alpha L)$ | $\frac{\alpha}{\pi}4L$ | $137\times10^{-3}$ | $152\times10^{-3}$ | $150\times10^{-3}$ | $165\times10^{-3}$ |
| $\mathcal{O}(\alpha)$ | $2\frac{1}{2}\frac{\alpha}{\pi}$ | $2.3\times10^{-3}$ | $2.3\times10^{-3}$ | $2.3\times10^{-3}$ | $2.3\times10^{-3}$ |
| $\mathcal{O}(\alpha^2 L^2)$ | $\frac{1}{2}\left(\frac{\alpha}{\pi}4L\right)^2$ | $9.4\times10^{-3}$ | $11\times10^{-3}$ | $11\times10^{-3}$ | $14\times10^{-3}$ |
| $\mathcal{O}(\alpha^2 L)$ | $\frac{\alpha}{\pi}\left(\frac{\alpha}{\pi}4L\right)$ | $0.31\times10^{-3}$ | $0.35\times10^{-3}$ | $0.35\times10^{-3}$ | $0.38\times10^{-3}$ |
| $\mathcal{O}(\alpha^3 L^3)$ | $\frac{1}{3!}\left(\frac{\alpha}{\pi}4L\right)^3$ | $0.42\times10^{-3}$ | $0.58\times10^{-3}$ | $0.57\times10^{-3}$ | $0.74\times10^{-3}$ |

Table 2: The canonical coefficients indicating the generic magnitude of various leading and subleading contributions up to third-order. The big-log $L=\ln(|t|/m_e^2)-1$ is calculated for $\theta_{min}=30$ mrad and $\theta_{min}=60$ mrad and for two values of the center of mass energy: at LEP1 ($\sqrt{s}=M_Z$), where the corresponding $|t|=(s/4)\theta_{min}^2$ are 1.86 and 7.53 GeV$^2$, and at LEP2 energy ($\sqrt{s}=200$ GeV), where the corresponding $|t|$ are 9 and 36 GeV$^2$, respectively.

For the SABH process, the smallness of the electron mass "ruins" the normal perturbative expansion order in the following sense: for instance, the $\mathcal{O}(\alpha^2)$ QED contributions can be expanded into $\mathcal{O}(\alpha^2L^2)$, $\mathcal{O}(\alpha^2L)$ and pure non-log $\mathcal{O}(\alpha^2)$. The non-log $\mathcal{O}(\alpha^2)$ corrections are completely uninteresting, while the $\mathcal{O}(\alpha^3L^3)$ corrections are as important as the $\mathcal{O}(\alpha^2L)$ corrections. Here $L=\ln(|t|/m_e^2)$ is the so-called big-log in the leading-logarithmic (LL) approximation, where $t$ is the momentum transfer in the $t$-channel (of the order of 1 GeV). This phenomenon is illustrated in Tab. 2. From this table, it is clear that for a precision of the order of 0.25% (for calorimetric ES's) it is enough to include the $\mathcal{O}(\alpha^1L)$, $\mathcal{O}(\alpha^1)$ and $\mathcal{O}(\alpha^2L^2)$. For a precision of the order of 0.1% or better, one has to add $\mathcal{O}(\alpha^3L^3)$ and $\mathcal{O}(\alpha^2L)$. These "scale coefficients" have to be kept in mind when discussing various QED calculations/programs. As we shall see, the higher order effects seen in the numerical results presented in the next sections generally conform to the above scale coefficients.

Table 2 demonstrates also the "scaling laws" for various QED corrections between LEP1 (Z peak) and LEP2 energies. If the angular range is kept the same, then $t$-channel transfer is proportional to $s=4E_{beam}^2$. Actually, at LEP2 experiments the luminosity measurement will rely more on the SABH process at larger angles, above 3°, and this is why we also included in the table another two columns for this angular range. As we see, photonic corrections do not change very much due to the increase of $\sqrt{s}$ from Z-peak energy to LEP2 energy (200 GeV) and due to going to twice larger angles. Actually, the change in canonical coefficients is negligible. One has only to pay attention to the $\mathcal{O}(\alpha^3 L^3)$ corrections, which in the worst case increase by a factor 1.75 (however, as we shall see they are under good control).

One has to remember that, as it was shown explicitly in ref. [11], the radiative corrections to the SABH process with the typical "double tag" detection are proportional to $\ln((\theta_{\max}/\theta_{\min})-1)$, i.e. they are bigger for "narrower" angular acceptance and smaller for "wider" angular acceptance. This has to be remembered, because at LEP2 in some experiments the angular range might be "narrowed" by placing masks in front of the SABH detectors in order to eliminate machine background radiation. We conclude that the change for "narrower" angular acceptance is more dangerous from the point of view of the increase of the pure photonic corrections, and we shall address this problem with a separate numerical exercise.

In ref. [11] it was also shown (numerically), using an $\mathcal{O}(\alpha)$ calculation, that for the purpose of the SABH process below 6° we may neglect the real and virtual QED interference contributions between photon emission from the electron and positron lines, the so called "up-down interference". In the numerical example in ref. [11] it was shown that, for the angular range $3.0^\circ - 4.24^\circ$, the "up-down interference" is below 0.015%. It is even smaller for smaller angles. It means that it is negligible for all practical purposes in the luminosity measurements. This phenomenon was also discussed in ref. [12] beyond $\mathcal{O}(\alpha)$ in the framework of the eikonal approximation.

## 2.3 Light pairs and other small contributions

To calculate pair corrections to the SABH two approaches have been used. **(1)** The first one is based on direct analysis of Feynman graphs and analytical extraction of graphs and terms contributing to the SABH within the $\mathcal{O}(0.1\%)$ accuracy. Both leading and next-to-leading terms are considered. **(2)** The other method uses the LL approximation to find the dominant pair contributions to SABH and to discard the negligible ones. Having isolated the dominant mechanism, an actual MC program for this particular mechanism is constructed.

**(1)** The dominant pair production corrections (enhanced by factors of $L^2$ and $L$) arise from kinematical configurations where one (or both) of the produced leptons is almost collinear with the incoming or outgoing $e^\pm$. These contributions have been calculated analytically [13, 14]. The analytical calculation [7, 8, 13–15] of the real hard pair production cross-section within logarithmic accuracy takes into account the contributions of the collinear and semi-collinear kinematical regions. All possible mechanisms for pair creation (Singlet and Non-Singlet), as well as the identity of the particles in the final state, are taken into account[3]. In the case of SABH only a part of the total 36 Feynman diagrams are relevant, i.e. the scattering diagrams[4] shown in Fig. 1.

[3]Here we have taken into account only $e^+e^-$ pair production. An estimate of the muon pair contribution gives less than 0.05% since $\ln(Q^2/m^2) \sim 3\ln(Q^2/m_\mu^2)$. Contributions of pion and tau-lepton pairs give still smaller corrections. Therefore, within the 0.1% accuracy, one may omit any pair production contribution except the

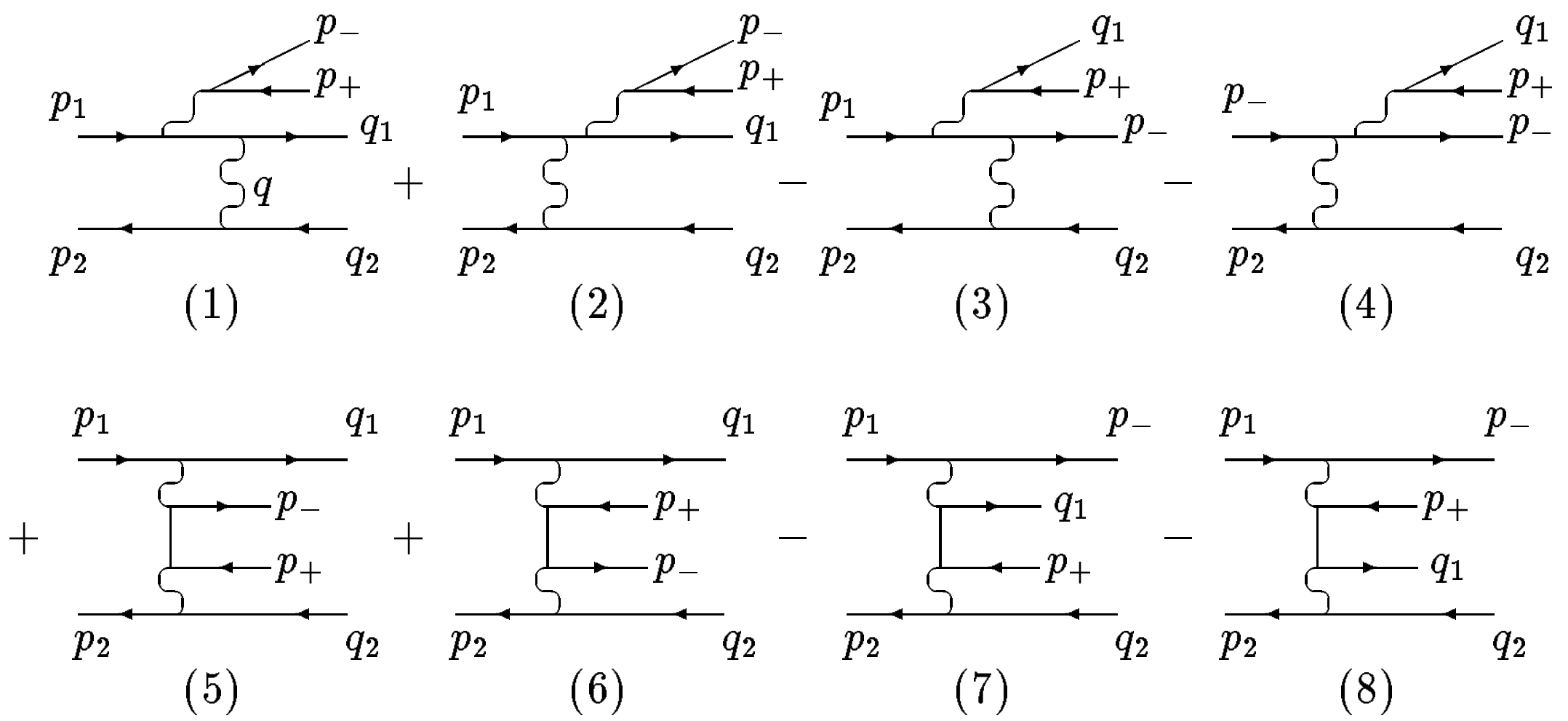

Figure 1: The Feynman diagrams giving logarithmically enhanced contributions in the kinematical region where the created pair goes along the electron direction. The signs represent the Fermi-Dirac statistics of the interchanged fermions.

| Channel | e e | $\mu\mu$ | $\tau\tau$ | $c\bar{c}$ | $u\bar{u}, d\bar{d}, s\bar{s}$ | total |
|---|---|---|---|---|---|---|
| $\sigma$ (nb) | 0.006 | 0.006 | 0.0008 | 0.0005 | 0.0011 | 0.0144 |

Table 3: Double Tag cross sections for fermion pair production from multiperipheral graphs. $\sqrt{s}$ = 91.2 GeV, 30 mrad $< \theta_{e^+}, \theta_{e^-} <$ 60 mrad. For $u, d, s$ quarks $W_{\gamma\gamma} > 4$ GeV. The uncorrected Born cross section $\sigma_{Born}$ is 104 nb.

The analytical formulae for virtual, soft, hard and total pair production contributions can be found in [7, 8, 15]. Numerical results for the pair contribution cross sections based on these formulae can be obtained by using the code `NLLBHA` (see below for a description of the code). The leading term can be described by the electron structure function $D_e^{\bar{e}}(x)$ [16–22]. Numerical results can be found in Refs. [2, 7, 8, 15, 23]. The contribution to SABH of the process of pair production accompanied by photon emission when both, pair and photons, may be real and virtual has also been analyzed and the relevant analytical formulae are given in [2, 23].

With the help of a Monte Carlo generator [24, 25], a dedicated study has been done for the contribution of the multiperipheral graphs, Fig. 1 (5–8), being for many kinematical setups the dominant mechanism of pair production. The total cross sections for the production of fermion pairs as detailed in Table 3 were obtained. The total contribution from the multiperipheral graphs is then estimated to be $1.4 \times 10^{-4} \sigma_{Born}$, with a relative error (from MC statistics) of

---

$e^+e^-$ one.

[4]It can be verified [7, 8, 15] that the interference between the amplitudes describing the production of pairs moving in the electron direction and the positron one cancels. This is known as up-down (interference) cancelation.

| $z_{min}$ | 0.3 | 0.4 | 0.5 | 0.6 | 0.7 | .3/175GeV | .7/175GeV |
|---|---|---|---|---|---|---|---|
| $10^4\sigma^{LL}_{NN}/\sigma_{Born}$ | −3.619 | −3.655 | −3.707 | −3.807 | −4.191 | −4.185 | −4.884 |
| $10^4\sigma^{LL}_{WW}/\sigma_{Born}$ | −2.748 | −2.798 | −2.883 | −3.175 | −3.771 | −3.177 | −4.405 |
| $10^4\sigma^{LL}_{NW}/\sigma_{Born}$ | −2.142 | −2.191 | −2.264 | −2.478 | −3.064 | −2.489 | −3.603 |

Table 4: LL Non-Singlet $e\bar{e}$ pair correction to SABH, SiCAL angular cuts WW: $1.5° - 3.15°$, NN: $1.61° - 2.8°$, in $10^4\times$ Born units, $\sqrt{s} = 91.1888$ GeV (175 GeV for last two entries), $z_{min} = s'_{min}/s$.

$\pm 50\%$. This correction, which still does not take into account a further reduction factor of $\simeq 20$ coming from a cut on the acoplanarity angle of the detected $e^\pm$, is thus negligible for SABH [26].

**(2)** The LL calculation of photonic corrections to SABH of Ref. [27] has been extended to pair corrections in [28]. Analytical formulae for arbitrary asymmetric angular cuts, for both Singlet and Non-Singlet corrections have been given in [28][5]. These formulae, based on [22], include both pairs and photons up to the exponentiated second or third order. The semianalytical program `BHPAIR` based on this calculation has been written [28]. Numerical results for the LCAL type angular cuts have been given in [28]. For the SiCAL type angular cuts the Singlet contribution is negligible (below $5\times 10^{-5}\sigma_{Born}$) and the Non-Singlet contribution (up to third order with exponentiation) is calculated in Table 4, also for the LEP2 energies. The strong dependence of the result on angular cuts (WW, NN or NW) may indicate significant effects due to more realistic ES's. This can only be analyzed with the MC simulation. Such a MC program has already been constructed [29]. This program, being an extension of the `BHLUMI` MC code [30], is based on the extension of the YFS resummation of soft photons to the resummation of infrared and collinear pairs, cf. [31]. Preliminary results [29] show that a calorimetric ES reduces further the pair correction of Table 4.

To summarize, numerical values of pair corrections as given in [7, 8, 15], [2, 23] and Table 4 agree within $4 \times 10^{-4}\sigma_{Born}$ for the NN and WW cuts. The total contribution from pairs and multiperipheral diagrams for the energy cut in the experimentaly interesting range $0.3 < x_c < 0.7$ is also at most $4 \times 10^{-4}\sigma_{Born}$. With the help of a MC simulation of a realistic ES, one should be able to control the pair contribution with an accuracy of $3 \times 10^{-4}\sigma_{Born}$, or better. A similar conclusion is to be expected also for the LEP2 energies.

## 2.4 Vacuum polarization

Vacuum polarization contributes about 5.3% and 4%, respectively, to the $e^+e^-$ cross-section in the angular region of the first and second generation of the luminosity detectors at LEP [3, 32]. The leptonic part of this contribution is known with excellent precision. The quark part, however, is more difficult since the quark masses are not unambiguously defined and perturbative QCD cannot be used for reliable calculations [33–35]. Therefore this part is calculated using a dispersion integral of $R_{had}$

$$R_{had} = \frac{\sigma(e^+e^- \to hadrons)}{\sigma(e^+e^- \to \mu^+\mu^-)} \tag{1}$$

measured experimentally.

[5] Extending further the analysis of Ref. [28], with the help of the 'parton-like' picture together with appropriate choices of structure functions and hard scattering cross-sections, one can calculate the other pair creation mechanisms, including the multiperipheral one, as well as other leptonic backgrounds to SABH resulting from the 'charge blindness' of the detectors. This analysis will appear elsewhere [29].

| $\Theta$ (rad) | $\lvert t\rvert$ ($GeV^2$) | Ref. [36] | (a) Ref. [33] | (b) Ref. [34] | (b)–(a) |
|---|---|---|---|---|---|
| .020 | .83 | −.00345 | $-.00340 \pm .00008(2.5\%)$ | $-.00339 \pm .00013(3.9\%)$ | .00002 |
| .030 | 1.87 | −.00505 | $-.00494 \pm .00014(2.8\%)$ | $-.00493 \pm .00020(4.1\%)$ | .00001 |
| .040 | 3.33 | −.00629 | $-.00612 \pm .00019(3.1\%)$ | $-.00613 \pm .00025(4.1\%)$ | −.00001 |
| .050 | 5.20 | −.00729 | $-.00711 \pm .00024(3.4\%)$ | $-.00714 \pm .00030(4.2\%)$ | −.00003 |
| .060 | 7.48 | −.00812 | $-.00795 \pm .00027(3.5\%)$ | $-.00801 \pm .00034(4.3\%)$ | −.00006 |
| .070 | 10.18 | −.00889 | $-.00869 \pm .00030(3.5\%)$ | $-.00876 \pm .00038(4.4\%)$ | −.00006 |
| .080 | 13.30 | −.00963 | $-.00936 \pm .00033(3.5\%)$ | $-.00941 \pm .00040(4.3\%)$ | −.00005 |
| .090 | 16.83 | −.01029 | $-.00997 \pm .00035(3.5\%)$ | $-.01000 \pm .00043(4.3\%)$ | −.00003 |
| .100 | 20.77 | −.01089 | $-.01052 \pm .00037(3.5\%)$ | $-.01058 \pm .00045(4.3\%)$ | −.00006 |
| .110 | 25.13 | −.01144 | $-.01103 \pm .00039(3.5\%)$ | $-.01110 \pm .00047(4.2\%)$ | −.00007 |
| .120 | 29.90 | −.01194 | $-.01150 \pm .00040(3.5\%)$ | $-.01157 \pm .00049(4.2\%)$ | −.00008 |
| .130 | 35.08 | −.01241 | $-.01193 \pm .00042(3.5\%)$ | $-.01201 \pm .00050(4.2\%)$ | −.00008 |

Table 5: The hadronic part of the vacuum polarization contribution to the small-angle Bhabha scattering as a function of the scattering angle (and corresponding momentum transfer t). In column 4 and 5 also the ratio of the error to the value of the hadronic contribution is given in brackets. The last column gives the difference between the results of Refs. [34] and [33].

Recently, several reevaluations of the hadronic contribution to the QED vacuum polarization have been performed, mainly to determine the effective QED coupling $\alpha(m_Z^2)$ [33, 37–42] and the anomalous magnetic moment (g-2) of the leptons [33]. At the same time, the vacuum polarization contribution to the small-angle Bhabha scattering has been recalculated [33, 34]. Table 5 compares the results of these two calculations of the hadronic contribution in the angular region of small-angle Bhabha scattering used at LEP for the luminosity measurements. They are in excellent agreement, as is evident from the very small differences listed in the last column. In brackets, the error is given as a percentage of the total hadronic contribution. We see that the error of Ref. [33] varies between 63% and 83% of that of Ref. [34] in the angular region presented here. Numbers have been obtained with the help of FORTRAN routines HADR5 [33] and REPI [34] available from the authors. Finally the values of the previously used hadronic contribution from Ref. [36] are also shown.

Fig. 2 from Ref. [34] shows the contribution of different energy regions of R to the value of the hadronic contribution and its error while the Fig. 3 from Ref. [33] shows the uncertainty of the hadronic vacuum polarization contribution to the calculation of the small-angle Bhabha scattering as a function of the momentum transfer.

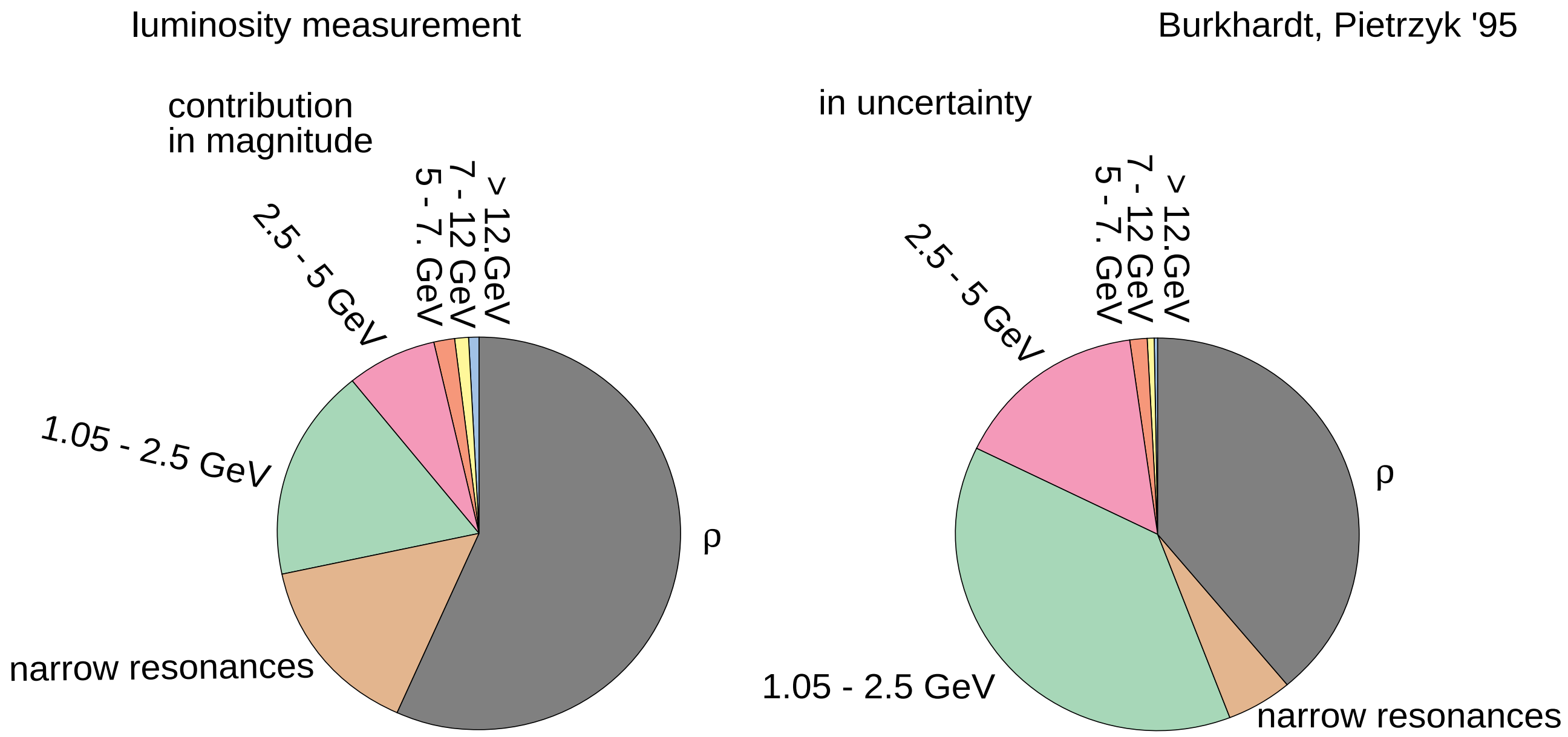

Figure 2: Relative contributions to $\Delta\alpha(t = -1.424\ \mathrm{GeV}^2)$ in magnitude and uncertainty from the Ref. [34].

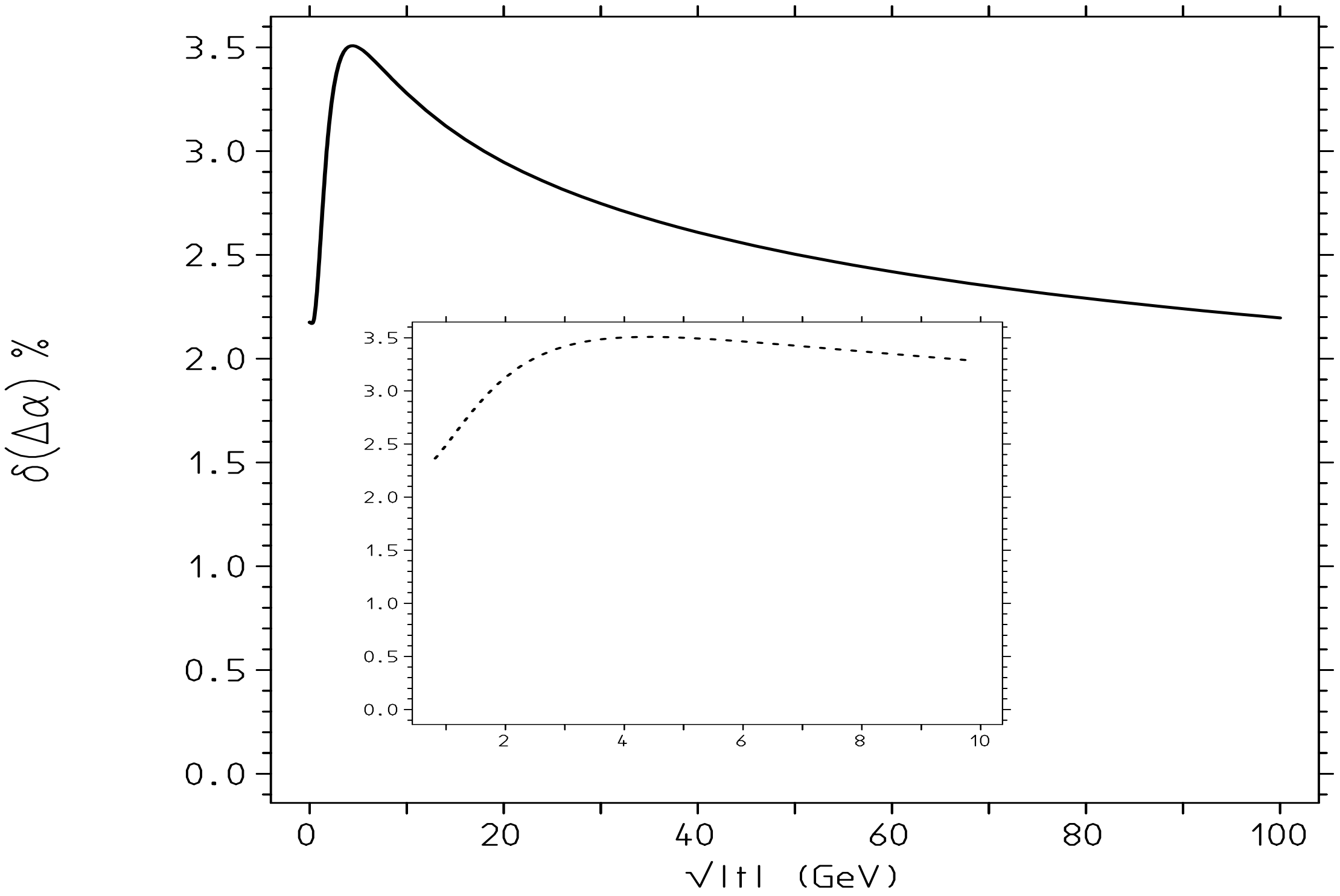

Figure 3: Relative uncertainty in percent of the hadronic vacuum polarization contribution as a function of the momentum transfer in the small-angle Bhabha scattering calculation from the Ref. [33].

| $\Theta$ (rad) | Ref. [36] | Ref. [33] | Ref. [34] | $\frac{hadronic}{total}$ (%) |
|---|---|---|---|---|
| .020 | −.01590 | −.01585 ± .00008 | −.01583 ± .00013 | 21 |
| .030 | −.01877 | −.01866 ± .00014 | −.01865 ± .00020 | 26 |
| .040 | −.02095 | −.02078 ± .00019 | −.02079 ± .00025 | 30 |
| .050 | −.02271 | −.02252 ± .00024 | −.02255 ± .00030 | 32 |
| .060 | −.02418 | −.02400 ± .00027 | −.02406 ± .00034 | 33 |
| .070 | −.02551 | −.02531 ± .00030 | −.02537 ± .00038 | 35 |
| .080 | −.02674 | −.02647 ± .00033 | −.02652 ± .00040 | 36 |
| .090 | −.02785 | −.02753 ± .00035 | −.02756 ± .00043 | 36 |
| .100 | −.02886 | −.02849 ± .00037 | −.02855 ± .00045 | 37 |
| .110 | −.02979 | −.02938 ± .00039 | −.02945 ± .00047 | 38 |
| .120 | −.03064 | −.03020 ± .00040 | −.03028 ± .00049 | 38 |
| .130 | −.03144 | −.03096 ± .00042 | −.03104 ± .00050 | 39 |

Table 6: The vacuum polarization contribution to the small-angle Bhabha scattering as a function of the scattering angle. The last column gives the ratio of the hadronic part to the total vacuum polarization contribution.

| Generation | typical $\Theta$ (rad) | Ref. [33] | Ref. [34] |
|---|---|---|---|
| first | .060 | .0003 | .0004 |
| second | .030 | .0005 | .0007 |

Table 7: Summary of the uncertainty of the vacuum polarization calculation for the first and second generation of the luminosity detectors of LEP according to Ref. [33, 34].

The total vacuum polarization contribution is obtained as sum of the leptonic contribution and the hadronic one. It is shown in Table 6. The contribution of the vacuum polarization error to the total error of the luminosity measurement is about twice the error given in the Table 6. The typical angular region of the first and second generation of the LEP luminosity detectors is 60 and 30 mrad, respectively [3]. The contribution of the vacuum polarization error to the luminosity calculation for the LEP detector is given in Table 7.

The vacuum polarization correction and its uncertainty are smaller for the lower angles covered by the second generation of luminosity detectors.

In conclusion, the error of the hadronic contribution of Ref. [34] makes a negligible contribution to the total error of the calculation of the small-angle Bhabha scattering. The error of Ref. [33] is even smaller. Thus the error of Ref. [34] can be considered as a conservative one.

## 2.5 Brief characteristics of the programs/calculations

Here we will very briefly summarize the basic features of the codes involved in the SABH comparisons. The only aim of the following is to just settle the frame, and not to give an exhaustive description of the codes, which can be found in the original literature and/or in the dedicated write-up's at the end of the present report.

BHAGEN95 [43] – It is a Monte Carlo integrator for both small- and large-angle Bhabha scattering. It is a structure function based program for all orders resummation, including complete photonic $\mathcal{O}(\alpha)$ and leading logarithmic $\mathcal{O}(\alpha^2 L^2)$ corrections in all channels.

BHLUMI [44] – Full scale Monte Carlo event generator for small-angle Bhabha scattering. It includes multiphoton radiation in the framework of YFS *exclusive exponentiation*. Its matrix element includes complete $\mathcal{O}(\alpha)$ and $\mathcal{O}(\alpha^2 L^2)$. The program provides the full event in terms of particle flavors and their four-momenta with an arbitrary number of radiative photons.

LUMLOG – It is a Monte Carlo event generator for SABH (part of BHLUMI, see [44]). Photonic corrections are treated at the leading logarithmic level at the *strictly collinear* and inclusive way. Structure functions exponentiated up to $\mathcal{O}(\alpha^3 L^3)$ are included (and without exponentiation up to $\mathcal{O}(\alpha^2 L^2)$).

NLLBHA [2,23] – It is the FORTRAN translation of a fully analytical up to $\mathcal{O}(\alpha^2)$ calculation, including all the next-to-leading corrections. It is also able to provide $\mathcal{O}(\alpha^3 L^3)$ photonic corrections and light pair corrections including simultaneous photon and light pair emission. *Not an event generator.*

OLDBIS – Classical Monte Carlo event generator for SABH from PETRA times [45] (the modernized version is incorporated in the BHLUMI set [44]). It includes photonic corrections at the exact $O(\alpha)$.

OLDBIS+LUMLOG – It is the well known "tandem" developed in order to take into account higher order corrections (LUMLOG) on top of the exact $O(\alpha)$ result (OLDBIS). The matching between $O(\alpha)$ and higher orders is realized in an *additive* form.

SABSPV [46] – It is a new Monte Carlo integrator, designed for small-angle Bhabha scattering. It is based on a proper matching of the exact $O(\alpha)$ cross section for $t$-channel photon exchange and of the leading logarithmic results in the structure function approach. The matching is performed in a *factorized* form, in order to preserve the classical limit.

## 2.6 Experimental event selection and theory uncertainty in luminosity measurements

In this section we discuss the interplay between experimental selection and higher-order radiative corrections. All numerical examples are for LEP1 at Z peak energy. The discussion of the results is generally limited to LEP1 but using "scaling rules" from the introduction one may easily extend it LEP2. In particular one has to remember that third order LL corrections have the strongest energy dependence, and going from the Z-peak to the highest LEP2 energy introduces in them a factor of almost two.

In this subsection three different event generators are used: **i)** a generator based on a complete first-order calculation OLDBIS [6, 30], which has at most one photon radiated; it includes $\mathcal{O}(\alpha)$ and $\mathcal{O}(\alpha L)$; **ii)** a generator based on a leading-logarithmic third-order exponentiated calculation LUMLOG [6, 30]; it includes $\mathcal{O}(\alpha L)$, $\mathcal{O}(\alpha^2 L^2)$, $\mathcal{O}(\alpha^2 L^2)$ in strictly collinear approximation; the 4-momenta of the final state photons are added to the electrons; **iii)** a truly multi-photon generator based on an exponentiated calculation (BHLUMI) [6, 30]; it includes complete $\mathcal{O}(\alpha)$, $\mathcal{O}(\alpha L)$ and $\mathcal{O}(\alpha^2 L^2)$ while $\mathcal{O}(\alpha^2 L)$ and $\mathcal{O}(\alpha^3 L^3)$ are incomplete; it generates explicitly 4-momenta of all photons above an arbitrary (user-defined) energy threshold, typically a fraction k° (typically $10^{-4}$) of the beam energy. The Bhabha cross section calculated with BHLUMI will be compared to the one calculated with the hybrid calculation consisting of OLDBIS plus higher-order contributions from LUMLOG (LUMLOG$_{HO}$). The cross section differences BHLUMI − OLDBIS and BHLUMI − (OLDBIS + LUMLOG$_{HO}$) are studied as a function of variations in the event selection parameters. Note that BHLUMI − OLDBIS is dominated by $\mathcal{O}(\alpha^2 L^2)$, $\mathcal{O}(\alpha^2 L)$ and $\mathcal{O}(\alpha^3 L^3)$ while BHLUMI − (OLDBIS + LUMLOG$_{HO}$) is dominated by $\mathcal{O}(\alpha^2 L)$ and $\mathcal{O}(\alpha^3 L^3)$.

Only the QED t-channel part of the generators is used, with photon vacuum-polarization switched off. We use an improved version of the BHLUMI event generator as discussed in Ref. [44]. BHLUMI − OLDBIS is used to estimate the higher-order contributions. We choose BHLUMI because the BHLUMI Monte Carlo distributions are in excellent agreement with the data distributions for all LEP experiments [47–52] A quantitative measurement of doubly radiative events [53] has shown consistency with the BHLUMI expectations and also with OLDBIS + LUMLOG$_{HO}$ expectations, while OLDBIS alone fails to describe this contribution, as expected. However, although the MC differential distributions agree with the data, the absolute scale of the integrated cross section remains uncertain, since the bulk of the radiative corrections are either virtual or involve soft ($< 5$ MeV) photons.

In order to set the scale for the following numerical investigation let us remind the reader that the LEP1 experiments have reached in 1993-94 a systematic experimental uncertainty in the measuring the SABH luminosity cross section better than 0.10% [3–5].

### 2.6.1 Reference event selections

We define an imaginary detector, consisting in a pair of cylindrical calorimeters covering the region between 62 and 142 mm radially out from the beam pipe centre and located at 2460 mm from the interaction point, at opposite sides of it. The beams are pointlike and centered within the beam pipe. The calorimeters are each divided into 32 azimuthal segments, subdivided into 32 radial pads. A parton (electron or photon) deposits all its energy in the pad it hits. Photons and electrons from Bhabha events that hit the detector within a region of ±16 radial pads and 5 azimuthal segments centered on the pad struck by the largest energy parton are combined into a cluster. The cluster energy is the pad energy sum. Coordinates of the cluster centroid are the energy weighted average polar coordinates $(R, \phi)$, summing over all pads in the cluster. Partons falling outside the principal cluster can originate secondary clusters, with no overlap. Only one cluster, the most energetic of all clusters, is used. Bhabha events are selected using the cluster energy $E_{cluster}$ and the radial coordinate of the cluster centroid in both calorimeters.

We then define a reference small-angle selection for Bhabha events (RSA selection). The radial acceptance edges for Bhabha events are set at pad boundaries. The "Wide" acceptance boundary extends up to two pads away from the detector inner and outer edges ($27.236 < \theta < 55.691$ mrad). The "Narrow" acceptance boundary extends up to six pads away from the detector inner and outer edges ($31.301 < \theta < 51.626$ mrad). A similar angular range is covered by the OPAL, L3, ALEPH luminometers [54–56]. An event is selected when the cluster coordinates are within the Wide acceptance at one side (side 1) and within the narrow acceptance at the opposite side (side 2). Events must satisfy the criterion $0.5(x_1 + x_2) > 0.75$, with $x = E_{cluster}/E_{beam}$. Selection criteria are also applied on the acoplanarity (0.2 rad) and the acollinearity (10 mrad) between the electron and positron clusters.

Another selection is also considered, similar to the previous one but extending over the angular range covered by the DELPHI luminometer [57] (RLA: reference large-angle selection). The calorimeters are located at 2200 mm from the interaction point and cover radially the region between 6.5 and 41.7 cm. A cluster is formed starting from the most energetic particle hitting the calorimeter and considering all particles whose angular distance $(\Delta\theta, \Delta\phi)$ (in radians) from the initial one satisfies the two shower separation condition (determined from the comparison with the data) $(\Delta\theta/0.03)^2 + (\Delta\phi/0.87)^2 < 1$. The cluster energy is the sum over the energies of all particles inside the cluster, while the cluster coordinates are given by the energy weighted sum of their polar coordinates. Bhabha events are selected by cutting on the minimum cluster energy $\min(x_1, x_2)$, on the acoplanarity ($20^0$) and on the cluster radial coordinate. The radial acceptance is defined on the Narrow side by the condition $43.502 < \theta < 113.151$ mrad and on the Wide side by the condition $38.629 < \theta < 126.592$ mrad.

### 2.6.2 Comparison of exponentiated and order-by-order calculations

**First-Order Calculation**

The Bhabha cross-section for the RSA and RLA selections has been calculated with OLDBIS. The results are shown in figure 4 for the RSA selection, where the cross section is subdivided into $x$-bins, separately for the narrow acceptance side and for the large acceptance side ($x_{Narrow}, x_{Wide}$). A sample of $3 \times 10^9$ events is used. The total Bhabha cross section within the RSA acceptance is $75.589 \pm 0.009$ nb. Displacing the generation minimum angle $\theta_{min}^{gen}$ from 10.4 mrad as recommended in [6, 30] to 5.2 mrad changes the accepted cross section by 0.0039(6) nbarns. No sizeable k° ($=E_\gamma/E_{beam}$) dependence is observed when varying k° from $10^{-4}$ to $10^{-5}$.

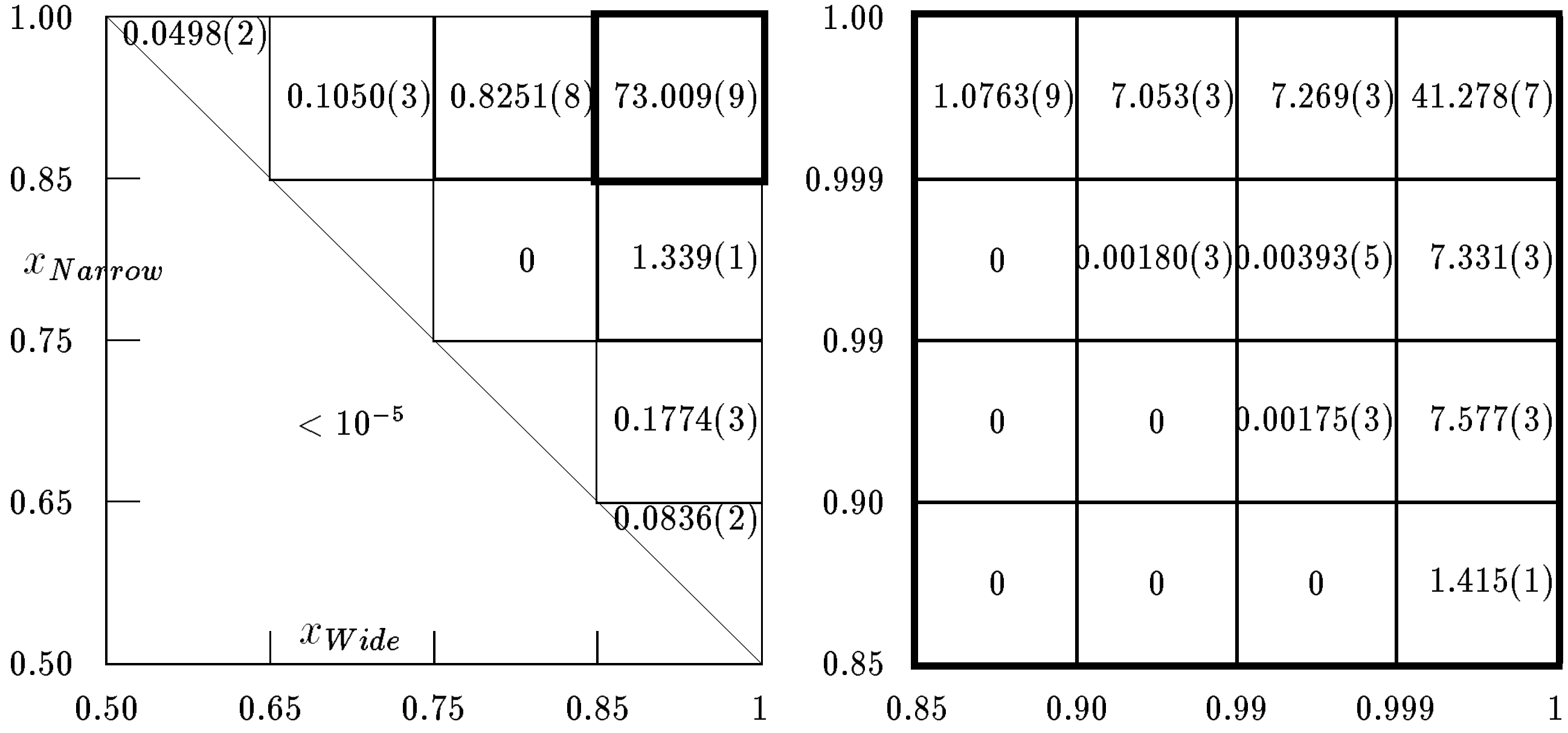

Figure 4: OLDBIS Bhabha cross section (nb) in phase space bins for the RSA selection (see text).

**Higher-Order Leading-Log Contribution**

The cross section difference $\mathrm{LUMLOG}_{HO}$= LUMLOG(all orders) − LUMLOG(first order) is used to estimate the higher-order leading-logarithmic contribution (figure 5) for the RSA selection. In LUMLOG only the initial state radiation has an impact on the measured cluster energies and angles, because the 4-momenta of the final state photons are combined together with the electrons. A sample of $2.1 \times 10^9$ events is used. There is a total higher-order leading-log contribution of $0.144 \pm 0.008$ nb to the Bhabha cross section within the RSA acceptance: the higher-order contribution is negative in the phase-space region dominated by singly radiative events; it is positive in the non radiative Bhabha peak and in the phase-space region of hard doubly radiative events.

**Exponentiated Calculation**

The Bhabha cross-section in phase-space bins for the RSA selection obtained with BHLUMI is presented in figure 6. A sample of $1.6 \times 10^9$ events is used. The total Bhabha cross section accepted by the RSA selection is $75.712 \pm 0.006$ nb. The accepted cross section changes by $< 10^{-5}$ when decreasing the $t^{gen}_{min}$ (minimum generated four-momentum transfer squared) value as recommended in [6,30] to half of it.

**Comparison of Exponentiated and Order-by-Order Calculations**

The BHLUMI and OLDBIS cross sections differ for the RSA selection by $(0.16 \pm 0.01)\%$, showing that the estimated contribution to the accepted cross section from higher-order radiative effects is very small. This estimate is also in reasonable agreement with the $\mathrm{LUMLOG}_{HO}$ expectation of $(0.19 \pm 0.01)\%$.

A similar study for the RLA selection results in a BHLUMI − OLDBIS relative difference

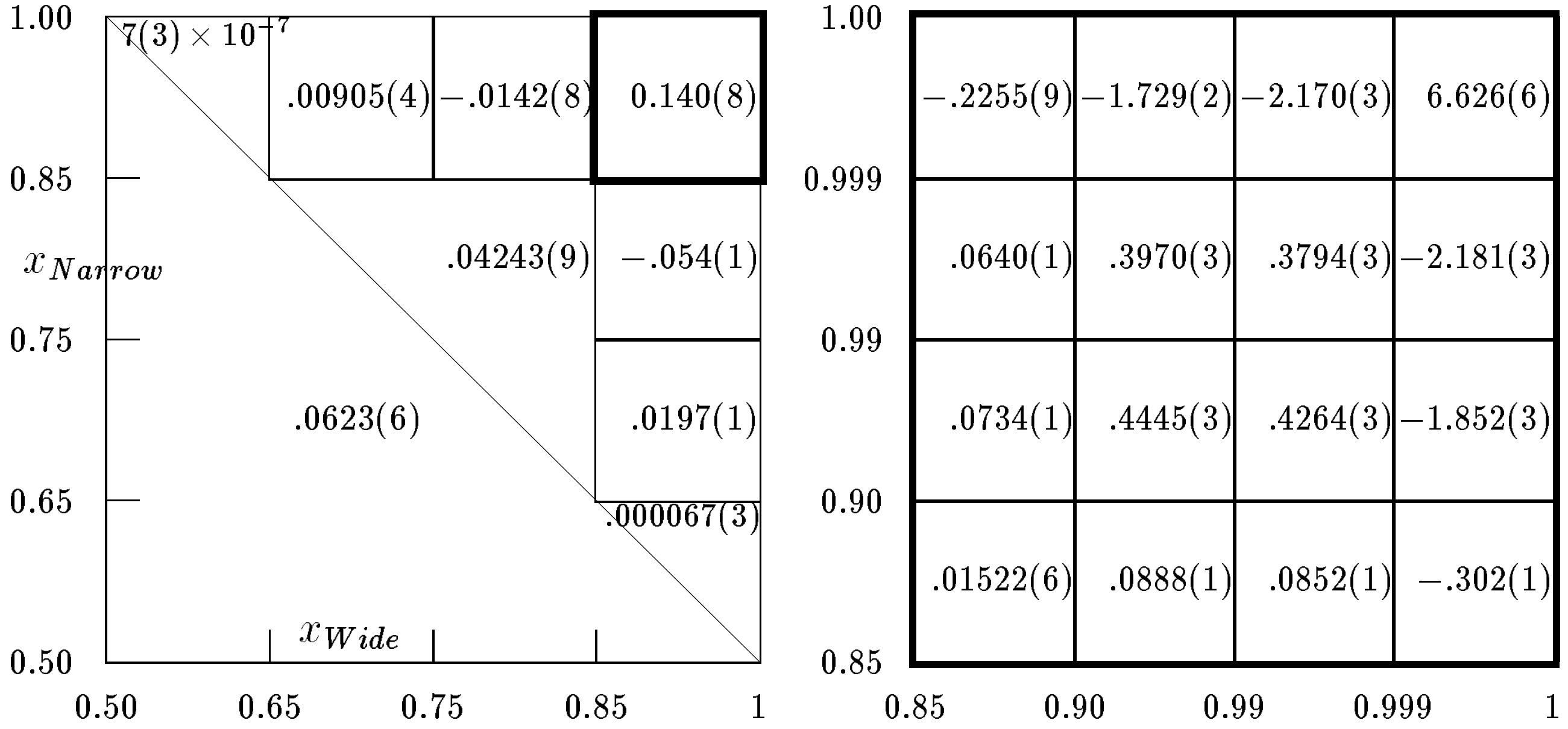

Figure 5: LUMLOG higher-order contribution to the Bhabha cross section (nb) in phase space bins for the RSA selection (see text).

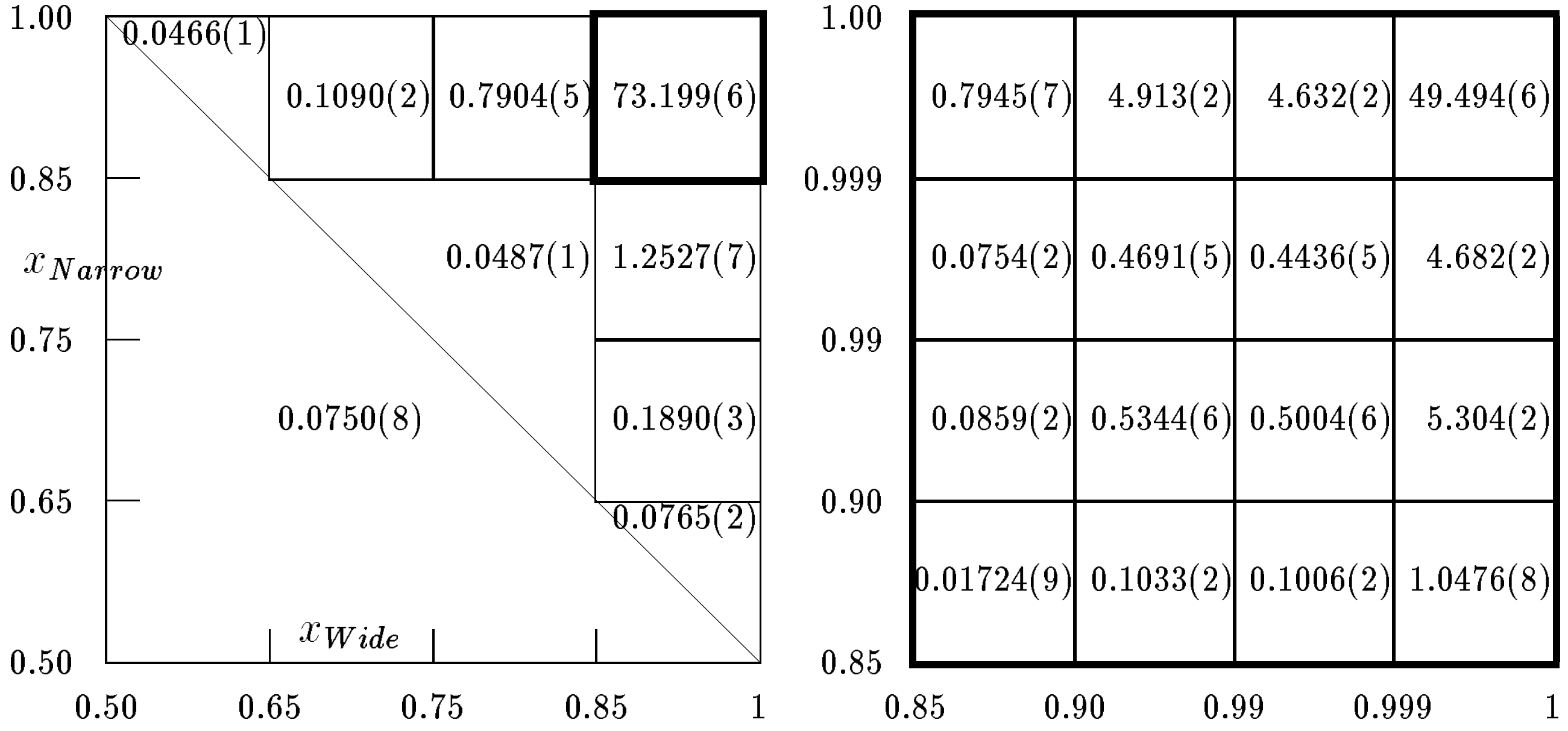

Figure 6: BHLUMI Bhabha cross section (nb) in phase space bins for the RSA selection (see text).

of $(-0.08 \pm 0.01)\%$ to be compared with a LUMLOG$_{HO}$ expectation of $(-0.03 \pm 0.01)\%$.

### 2.6.3 Dependence on energy and acollinearity cuts

The cross section relative difference (BHLUMI − OLDBIS)/BHLUMI(RSA), where BHLUMI(RSA) refers to the RSA selection, is studied in table 8 for several selection criteria on energy and acollinearity. With $x^{cut}_{min}$ we mean that the energy cut $min(x_1, x_2) > x^{cut}_{min}$ is applied. Through transverse momentum conservation, energy and acollinearity cuts are strongly correlated in events with initial state radiation. The relative difference BHLUMI − OLDBIS is indicative of the higher-order contribution, which clearly appears in table 8 to be huge for large $x^{cut}_{min}$. It becomes progressively smaller for smaller $x^{cut}_{min}$. It should be stressed that the h.o. corrections are small (at the per mille level) over a very broad region of $x^{cut}_{min}$ and acollinearity.

A second estimate of the Bhabha cross section with higher-order radiative corrections can be obtained with OLDBIS + LUMLOG$_{HO}$. The three generator relative difference (BHLUMI − (OLDBIS + LUMLOG$_{HO}$))/BHLUMI(RSA) in table 8 shows that the h.o. corrections in BHLUMI and in LUMLOG track each other very well, giving confidence that the h.o. contributions are in fact small when they are estimated to be so. The unstable region is limited to very large $x^{cut}_{min}$. The BHLUMI and OLDBIS + LUMLOG$_{HO}$ Bhabha cross sections agree at the 0.1% level over an extremely broad range of energy and acollinearity cuts.

The cross section differences BHLUMI − OLDBIS and BHLUMI − (OLDBIS + LUMLOG$_{HO}$) for the RSA selection change by $(-0.013 \pm 0.009)\%$ when the acoplanarity cut is not applied.

For the RLA selection the cross section differences BHLUMI − OLDBIS and BHLUMI − (OLDBIS + LUMLOG$_{HO}$) normalized to the BHLUMI result are shown in table 9 as a function of the cut on $x^{cut}_{min}$. The higher-order contribution to the Bhabha cross section for the RLA selection both in BHLUMI and in LUMLOG is very small over a broad range of $x^{cut}_{min}$.

BHLUMI−OLDBIS

| | *Acollinearity cut (rad)* | | |
|---|---|---|---|
| $x^{cut}_{min}$ | 0.005 | 0.010 | no cut |
| 0.999 | 11.35(1)% | 10.86(1)% | 10.61(1)% |
| 0.99 | 4.65(1)% | 4.45(1)% | 4.35(1)% |
| 0.90 | 0.69(2)% | 0.60(1)% | 0.58(1)% |
| 0.85 | 0.68(1)% | 0.25(1)% | 0.24(1)% |
| 0.75 | 0.75(1)% | 0.12(1)% | -0.00(1)% |
| triang. | 0.78(1)% | 0.16(1)% | -0.09(1)% |
| 0.50 | 0.83(1)% | 0.26(1)% | 0.06(1)% |

BHLUMI−(OLDBIS+LUMLOG$_{HO}$)

| | *Acollinearity cut (rad)* | | |
|---|---|---|---|
| $x^{cut}_{min}$ | 0.005 | 0.010 | no cut |
| 0.999 | 2.19(2)% | 2.10(1)% | 2.05(1)% |
| 0.99 | 0.98(2)% | 0.94(2)% | 0.92(2)% |
| 0.90 | 0.19(2)% | 0.15(2)% | 0.14(2)% |
| 0.85 | 0.15(2)% | 0.06(2)% | 0.06(2)% |
| 0.75 | 0.13(2)% | -0.00(2)% | -0.03(2)% |
| triang. | 0.12(2)% | -0.03(2)% | -0.09(2)% |
| 0.50 | 0.18(2)% | -0.02(2)% | -0.07(2)% |

Table 8: Cross section differences BHLUMI−OLDBIS and BHLUMI−(OLDBIS+LUMLOG$_{HO}$) normalized to the BHLUMI Bhabha cross section for the RSA selection. The label "triangular" stands for the cut $0.5(x_1 + x_2) > 0.75$.

| $x_{min}^{cut}$ | BHL−OB |
|---|---|
| 0.9 | 0.76(1)% |
| 0.8 | 0.10(1)% |
| 0.7 | -0.06(1)% |
| 0.6 | -0.08(1)% |
| 0.5 | -0.05(1)% |

| $x_{min}^{cut}$ | BHL−(OB+LL$_{HO}$) |
|---|---|
| 0.9 | 0.21(1)% |
| 0.8 | 0.03(1)% |
| 0.7 | -0.03(1)% |
| 0.6 | -0.05(1)% |
| 0.5 | -0.06(1)% |

Table 9: Cross section differences BHLUMI−OLDBIS and BHLUMI−(OLDBIS+LUMLOG$_{HO}$) normalized to the BHLUMI Bhabha cross section for the RLA selection.

### 2.6.4 Wide-Wide, Narrow-Narrow versus Wide-Narrow acceptance

In the reference selections (RFA and RLA) an asymmetric acceptance (Wide on one side and Narrow on the opposite side) is used. All 4 LEP experiments use an asymmetric acceptance for the LEP luminosity measurement. We study in table 10 how the results change when using a symmetric (Wide-Wide or Narrow-Narrow). The BHLUMI − OLDBIS cross section difference becomes large (0.77(1)% for the Narrow-Narrow acceptance). A similar result is also obtained using LUMLOG$_{HO}$ and then the BHLUMI − (OLDBIS + LUMLOG$_{HO}$) difference is small. We thus conclude that the higher-order contributions to the accepted Bhabha cross section, as estimated with BHLUMI or LUMLOG, are largely reduced when using an asymmetric Wide-Narrow acceptance.

| | WN | WW | NN |
|---|---|---|---|
| BHLUMI | 75.712(5)$nb$ | 117.918(6)$nb$ | 73.344(5)$nb$ |
| OLDBIS | 75.589(8)$nb$ | 117.219(9)$nb$ | 72.781(8)$nb$ |
| LUMLOG$_{HO}$ | 0.144(8)$nb$ | 0.568(9)$nb$ | 0.465(8)$nb$ |
| (BHL−OB)/BHL | 0.16(1)% | 0.59(1)% | 0.77(1)% |
| (BHL−OB−LL$_{HO}$)/BHL | -0.03(2)% | 0.11(2)% | 0.13(2)% |

Table 10: Comparison of BHLUMI, OLDBIS and LUMLOG$_{HO}$ Bhabha cross sections for Wide-Narrow, Wide-Wide, Narrow-Narrow event selections. All other cuts as in the RSA selection.

### 2.6.5 Multiple photon radiation

A very relevant property of exclusive exponentiation is that there are many more multi-photon events than expected from perturbation theory at a fixed order in $\alpha$. In a sample of $10^6$ BHLUMI Bhabha events, the events have up to eight photons with energy larger than $k^\circ E_{beam}(\approx 5$ MeV), as shown in figure 7. This may enhance the difference between cross section calculations performed with BHLUMI and with OLDBIS + LUMLOG$_{HO}$. In the following we study the stability of the BHLUMI − OLDBIS and BHLUMI − OLDBIS − LUMLOG(ho) differences in table 8 and in table 9 when varying those parameters in the experimental selection which are sensitive to the presence of many photons.

**Lower Energy Photon Cut-off**

We define a $K_c$ parameter (in MeV) expressing the sensitivity to soft photons: the detector is fully efficient for photons of energy larger than $K_c$. An implicit $K_c$ cut-off is present in BHLUMI at $K_c = k^\circ E_{beam}$ (5 MeV) for the cross sections calculations presented above. The relative variation of the BHLUMI Bhabha cross section when varying $K_c$ is reported in table 11 for the RSA selection and in table 12 for the RLA selection. The effect is at most of $-0.03\%$ for the RSA acceptance in the extreme case of $K_c$=500 MeV. The relative changes in the BHLUMI and OLDBIS cross sections are compared in figures 8 and 9. The large-x region is very different; most of the difference has already disappeared for $x^{cut}_{min}$=0.9. LUMLOG remains unaffected: it has in the output only the electron and positron 4-momenta with the final state photons combined with the electrons/positrons. Hence, the effect on the relative cross section differences BHLUMI − OLDBIS and BHLUMI − (OLDBIS + $\text{LUMLOG}_{HO}$) for the RSA selection is at most $-0.030(4)\%$.

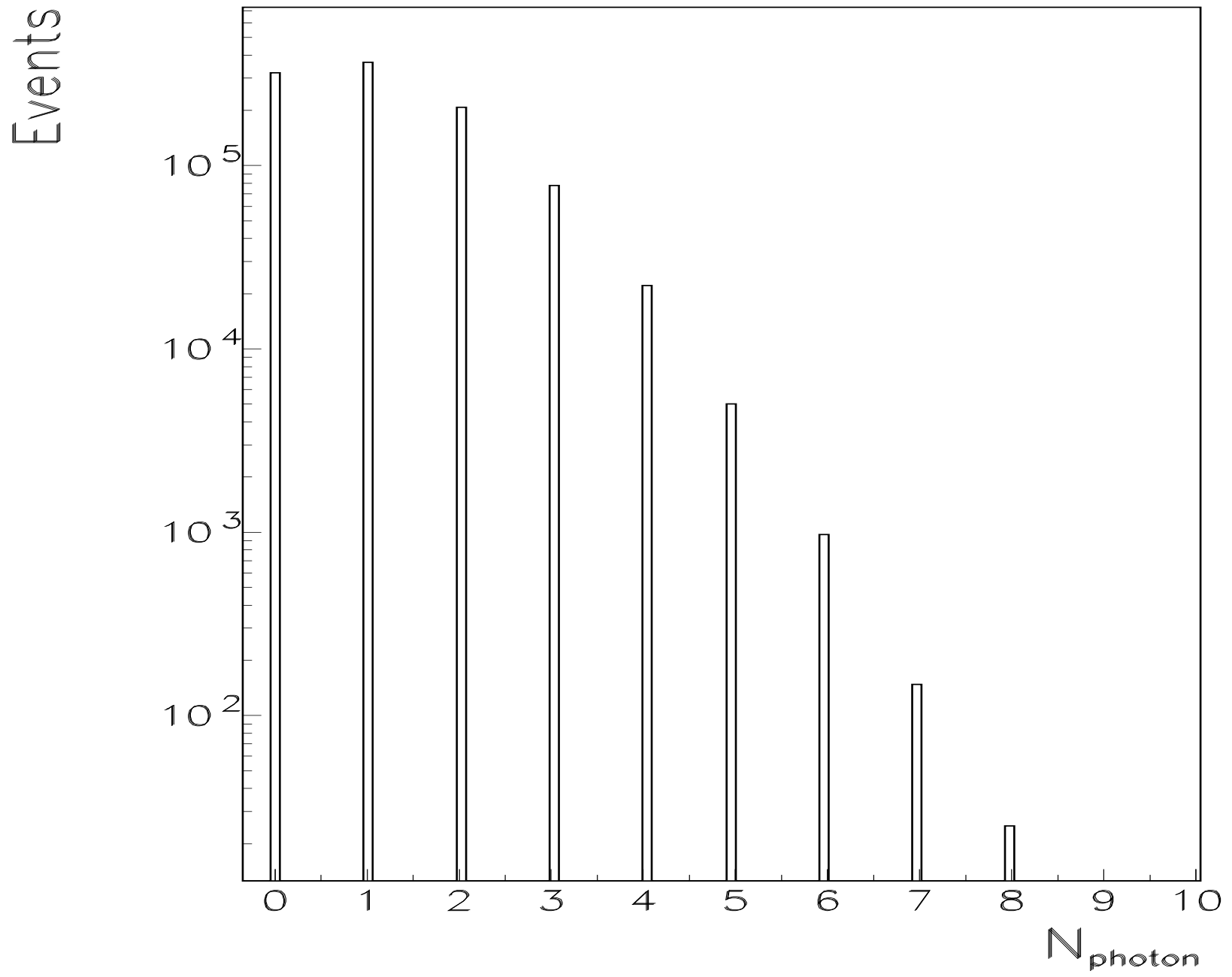

Figure 7: Distribution in number of emitted photons for a sample of $10^6$ unweighted BHLUMI events. (The Removal flag is switched on in BHLUMI, with $K_c = k^\circ E_{beam} = 5$ MeV).

**Cluster Size**

The relative variation of the accepted Bhabha cross section with respect to the RSA selection when changing the cluster size is studied in figure 10 using BHLUMI generated events and using OLDBIS generated events. For large cluster sizes BHLUMI and OLDBIS track each other very well and the BHLUMI − OLDBIS relative difference observed for the RSA selection remains unchanged. On the contrary, for small cluster sizes, the effect of many photons in BHLUMI generated events shows up strongly. The LUMLOG result remains unaffected. Thus, for the RSA selection we can exclude an effect larger than $\pm 0.007\%$ on the BHLUMI − OLDBIS and on the BHLUMI − (OLDBIS + LUMLOG$_{HO}$) cross section differences.

| $x_{min}^{cut}$ | $K_c$ (MeV) | | | |
|---|---|---|---|---|
| | 10 | 50 | 100 | 500 |
| 0.999 | -0.025(4)% | -0.75(2)% | -3.51(5)% | -9.45(8)% |
| 0.90 | -0.001(1)% | -0.002(1)% | -0.004(2)% | -0.029(4)% |
| 0.85 | $< 10^{-5}$ | -0.003(1)% | -0.003(1)% | -0.012(3)% |
| triangular | -0.004(2)% | -0.015(3)% | -0.018(3)% | -0.030(4)% |

Table 11: Variation of the BHLUMI Bhabha cross section when changing the photon minimum detectable energy $K_c$. Normalization is with respect to the RSA selection with $K_c = k^\circ E_{beam} = 5$ MeV. The label "triangular" stands for the cut $0.5(x_1 + x_2) > 0.75$.

| $x_{min}^{cut}$ | $K_c$ (MeV) | | | |
|---|---|---|---|---|
| | 10 | 50 | 100 | 500 |
| 0.9 | -0.0005(2)% | -0.0032(5)% | -0.0067(7)% | -0.033(2)% |
| 0.7 | -0.0003(2)% | -0.0010(3)% | -0.0015(3)% | -0.0085(8)% |
| 0.5 | $< 10^{-6}$ | -0.0002(1)% | -0.0005(2)% | -0.0025(5)% |

Table 12: Variation of the BHLUMI Bhabha cross section when changing the photon minimum detectable energy $K_c$ from $K_c = 5$ MeV for the RLA selection.

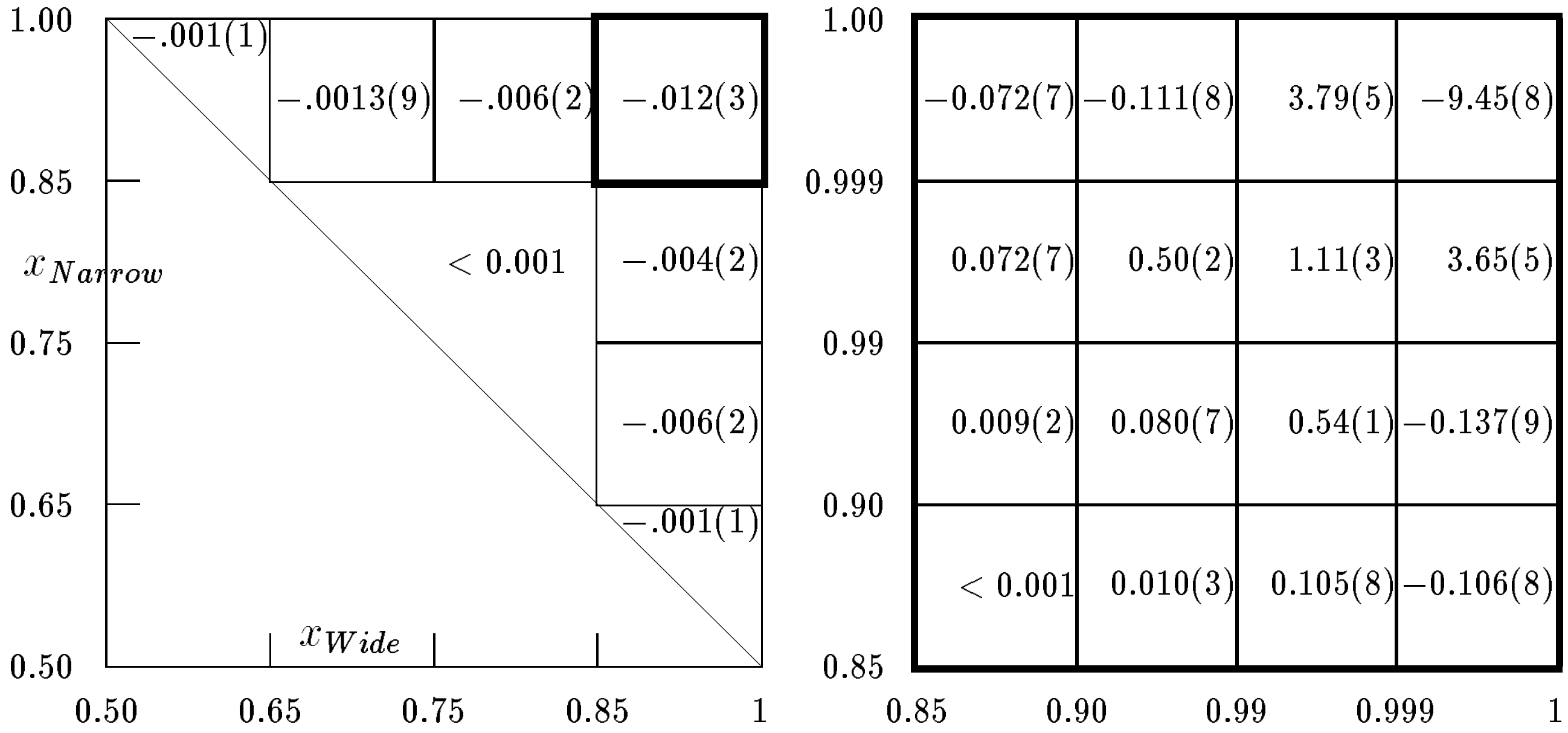

Figure 8: Percentage variation of the BHLUMI Bhabha cross section when setting the photon minimum detectable energy $K_c$ to 500 MeV (see also figure 3) instead of $K_c = 5$ MeV.

| $x_{Narrow}$ \ $x_{Wide}$ | 0.85–0.90 | 0.90–0.99 | 0.99–0.999 | 0.999–1 |
|---|---|---|---|---|
| 0.999–1.00 | $< 0.001$ | 0.35(1) | 7.80(5) | −16.32(8) |
| 0.99–0.999 | 0 | $< 0.001$ | $< 0.001$ | 7.85(5) |
| 0.90–0.99 | 0 | 0 | $< 0.001$ | 0.39(1) |
| 0.85–0.90 | 0 | 0 | 0 | $< 0.001$ |

Figure 9: Percentage variation of the OLDBIS Bhabha cross section when setting the photon minimum detectable energy $K_c$ to 500 MeV (see also figure 1) instead of $K_c = 5$ MeV.

**Cluster Coordinate**

The energy weighting algorithm for extracting the cluster coordinates couples the coordinates to the cluster size. A different coordinate reconstruction algorithm (PADMAX) is then used: we select the pad with the largest energy deposit and use the 4-momentum sum of the partons which enter that pad to calculate an impact point in the pad; the impact point so calculated defines the cluster coordinates, independent of the cluster dimensions. The BHLUMI cross section when changing from $\theta$ ($\phi$) energy weighted coordinates to PADMAX coordinates in the RSA selection changes by $(-0.088 \pm 0.003)\%$. The OLDBIS cross section when changing from $\theta$ ($\phi$) energy weighted coordinates to PADMAX coordinates changes by $(-0.091 \pm 0.005)\%$. The LUMLOG result is unaffected. The effect on the BHLUMI − OLDBIS and on the BHLUMI − (OLDBIS + LUMLOG$_{HO}$) cross section differences in the RSA selection when using the PADMAX coordinates instead of the energy weighted coordinates is $(0.003 \pm 0.006)\%$.

### 2.6.6 Summary

We have shown that there is a strong correlation between the magnitude of the $O(\alpha^2)$ radiative corrections to the Bhabha cross section and distinctive characteristics of the experimental Bhabha event selection. In particular, we have shown that the Bhabha selections used by the LEP experiments to measure the accelerator luminosity minimize the sensitivity to $O(\alpha^2)$ radiative corrections.

BHLUMI

| $\Delta_{PAD}$ \ $N_{SEG}$ | 0 | 1 | 2 | 4 | 8 | 12 | 16(all) |
|---|---|---|---|---|---|---|---|
| 32(all) | −.071(3)% | | +.075(3)% | +.160(4)% | +.270(5)% | | +.310(5)% |
| 16 | −.086(3)% | −.020(1)% | 0 | +.044(2)% | +.109(3)% | +.117(3)% | +.119(3)% |
| 8 | | | −.071(3)% | | | | |
| 4 | | | −.099(3)% | | | | |
| 0 | −.237(5)% | | −.224(5)% | | | | −.221(5)% |

OLDBIS

| $\Delta_{PAD}$ \ $N_{SEG}$ | 0 | 1 | 2 | 4 | 8 | 12 | 16(all) |
|---|---|---|---|---|---|---|---|
| 32(all) | −.028(4)% | | +.074(6)% | +.156(9)% | +.261(12)% | | +.311(13)% |
| 16 | −.053(5)% | −.017(3)% | 0 | +.042(5)% | +.101(7)% | +.118(8)% | +.126(8)% |
| 8 | | | −.069(6)% | | | | |
| 4 | | | −.073(6)% | | | | |
| 0 | −.073(6)% | | −.072(6)% | | | | −.066(6)% |

Figure 10: Relative variation of the accepted Bhabha cross section with respect to the RSA selection when changing cluster radial (PAD's) and azimuthal (SEGments) dimensions. A cluster extends for $\pm\Delta_{PAD}$ pads and $\pm N_{SEG}$ segments around the pad containing the largest energy deposit. A pad subtends a polar angle of about 1 mrad; a segment covers azimuthally an angle of 11.25 degrees. The RSA selection has $\Delta_{PAD} = 16$ and $N_{SEG} = 2$.

The $O(\alpha^2)$ contributions have been estimated using BHLUMI − OLDBIS and LUMLOG$_{HO}$= LUMLOG$_{all-orders}$−LUMLOG$_{first-order}$. The cross section differences BHLUMI − OLDBIS and BHLUMI − (OLDBIS + LUMLOG$_{HO}$) are very small (at the per mille level) in a broad region of phase space around the experimental selections. We have considered two angular ranges $27 < \theta < 57$ mrad and $44 < \theta < 113$ mrad, with a variety of energy and acollinearity cuts. The sensitivity to the possible presence of many photons, predicted by exclusive exponentiation, the effect of small or large cluster sizes and different ways of reconstructing the cluster coordinates have been investigated. Large cluster sizes, rather soft energy cuts and a Wide-Narrow method are very effective in minimizing the cross section differences BHLUMI − OLDBIS and BHLUMI − (OLDBIS + LUMLOG$_{HO}$). Vice versa, these same effects could be used to enhance the sensitivity to the $O(\alpha^2)$ radiative corrections in order to perform measurements and test the theory predictions.

## 2.7 Comparisons of event generators for small-angle Bhabha scattering

In contrast to the previous section, where we have seen results from many variants of ES's with varying cut parameters but for only three types of QED calculations, here we shall limit ourselves to "only" four ES's (two of which very close to realistic experimental situations), but we shall discuss *all the available* theoretical calculations. The outline of this section is the following: the actual comparisons will be presented first at the $\mathcal{O}(\alpha^1)$ level, in order to determine the basic technical precision, and later for more advanced QED matrix elements beyond $\mathcal{O}(\alpha^1)$, in order to explore physical precision. These comparisons will be done first for LEP1 energy and later will be also extended to LEP2 energies.

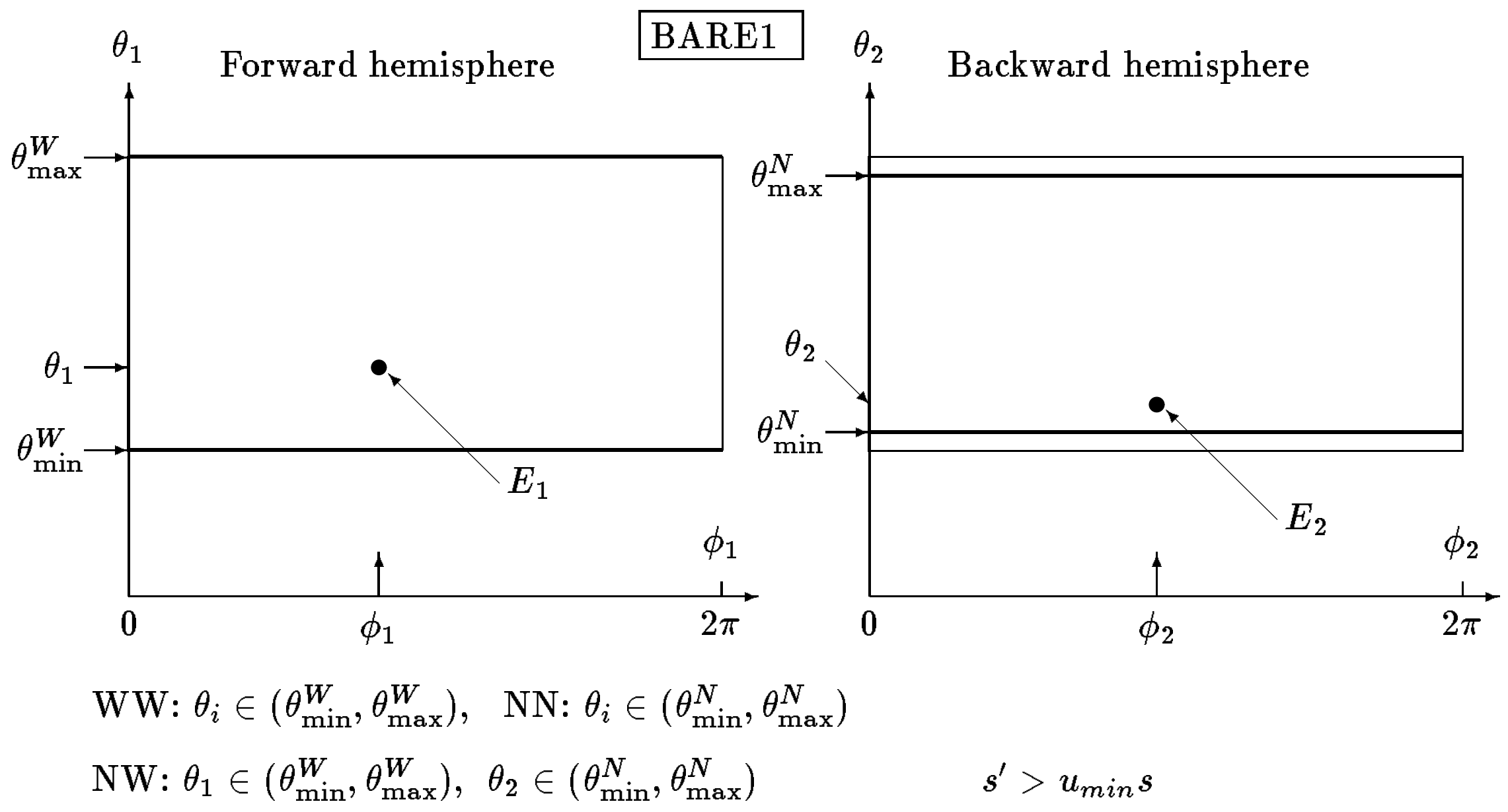

Figure 11: Geometry and acceptance of the simple (non-calorimetric) ES BARE1. This ES restricts polar *angles* $\theta_i$ in the forward/backward hemispheres and requires a certain minimum *energy* to be detected simultaneously in both hemispheres. Photon momentum is not constrained at all. The entire "fiducial" $\theta$-range, i.e. wide (W) range, is $(\theta^W_{\min}, \theta^W_{\max}) = (0.024, 0.058)$ rad and the narrow (N) range is $(\theta^N_{\min}, \theta^N_{\max})$, where $\theta^N_{\min} = \theta^W_{\min} + \delta_\theta$, $\theta^N_{\max} = \theta^W_{\max} - \delta_\theta$ and $\delta_\theta = (\theta^W_{\max} - \theta^W_{\min})/16$. This ES can be symmetric Wide-Wide (WW) or Narrow-Narrow (NN), or asymmetric Narrow-Wide (NW), see the description in the figure. The energy cut $s' > u_{min} s$ involves momenta of outgoing $e^\pm$ ($s' = (q^+ + q^-)^2$) only.

### 2.7.1 Event selections

One cannot talk about the cross section for the small-angle Bhabha (SABH) process without defining precisely all cuts, or, in other terms, without specifying the ES. The most interesting ES is that of the actual experiment. LEP1 and LEP2 experiments employ in the measurement of the small-angle Bhabha scattering cross section a rich family of ES's. They do, however,

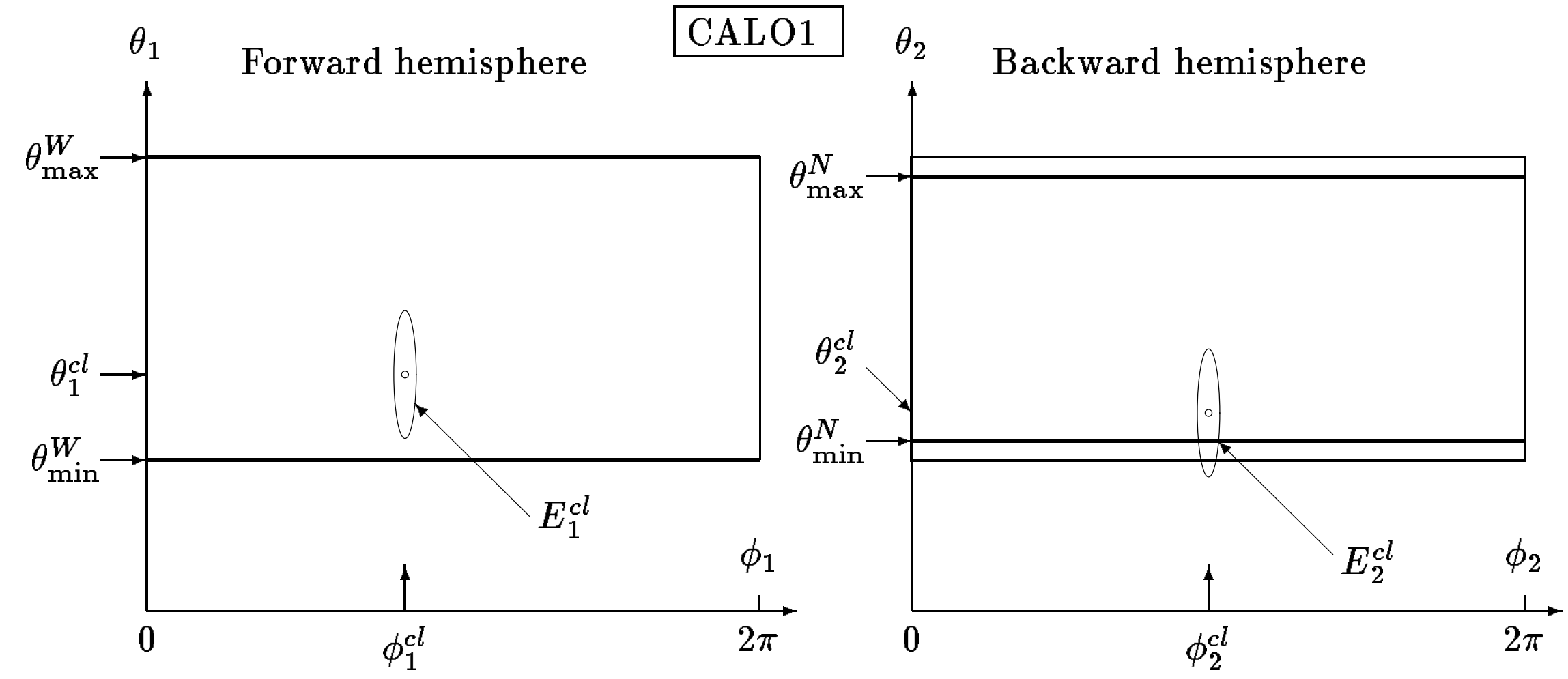

WW: $\theta_i^{cl} \in (\theta^W_{\rm min}, \theta^W_{\rm max})$, NN: $\theta_i^{cl} \in (\theta^N_{\rm min}, \theta^N_{\rm max})$

NW: $\theta_1^{cl} \in (\theta^W_{\rm min}, \theta^W_{\rm max})$, $\theta_2^{cl} \in (\theta^N_{\rm min}, \theta^N_{\rm max})$ $\qquad E_1^{cl} E_2^{cl} > z_{min} E^2_{beam}$

Figure 12: Geometry and acceptance of the calorimetric ES CALO1. This ES restricts polar *angles* $\theta_i$ in the forward/backward hemispheres and requires a certain minimum *energy* to be detected simultaneously in both hemispheres. The entire "fiducial" $\theta$-range, i.e. wide (W) range, is $(\theta^W_{\rm min}, \theta^W_{\rm max}) = (0.024, 0.058)$ rad and the narrow (N) range is $(\theta^N_{\rm min}, \theta^N_{\rm max})$, where $\theta^N_{\rm min} = \theta^W_{\rm min} + \delta_\theta$, $\theta^N_{\rm max} = \theta^W_{\rm max} - \delta_\theta$ and $\delta_\theta = (\theta^W_{\rm max} - \theta^W_{\rm min})/16$. This ES can be symmetric Wide-Wide (WW) or Narrow-Narrow (NN), or asymmetric Narrow-Wide (NW), see the description in the figure. The energy cut involves the definition of the *cluster*: the cluster center $(\theta^{cl}_i, \phi^{cl}_i)$, $i = 1, 2$, is identical to the angular position of the positron in the forward and the electron in the backward hemisphere. The angular "cone" of radius $\delta = 0.010$ rad around $e^\pm$ is called cluster. The cone/cluster in the $\theta, \phi$ plane is an elongated ellipsis, due to smallness of *theta*. The total energy registered in the cluster is denoted by $E^{cl}_i$. (Note that $\phi_1 = \phi_2$ for back-to-back configuration.)

have essential common features. The most important is the "double tag". It means that $e^+$ and $e^-$ are *both* detected with a certain minimum energy and minimum scattering angle in the forward and backward direction, close to the beams. The other important feature of the typical experimental ES is that (except for rare cases) the photons and $e^\pm$ cannot be distinguished – only the combined energy and angle is registered. It is said that the typical experimental ES is *calorimetric*. On the other hand, for comparing theoretical calculations it is useful to deal with simplified ES's, in which only $e^\pm$ are measured and the accompanying bremsstrahlung photons ($e^\pm$ pairs) are ignored. The "double tag" is done on "bare $e^\pm$". Actually, in order to compare efficiently numerical results from the various programs, we employed the family of four ES's connecting in an almost continuous way the experimentally unrealistic (but useful for theorists) examples of ES's to experimentally realistic (but difficult for some class of theoretical calculations) ones. In order to compare theoretical results for SABH, we use one simple non-calorimetric ES called BARE1, see Figs. (11), and three calorimetric ES's called CALO1, CALO2 and SICAL2, with increasing degrees of sophistication. They are defined in Figs. (12,13) and Fig. (14). The last one, SICAL2 of Fig. (14), corresponds very closely to the ES of the real silicon detector of OPAL or ALEPH.

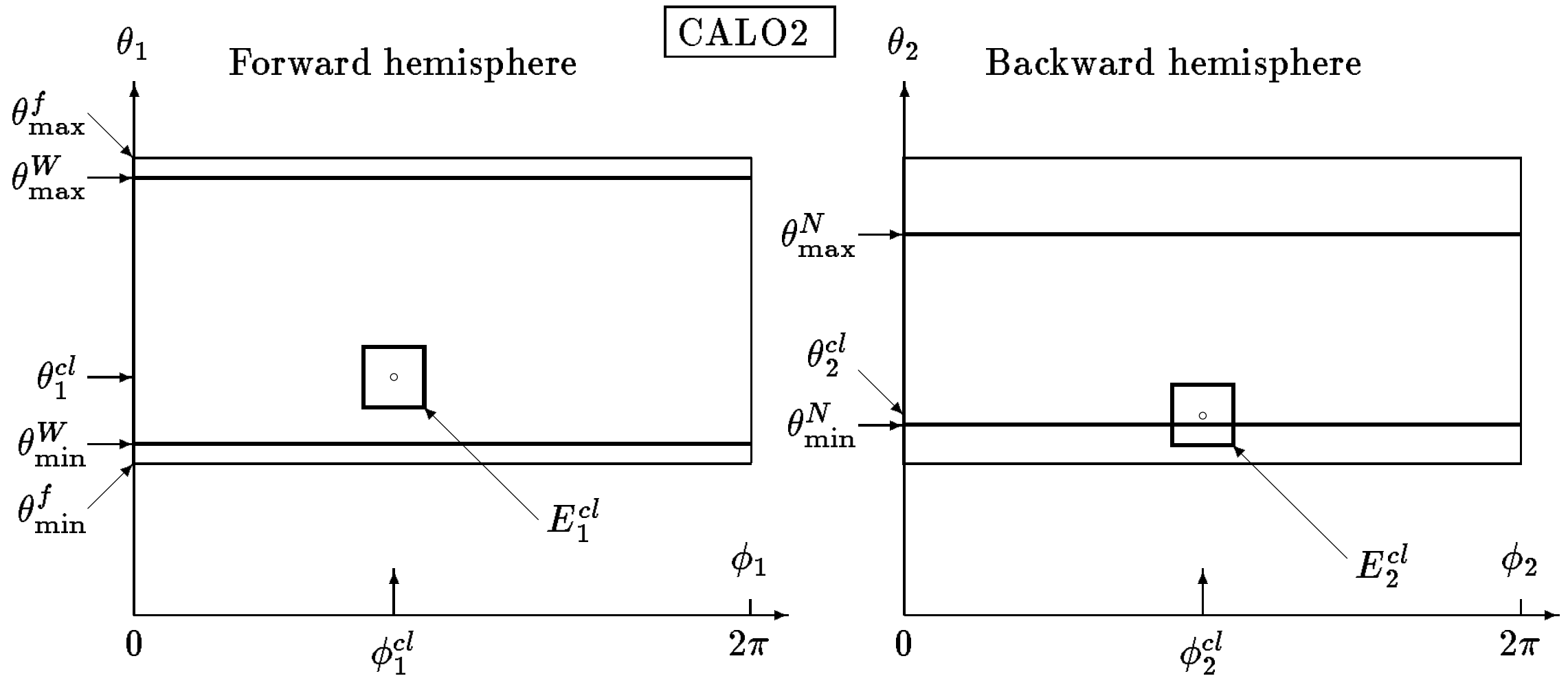

WW: $\theta_i^{cl} \in (\theta^W_{\min}, \theta^W_{\max})$, NN: $\theta_i^{cl} \in (\theta^N_{\min}, \theta^N_{\max})$

NW: $\theta_1^{cl} \in (\theta^W_{\min}, \theta^W_{\max})$, $\theta_2^{cl} \in (\theta^N_{\min}, \theta^N_{\max})$ $\qquad E_1^{cl} E_2^{cl} > z_{min} E^2_{beam}$

Figure 13: Geometry and acceptance of the calorimetric ES CALO2. This ES restricts polar *angles* $\theta_i$ in the forward/backward hemispheres and requires a certain minimum *energy* to be detected simultaneously in both hemispheres. The entire "fiducial" $\theta$-range, $(\theta^f_{\min}, \theta^f_{\max}) = (0.024, 0.058)$ rad, includes the wide (W) range $(\theta^W_{\min}, \theta^W_{\max})$ and the narrow (N) range $(\theta^N_{\min}, \theta^N_{\max})$, where $\theta^W_{\min} = \theta^f_{\min} + \delta_\theta$, $\theta^W_{\max} = \theta^f_{\max} - \delta_\theta$, $\delta_\theta = (\theta^f_{\max} - \theta^f_{\min})/16$, and $\theta^N_{\min} = \theta^f_{\min} + 2\delta_\theta$, $\theta^N_{\max} = \theta^f_{\max} - 4\delta_\theta$. This ES can be symmetric Wide-Wide (WW) or Narrow-Narrow (NN), or asymmetric Narrow-Wide (NW), see the description in the figure. The energy cut involves the definition of the *cluster*: the cluster center $(\theta_i^{cl}, \phi_i^{cl})$, $i = 1, 2$, is identical to the angular position of the positron in the forward and electron in the backward hemisphere. The angular "plaquette" $(\theta_i^{cl} + 1.5\delta_\theta,\ \theta_i^{cl} - 1.5\delta_\theta) \times (\phi_i^{cl} + 1.5\delta_\phi,\ \phi_i^{cl} - 1.5\delta_\phi)$, where $\delta_\phi = 2\pi/32$, around $e^\pm$ is called cluster. The total energy registered in the cluster is denoted by $E_i^{cl}$. (Note that $\phi_1 = \phi_2$ for back-to-back configuration.)

### 2.7.2 First order - technical precision

We start the numerical comparisons of the various theoretical calculations with the calibration exercise in which we limit ourselves to strict $\mathcal{O}(\alpha^1)$ with Z exchange, up-down interference and vacuum polarization switched off, i.e. we examine pure photonic corrections without up-down interferences. We calculate the corresponding total cross section for all our four ES's at the LEP1 energy, $\sqrt{s} = 92.3$ GeV. The purpose of this exercise is to eliminate possible trivial normalization problems in the core MC programs and in the testing programs which implement our ES's. Since $\mathcal{O}(\alpha^1)$ is unique and common, the difference of the results will be entirely due to numerical/technical problems and, following ref. [11] where the analogous exercise of this type was done for the first time, we call it the "technical precision" of the involved calculations/programs. The results are shown in Tab. 13. Since tables are hard to read, we always include a figure which contains exactly the same result in the pictorial way. In the figure, one of the cross sections is used as a reference cross section and is subtracted from the other ones. It is plotted however on the horizontal line with its true statistical error.

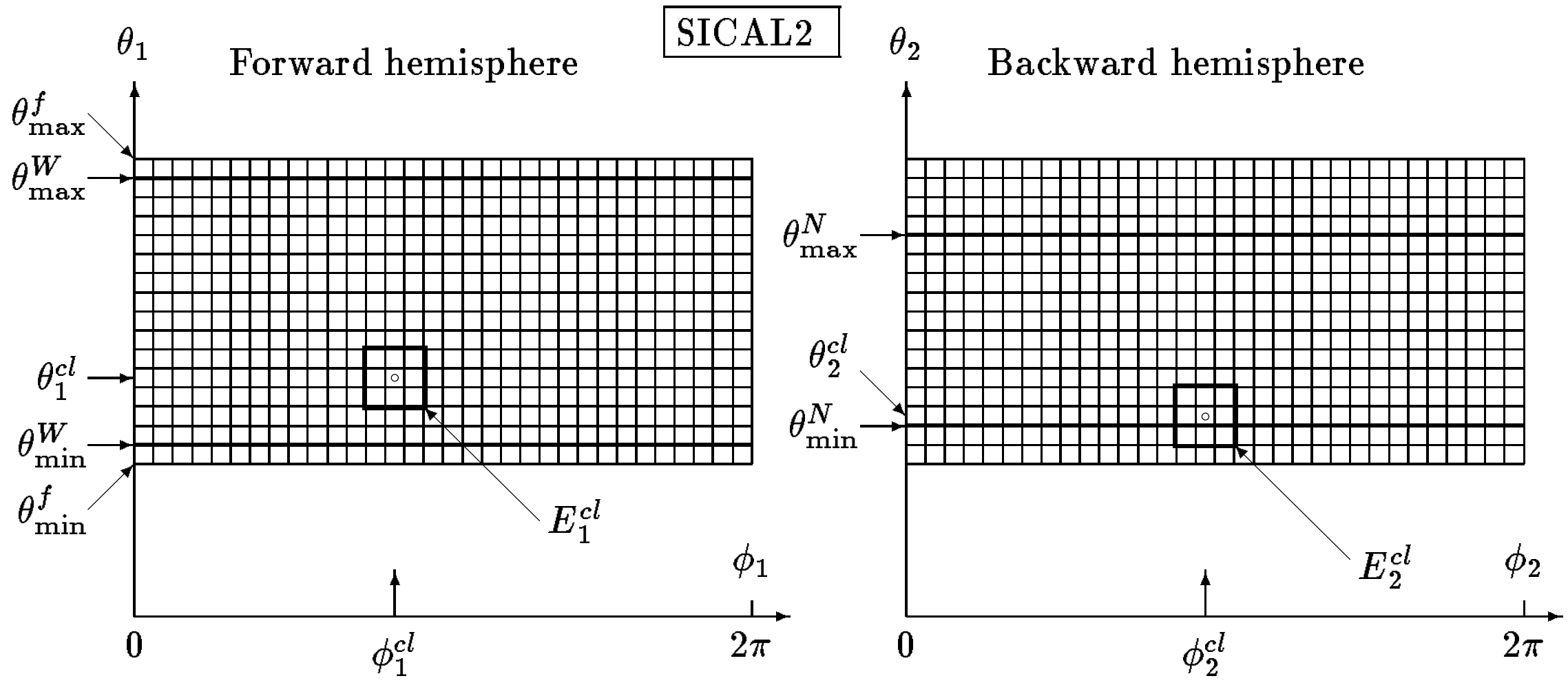

WW: $\theta_i^{cl} \in (\theta^W_{\min}, \theta^W_{\max})$, NN: $\theta_i^{cl} \in (\theta^N_{\min}, \theta^N_{\max})$

NW: $\theta_1^{cl} \in (\theta^W_{\min}, \theta^W_{\max})$, $\theta_2^{cl} \in (\theta^N_{\min}, \theta^N_{\max})$ $\qquad E_1^{cl} E_2^{cl} > z_{min} E^2_{beam}$

Figure 14: Geometry and acceptance of the calorimetric ES SICAL2. This ES restricts polar *angles* $\theta_i$ in the forward/backward hemispheres and requires a certain minimum *energy* to be detected simultaneously in both hemispheres. No restrictions on azimuthal angles $\phi_i$ are there. The entire "fiducial" $\theta$-range, $(\theta^f_{\min}, \theta^f_{\max}) = (0.024, 0.058)$ rad, includes the wide (W) range $(\theta^W_{\min}, \theta^W_{\max})$ and the narrow (N) range $(\theta^N_{\min}, \theta^N_{\max})$ exactly as depicted in the figure. This ES can be symmetric Wide-Wide (WW) or Narrow-Narrow (NN), or asymmetric Narrow-Wide (NW). The energy cut and $\theta$-cuts involve the definition of the *cluster*. Eeach side detector consists of $16 \times 32$ equal *plaquetes*. A single plaquete registers the total energy of electrons and photons. The plaquete with the maximum energy, together with its $3 \times 3$ neighborhood, is called cluster. The total energy registered in the cluster is $E_i^{cl}$ and its angular position is $(\theta_i^{cl}, \phi_i^{cl})$, $i = 1, 2$. More precisely the angular position of a cluster is the average position of the *centers* of all $3 \times 3$ plaquetes, weighted by their energies (the definitions of $\phi$'s are adjusted in such a way that $\phi_1 = \phi_2$ for back-to-back configuration). The plaquetes of the cluster which spill over the angular range (outside thick lines) are also used to determine the total energy and the average position of the cluster (see backward hemisphere).

Here Tab. 13 is visualized in Fig. 15. In this figure, the cross sections from the Monte Carlo OLDBIS (an improved version of the MC program written originally by Berends and Kleiss in PETRA times, now part of BHLUMI) is used as a reference. As we see, all calculations agree well within $3 \times 10^{-4}$ relative deviation. The apparent discrepancy of the $\mathcal{O}(\alpha^1)$ SABSPV for the SICAL2 ES is not statistically significant. The cross section from the non-Monte-Carlo type of calculation NLLBHA is available only for the simplest BARE1. As we have already discussed, the photonic radiative corrections for the SABH process scale smoothly with energy, so we regard this test to be valid for LEP2 energies within a factor two, i.e. within $6 \times 10^{-4}$.

### 2.7.3 Beyond first order - physical precision

Having found good agreement of the various calculations at the first order level, we now reinstall the photonic corrections beyond first order. More precisely we keep again Z exchange, up-down interference and vacuum polarization switched off, but compare numerical results which include $\mathcal{O}(\alpha^2 L^2)$, $\mathcal{O}(\alpha^2 L)$ and $\mathcal{O}(\alpha^3 L^3)$ contributions due to photon bremsstrahlung. We do not include production of light fermion pairs unless stated otherwise. The numerical results are shown in Tab. 14 and Fig. 16. In the figure, the cross section from the second order exponentiated Monte Carlo BHLUMI is used as a reference cross section. The differences between various calculations now represent not only technical precision, but also physical precision because the cross sections are calculated using different QED matrix elements.

| $z_{\rm min}$ | OLDBIS [nb] | SABSPV [nb] | BHAGEN95 [nb] | NNLBHA [nb] | BHLUMI [nb] |
|---|---|---|---|---|---|
| | (a) BARE1 | | | | |
| .100 | 166.079 ± .013 | 166.070 ± .024 | .000 ± .000 | 166.070 ± .017 | 166.046 ± .021 |
| .300 | 164.772 ± .013 | 164.762 ± .012 | 164.756 ± .012 | 164.767 ± .016 | 164.740 ± .021 |
| .500 | 162.277 ± .013 | 162.263 ± .012 | 162.258 ± .012 | 162.265 ± .016 | 162.241 ± .021 |
| .700 | 155.465 ± .013 | 155.452 ± .012 | 155.444 ± .012 | 155.453 ± .015 | 155.431 ± .020 |
| .900 | 134.417 ± .012 | 134.401 ± .023 | 134.394 ± .012 | 134.393 ± .014 | 134.390 ± .020 |
| | (b) CALO1 | | | | |
| .100 | 166.361 ± .013 | 166.353 ± .024 | .000 ± .000 | .000 ± .000 | 166.329 ± .021 |
| .300 | 166.081 ± .013 | 166.071 ± .021 | 166.074 ± .013 | .000 ± .000 | 166.049 ± .021 |
| .500 | 165.319 ± .013 | 165.311 ± .012 | 165.312 ± .013 | .000 ± .000 | 165.287 ± .021 |
| .700 | 161.823 ± .013 | 161.817 ± .024 | 161.818 ± .013 | .000 ± .000 | 161.794 ± .021 |
| .900 | 149.942 ± .013 | 149.934 ± .023 | 149.934 ± .013 | .000 ± .000 | 149.925 ± .020 |
| | (c) CALO2 | | | | |
| .100 | 131.061 ± .012 | 131.070 ± .022 | 131.051 ± .010 | .000 ± .000 | 131.032 ± .019 |
| .300 | 130.769 ± .012 | 130.778 ± .022 | 130.758 ± .010 | .000 ± .000 | 130.739 ± .019 |
| .500 | 130.206 ± .012 | 130.214 ± .022 | 130.194 ± .010 | .000 ± .000 | 130.176 ± .019 |
| .700 | 127.555 ± .012 | 127.565 ± .022 | 127.546 ± .010 | .000 ± .000 | 127.528 ± .019 |
| .900 | 117.557 ± .011 | 117.572 ± .025 | 117.543 ± .010 | .000 ± .000 | 117.541 ± .018 |
| | (d) SICAL2 | | | | |
| .100 | 132.011 ± .012 | 131.965 ± .023 | 132.004 ± .028 | .000 ± .000 | 131.984 ± .019 |
| .300 | 131.900 ± .012 | 131.862 ± .021 | 131.893 ± .027 | .000 ± .000 | 131.872 ± .019 |
| .500 | 131.587 ± .012 | 131.539 ± .018 | 131.581 ± .027 | .000 ± .000 | 131.559 ± .019 |
| .700 | 128.363 ± .012 | 128.306 ± .016 | 128.364 ± .027 | .000 ± .000 | 128.338 ± .019 |
| .900 | 117.843 ± .011 | 117.795 ± .012 | 117.811 ± .027 | .000 ± .000 | 117.828 ± .018 |

Table 13: Monte Carlo results for the symmetric Wide-Wide ES's BARE1, CALO1, CALO2 and SICAL2, for the $\mathcal{O}(\alpha^1)$ matrix element. Z exchange, up-down interference and vacuum polarization are switched off. The center of mass energy is $\sqrt{s} = 92.3$ GeV. Not available x-sections are set to zero.

The results shown in Tab. 14 and Fig. 16 have remarkable properties. For values of the energy-cut variable in the experimentally interesting range $0.25 < z_{\rm min} < 0.75$, the cross section from the programs BHLUMI and SABSPV agree throughout all the four ES's, from the unrealistic BARE1 to very realistic SICAL2, to within $1.0 \times 10^{-3}$ relative deviation. This agreement is definitely better than the difference between BHLUMI and OLDBIS+LUMLOG, which

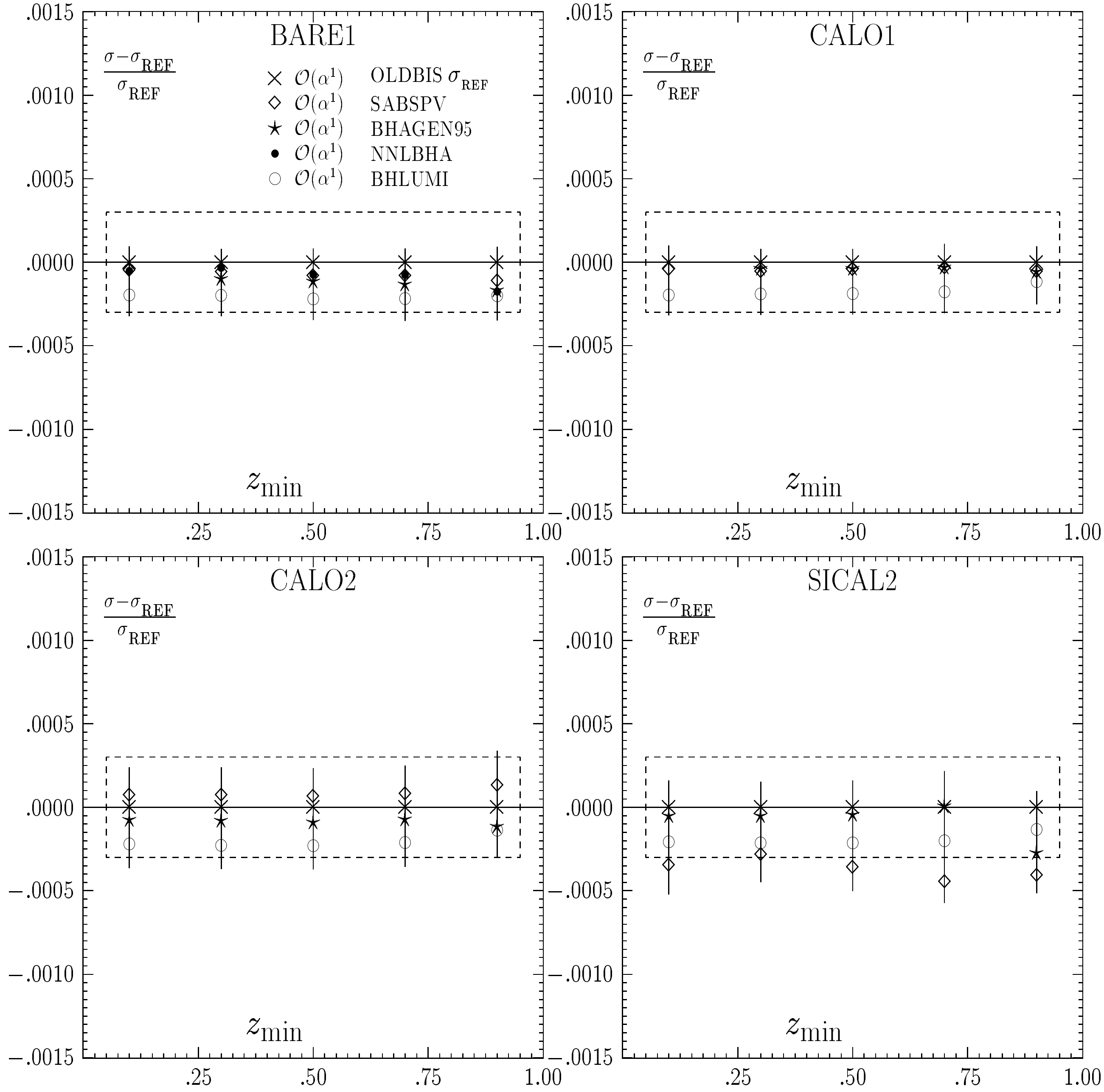

Figure 15: Monte Carlo results for the symmetric Wide-Wide ES's BARE1, CALO1, CALO2 and SICAL2, for the $\mathcal{O}(\alpha^1)$ matrix element. Z exchange, up-down interference and vacuum polarization are switched off. The center of mass energy is $\sqrt{s} = 92.3$ GeV. In the plot, the cross section from the program OLDBIS (part from BHLUMI 4.02.a, originally written by Berends and Kleiss) is used as a reference cross section.

in the last years was routinely used (see Refs. [6, 58]) in order to estimate missing higher order and subleading corrections. Remarkably, the OLDBIS+LUMLOG results coincide extremely well with BHAGEN95. Let us note that the OLDBIS+LUMLOG matrix element does not ex-

| $z_{min}$ | BHLUMI [nb] | SABSPV [nb] | BHAGEN95 [nb] | OBI+LMG [nb] | NLLBHA [nb] |
|---|---|---|---|---|---|
| | (a) BARE1 | | | | |
| .100 | $166.892 \pm .006$ | $166.795 \pm .028$ | $.000 \pm .000$ | $166.672 \pm .017$ | $166.948 \pm .000$ |
| .300 | $165.374 \pm .006$ | $165.323 \pm .028$ | $165.190 \pm .012$ | $165.187 \pm .017$ | $165.448 \pm .000$ |
| .500 | $162.530 \pm .006$ | $162.529 \pm .028$ | $162.330 \pm .012$ | $162.365 \pm .017$ | $162.581 \pm .000$ |
| .700 | $155.668 \pm .006$ | $155.751 \pm .026$ | $155.466 \pm .012$ | $155.519 \pm .017$ | $155.617 \pm .000$ |
| .900 | $137.342 \pm .006$ | $137.528 \pm .022$ | $137.188 \pm .011$ | $137.210 \pm .017$ | $137.201 \pm .000$ |
| | (b) CALO1 | | | | |
| .100 | $167.203 \pm .006$ | $167.106 \pm .028$ | $.000 \pm .000$ | $167.000 \pm .017$ | $.000 \pm .000$ |
| .300 | $166.795 \pm .006$ | $166.715 \pm .028$ | $166.618 \pm .012$ | $166.623 \pm .017$ | $.000 \pm .000$ |
| .500 | $165.830 \pm .006$ | $165.768 \pm .014$ | $165.661 \pm .014$ | $165.686 \pm .017$ | $.000 \pm .000$ |
| .700 | $162.237 \pm .006$ | $162.203 \pm .027$ | $162.048 \pm .014$ | $162.053 \pm .017$ | $.000 \pm .000$ |
| .900 | $151.270 \pm .006$ | $151.272 \pm .025$ | $150.823 \pm .014$ | $150.707 \pm .017$ | $.000 \pm .000$ |
| | (c) CALO2 | | | | |
| .100 | $131.835 \pm .006$ | $131.755 \pm .027$ | $131.658 \pm .007$ | $131.632 \pm .016$ | $.000 \pm .000$ |
| .300 | $131.450 \pm .006$ | $131.393 \pm .027$ | $131.285 \pm .012$ | $131.274 \pm .016$ | $.000 \pm .000$ |
| .500 | $130.727 \pm .006$ | $130.708 \pm .027$ | $130.575 \pm .012$ | $130.584 \pm .016$ | $.000 \pm .000$ |
| .700 | $127.969 \pm .006$ | $127.999 \pm .027$ | $127.802 \pm .014$ | $127.802 \pm .016$ | $.000 \pm .000$ |
| .900 | $118.792 \pm .006$ | $118.879 \pm .029$ | $118.293 \pm .013$ | $118.201 \pm .015$ | $.000 \pm .000$ |
| | (d) SICAL2 | | | | |
| .100 | $132.816 \pm .006$ | $132.612 \pm .026$ | $132.611 \pm .028$ | $132.582 \pm .016$ | $.000 \pm .000$ |
| .300 | $132.553 \pm .006$ | $132.427 \pm .025$ | $132.420 \pm .028$ | $132.405 \pm .016$ | $.000 \pm .000$ |
| .500 | $131.985 \pm .006$ | $131.966 \pm .022$ | $131.962 \pm .027$ | $131.965 \pm .016$ | $.000 \pm .000$ |
| .700 | $128.672 \pm .006$ | $128.691 \pm .019$ | $128.620 \pm .027$ | $128.610 \pm .016$ | $.000 \pm .000$ |
| .900 | $119.013 \pm .006$ | $119.075 \pm .015$ | $118.561 \pm .027$ | $118.488 \pm .015$ | $.000 \pm .000$ |

Table 14: Monte Carlo results for the symmetric Wide-Wide ES's BARE1, CALO1, CALO2 and SICAL2, for matrix elements beyond first order. Z exchange, up-down interference and vacuum polarization are switched off. The center of mass energy is $\sqrt{s} = 92.3$ GeV. Not available x-sections are set to zero.

ponentiate properly $\mathcal{O}(\alpha^2 L)$ corrections, i.e. they are wrong in the soft photon limit. This may explain why BHLUMI and SABSPV, which do not have such problems, agree better. According to the authors, BHAGEN95 does not suffer of the same problem as it has the soft photon limit properly treated by construction, but some corrections are expected due to the approximate treatment of two hard photon emission. The result from NLLBHA is present only for unrealistic BARE1 selection, and for $0.25 < z_{\rm min} < 0.75$ it agrees to within 0.1% with BHLUMI and SABSPV. It is an interesting result because NLLBHA features complete $\mathcal{O}(\alpha^2 L)$ corrections, while all the other programs have only incomplete $\mathcal{O}(\alpha^2 L)$ contributions. In Tab. 14 and Fig. 16 the results of BHLUMI, SABSPV and BHAGEN95 include exponentiation, and therefore they include necessarily $\mathcal{O}(\alpha^3 L^3)$ effects (incomplete). We therefore compare them with a version of NLLBHA which includes, besides $\mathcal{O}(\alpha^2 L)$, also $\mathcal{O}(\alpha^3 L^3)$ corrections. All the above results will be used as an input in our final estimate of the total theoretical uncertainty of SABH cross

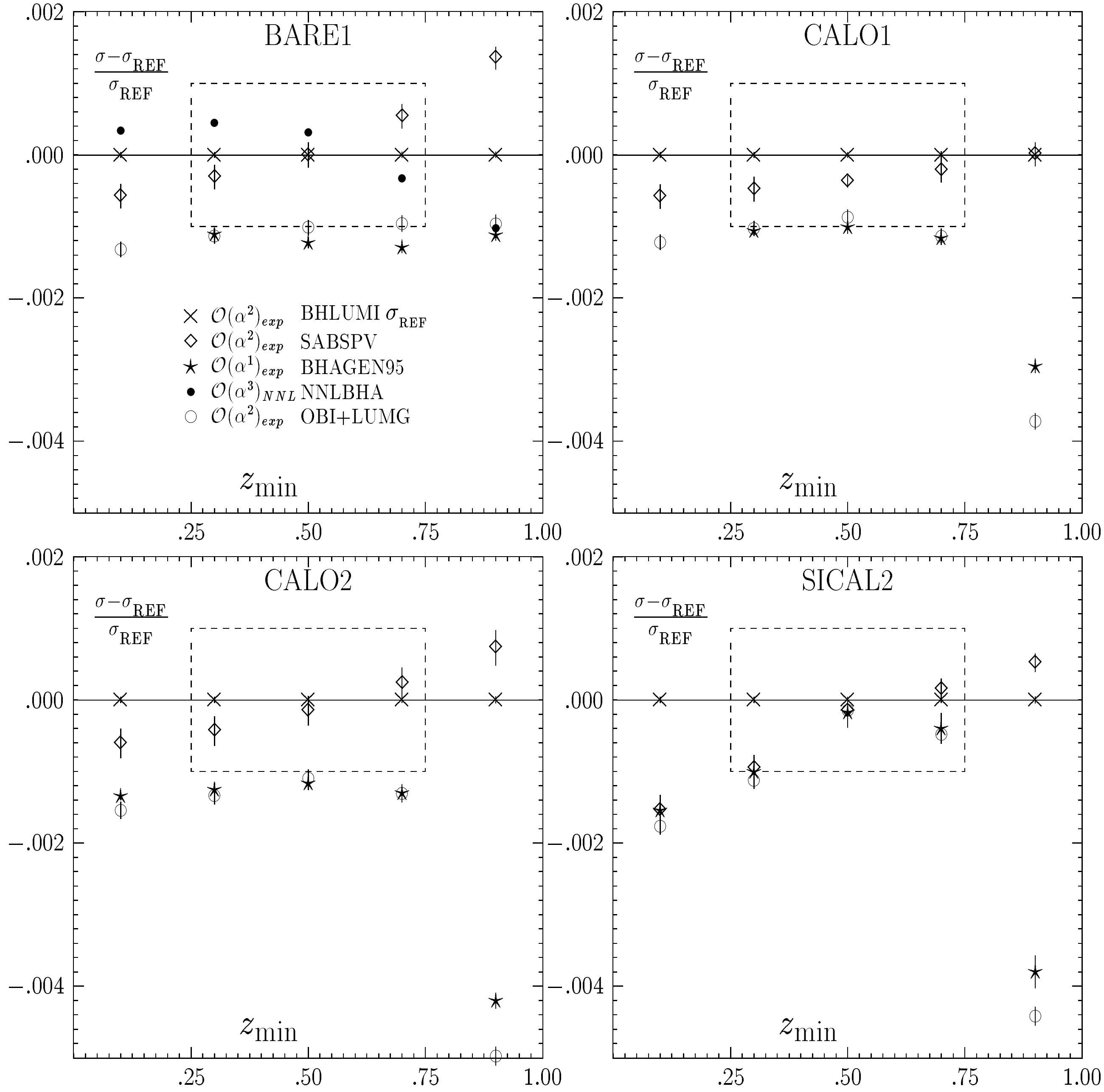

Figure 16: Monte Carlo results for the symmetric Wide-Wide ES's BARE1, CALO1, CALO2 and SICAL2, for matrix elements beyond first order. Z exchange, up-down interference and vacuum polarization are switched off. The center of mass energy is $\sqrt{s} = 92.3$ GeV. In the plot, the $\mathcal{O}(\alpha^2)^{YFS}_{exp}$ cross section $\sigma_{\rm BHL}$ from BHLUMI 4.02.a is used as a reference cross section.

section for LEP1/LEP2 energies.

Finally, we present similar numerical comparisons of the calculations beyond $\mathcal{O}(\alpha^1)$ at one LEP2 energy $\sqrt{s} = 176$ GeV. As before, since the tables are hard to read, we accompany the table with a figure which shows the same numerical result in a pictorial way (the caption is common for the table and figure). This way of presenting results in the form of the twin table/figure will be used often in the following. As before, in the figure one of the cross sections is used as a reference cross section and is subtracted from the other ones. The main result is shown in table/figure 15. Here, results are shown for the symmetric Wide-Wide variant of the CALO2 and SICAL2 ES's. As expected, the difference between the programs is almost the same! The higher order corrections at LEP2 are only slightly stronger. This result was already anticipated when analyzing "scaling rules" derived from Tab. 2. From the scaling rules we also know that this result will be essentially the same for the wider angular range $3^\circ - 6^\circ$. *The practical message is that, within 20-30%, the precision estimates derived from the numerical exercises for the SABH process at LEP1 should be valid also for LEP2.*

| $z_{min}$ | BHLUMI [nb] | SABSPV [nb] | BHAGEN95 [nb] | OBI+LUM [nb] |
|---|---|---|---|---|
| | (a) CALO2 LEP2 | | | |
| .100 | $36.123 \pm .003$ | $36.096 \pm .008$ | $.000 \pm .000$ | $36.060 \pm .006$ |
| .300 | $36.013 \pm .003$ | $35.992 \pm .008$ | $35.963 \pm .005$ | $35.958 \pm .006$ |
| .500 | $35.807 \pm .003$ | $35.796 \pm .008$ | $35.762 \pm .005$ | $35.761 \pm .006$ |
| .700 | $35.001 \pm .003$ | $35.005 \pm .008$ | $34.951 \pm .005$ | $34.948 \pm .006$ |
| .900 | $32.324 \pm .003$ | $32.341 \pm .008$ | $32.173 \pm .006$ | $32.145 \pm .006$ |
| | (b) SICAL2 LEP2 | | | |
| .100 | $36.394 \pm .003$ | $36.337 \pm .011$ | $.000 \pm .000$ | $36.322 \pm .006$ |
| .300 | $36.316 \pm .003$ | $36.284 \pm .010$ | $36.271 \pm .009$ | $36.270 \pm .006$ |
| .500 | $36.150 \pm .003$ | $36.147 \pm .009$ | $36.139 \pm .009$ | $36.142 \pm .006$ |
| .700 | $35.193 \pm .003$ | $35.203 \pm .008$ | $35.173 \pm .009$ | $35.171 \pm .006$ |
| .900 | $32.383 \pm .003$ | $32.405 \pm .006$ | $32.243 \pm .009$ | $32.224 \pm .006$ |

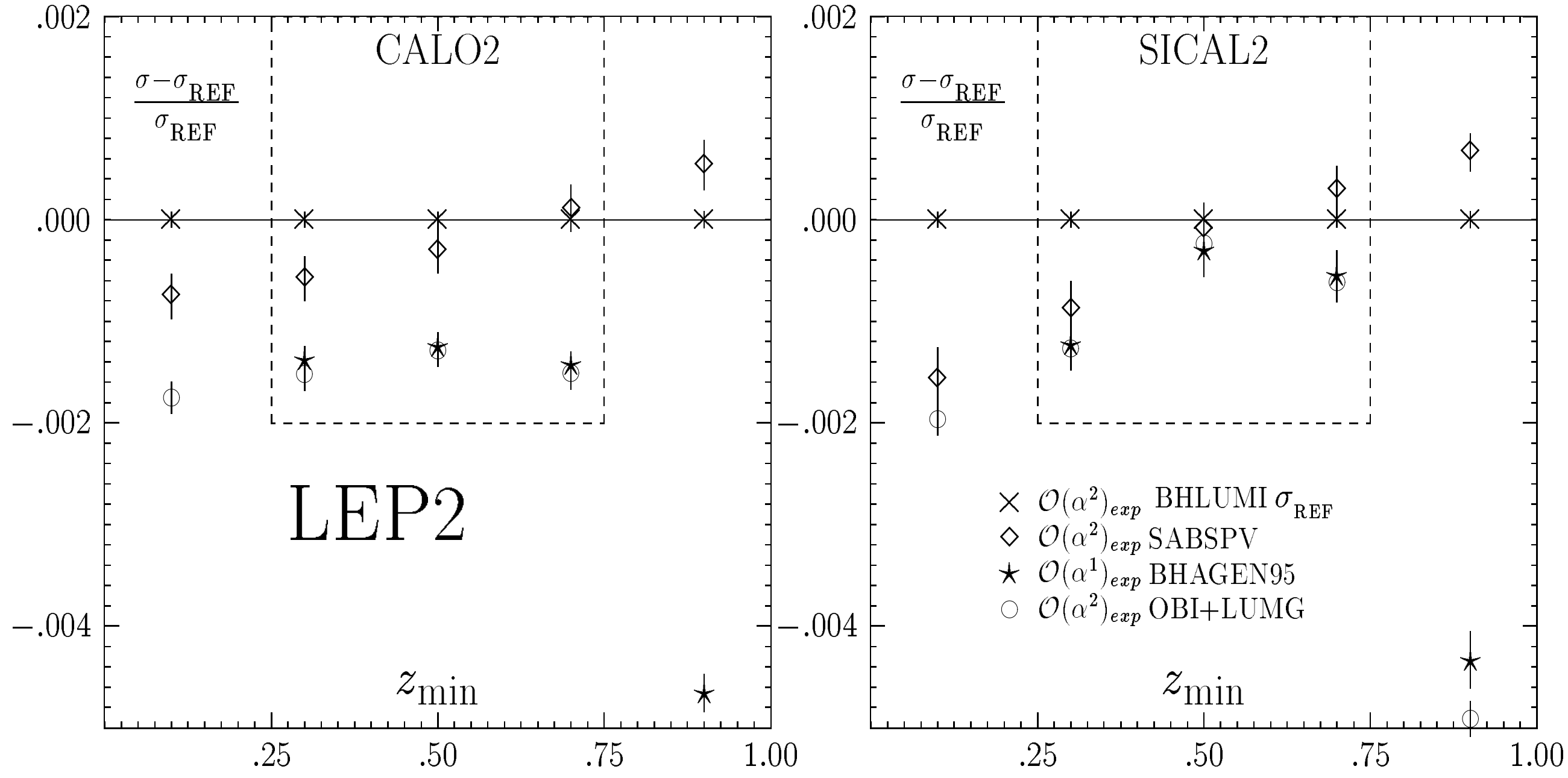

Table 15: In this table/figure we show cross sections for LEP2 center of mass energy, $\sqrt{s} = 176$ GeV. Monte Carlo results are shown for various symmetric Wide-Wide ES's and matrix elements beyond first order. Z exchange, up-down interference and vacuum polarization are switched off. Not available x-sections are set to zero. In the plot, the $\mathcal{O}(\alpha^2)_{exp}$ cross section $\sigma_{\rm BHL}$ from BHLUMI 4.02.a is used as a reference cross section.

Precision requirements at LEP2 are less stringent. In the figure, we draw a LEP2-type box which spans over 0.2% and extends over the experimentally interesting range $0.25 < z_{min} < 0.75$. All programs come together within the above range. The above 0.2% limit will be used as an input in our final estimate of the total theoretical uncertainty of the SABH cross section for LEP2 energies. This limit has obviously a large safety margin, close to a factor of two.

### 2.7.4 Asymmetric and very narrow event selections

The numerical comparisons shown in the previous section were done, for pure technical reasons (less chances for programming errors in the testing programs), for the *symmetric* Wide-Wide version of the ES. As we know very well (see the introduction), the higher order contributions are sensitive to the "asymmetricity" of the ES. In order to avoid any danger due to the above simplification, we have done another series of comparisons of the various calculations for the symmetric Narrow-Narrow and asymmetric Narrow-Wide versions of the ES's CALO2, which are defined in Fig. 13. Let us remind the reader that the variation of the difference BHLUMI$-$(OLDBIS+LUMLOG$_{HO}$) over the WW, NN and NW selection was the cornerstone of the previous estimates [6,58] of the size of uncontrolled higher order photonic corrections (together with technical precision). We believe that CALO2 is close enough to our most realistic ES SICAL2 and the results obtained for CALO2 are valid for SICAL2. Let us also recall that the typical experimental ES is of the asymmetric Narrow-Wide type. The corresponding results are shown in table/figure 16 for the matrix elements in the $\mathcal{O}(\alpha^2)$ class (we have checked that for the $\mathcal{O}(\alpha^1)$ level the same programs agree better than 0.03%, but we omit the corresponding table/plot due to lack of space).

As we see in tables/figures 16 and 14, for all the three types of the CALO2 ES (WW, NN and NW), BHLUMI and SABSPV stay within 0.1% from one another for all the values of the energy-cut variable in the experimentally interesting range $0.25 < z_{\rm min} < 0.75$. This is a new nontrivial result, which will be exploited to decrease the estimated error due to the higher order photonic corrections from 0.15% down to 0.1%. In a sense, we replace the old estimate based on BHLUMI $-$ (OLDBIS + LUMLOG$_{HO}$) with a new one based on BHLUMI$-$SABSPV. Hybrid Monte Carlo's (OLDBIS + LUMLOG$_{HO}$) and BHAGEN95 are off of about 0.2% in the NN case but, noticeably, they are on the same ground as BHLUMI and SABSPV for the most interesting NW case. The above exercise was done for the LEP1 energy, and in view of the results shown in table/figure 15 and our "scaling rules" (see the introduction), we do not foresee any problem with extending its validity to LEP2 energies.

As we already stressed in the introduction, for the purpose of LEP2 it is more important, however, to check if the change of the "narrowness", i.e. the ratio $\theta_{max}/\theta_{min} - 1$, to smaller values does not spoil the agreement of the table/figure 16. As we have already indicated, at LEP2 the decrease of the narrowness $\theta_{max}/\theta_{min} - 1$ may cause a significant increase in the photonic radiative corrections. The relevant cross-check is done in table/figure 17. It represents the *worst possible scenario* at LEP2. The results are shown for the narrower version of the CALO2 ES, which we call CALO3, in the symmetric and asymmetric versions. As we see, BHLUMI and SABSPV differ again for the above ES by less than 0.2%. This result will be used for estimating theoretical uncertainty of the SABH process at LEP2. Hybrid Monte Carlo's (OLDBIS + LUMLOG$_{HO}$) and BHAGEN95 are off of about 0.4% in the NN case but, noticeably, they are on the same ground as BHLUMI and SABSPV for the most interesting NW case.

| $z_{min}$ | BHLUMI [nb] | OBI+LMG [nb] | SABSPV [nb] | BHAGEN95 [nb] |
|---|---|---|---|---|
| | CALO2 *Symmetric Narrow-Narrow* | | | |
| .100 | 95.458 ± .005 | 95.259 ± .014 | 95.363 ± .013 | 95.287 ± .009 |
| .300 | 95.233 ± .005 | 95.048 ± .014 | 95.157 ± .016 | 95.065 ± .009 |
| .500 | 94.841 ± .005 | 94.672 ± .014 | 94.792 ± .016 | 94.680 ± .009 |
| .700 | 93.520 ± .005 | 93.347 ± .014 | 93.513 ± .019 | 93.354 ± .009 |
| .900 | 87.359 ± .005 | 86.899 ± .013 | 87.396 ± .012 | 86.958 ± .009 |
| | CALO2 *Asymmetric Narrow-Wide* | | | |
| .100 | 98.834 ± .003 | 98.809 ± .010 | 98.859 ± .017 | 98.804 ± .009 |
| .300 | 98.539 ± .003 | 98.535 ± .010 | 98.577 ± .017 | 98.515 ± .009 |
| .500 | 98.020 ± .003 | 98.038 ± .010 | 98.073 ± .019 | 98.006 ± .009 |
| .700 | 96.054 ± .003 | 96.061 ± .010 | 96.131 ± .018 | 96.033 ± .009 |
| .900 | 88.554 ± .003 | 88.220 ± .009 | 88.648 ± .015 | 88.263 ± .009 |

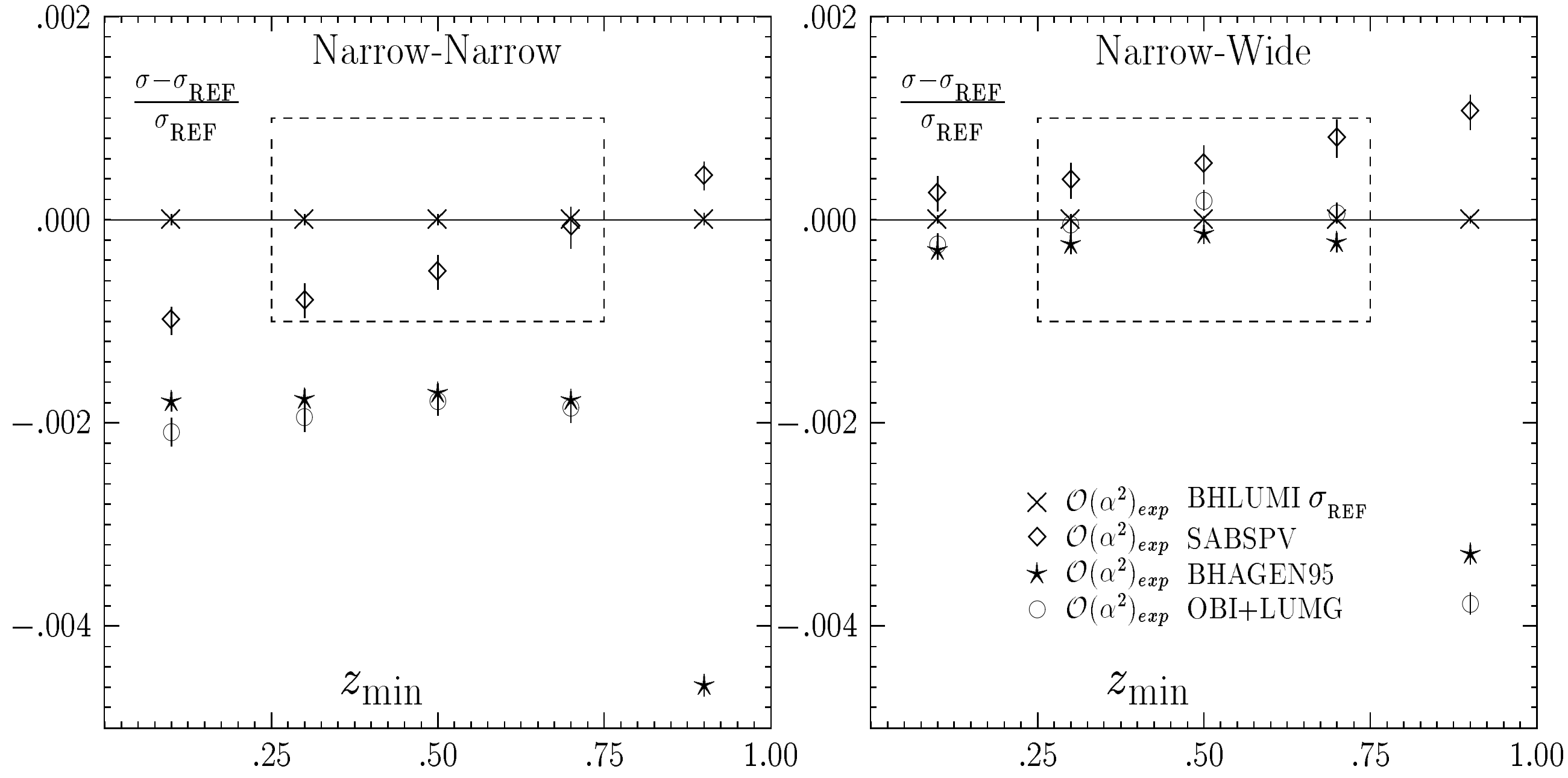

Table 16: In this table/figure we show cross sections for various symmetric/asymmetric versions of the CALO2 ES, for matrix elements beyond first order. Z exchange, up-down interference and vacuum polarization are switched off. The center of mass energy is $\sqrt{s} = 92.3$ GeV. Not available x-sections are set to zero. The wide range is defined by $\theta_{1w} = \theta_{1f} + \delta_{segm}$ and $\theta_{2w} = \theta_{2f} - \delta_{segm}$, and the narrow range by $\theta_{1n} = \theta_{1f} + 2\delta_{segm}$ and $\theta_{2n} = \theta_{2f} - 4\delta_{segm}$; $\delta_{segm} = (\theta_{2f} - \theta_{1f})/16$, $\theta_{1f} = 0.024$ and $\theta_{2f} = 0.058$ rad, respectively.

| $z_{min}$ | BHLUMI [nb] | OBI+LMG [nb] | SABSPV [nb] | BHAGEN95 [nb] |
|---|---|---|---|---|
| | CALO2 *Symmetric Narrow-Narrow LEP2* | | | |
| .100 | 8.088 ± .001 | 8.052 ± .003 | 8.074 ± .001 | 8.058 ± .001 |
| .300 | 8.074 ± .001 | 8.039 ± .003 | 8.061 ± .001 | 8.044 ± .001 |
| .500 | 8.048 ± .001 | 8.014 ± .003 | 8.039 ± .001 | 8.018 ± .001 |
| .700 | 7.989 ± .001 | 7.954 ± .003 | 7.986 ± .001 | 7.958 ± .001 |
| .900 | 7.574 ± .001 | 7.515 ± .003 | 7.582 ± .001 | 7.522 ± .001 |
| | CALO2 *Asymmetric Narrow-Wide LEP2* | | | |
| .100 | 8.523 ± .001 | 8.514 ± .002 | 8.518 ± .001 | 8.515 ± .001 |
| .300 | 8.501 ± .001 | 8.494 ± .002 | 8.499 ± .001 | 8.494 ± .001 |
| .500 | 8.464 ± .001 | 8.457 ± .002 | 8.465 ± .001 | 8.457 ± .001 |
| .700 | 8.374 ± .001 | 8.366 ± .002 | 8.380 ± .001 | 8.365 ± .001 |
| .900 | 7.755 ± .001 | 7.716 ± .002 | 7.769 ± .001 | 7.720 ± .001 |

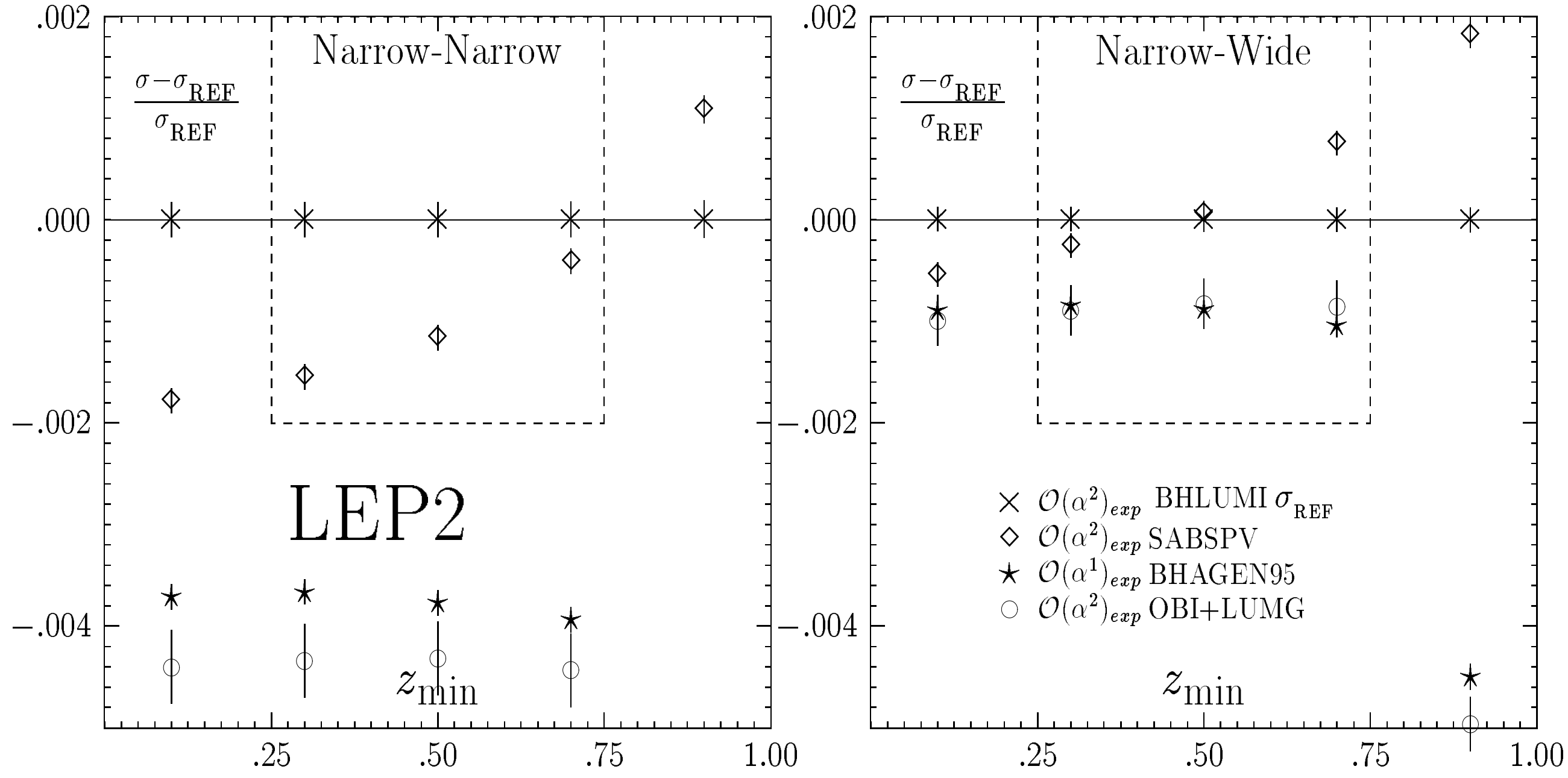

Table 17: In this table/figure we show cross sections for for the symmetric/asymmetric CALO3 ES's (the narrower version of CALO2) for matrix elements beyond first order. Z exchange, up-down interference and vacuum polarization are switched off. The center of mass energy is $\sqrt{s} = 176$ GeV. Not available x-sections are set to zero. The wide range is defined by $\theta_{1w} = \theta_{1f} + 6\delta_{segm}$ and $\theta_{2w} = \theta_{1f} + 16\delta_{segm}$, and the narrow range by $\theta_{1n} = \theta_{1f} + 8\delta_{segm}$ and $\theta_{2n} = \theta_{1f} + 15\delta_{segm}$; $\delta_{segm} = (\theta_{2f} - \theta_{1f})/16$, $\theta_{1f} = 0.024$ and $\theta_{2f} = 0.058$ rad, respectively.

| $z_{min}$ | BHLUMI [nb] | SABSPV [nb] | BHAGEN95 | VP+Z Bhlumi |
|---|---|---|---|---|
| | CALO2 *Symmetric Wide-Wide* | | | |
| .100 | 136.975 ± .010 | 136.831 ± .018 | 136.861 ± .008 | 5.140 ± .008 |
| .300 | 136.576 ± .010 | 136.453 ± .018 | 136.482 ± .008 | 5.126 ± .008 |
| .500 | 135.827 ± .010 | 135.742 ± .018 | 135.770 ± .008 | 5.100 ± .008 |
| .700 | 132.962 ± .010 | 132.928 ± .017 | 132.927 ± .008 | 4.994 ± .008 |
| .900 | 123.420 ± .009 | 123.430 ± .018 | 123.114 ± .009 | 4.627 ± .008 |
| | CALO2 *Symmetric Narrow-Narrow* | | | |
| .100 | 99.208 ± .009 | 99.074 ± .017 | 99.089 ± .011 | 3.751 ± .007 |
| .300 | 98.975 ± .009 | 98.851 ± .021 | 98.866 ± .011 | 3.742 ± .007 |
| .500 | 98.570 ± .009 | 98.477 ± .017 | 98.479 ± .011 | 3.728 ± .007 |
| .700 | 97.198 ± .008 | 97.147 ± .017 | 97.128 ± .011 | 3.678 ± .007 |
| .900 | 90.789 ± .008 | 90.791 ± .016 | 90.537 ± .011 | 3.430 ± .007 |
| | CALO2 *Asymmetric Narrow-Wide* | | | |
| .100 | 102.717 ± .006 | 102.703 ± .017 | 102.724 ± .010 | 3.883 ± .004 |
| .300 | 102.412 ± .006 | 102.411 ± .017 | 102.434 ± .010 | 3.873 ± .004 |
| .500 | 101.874 ± .006 | 101.894 ± .017 | 101.922 ± .010 | 3.854 ± .004 |
| .700 | 99.833 ± .005 | 99.878 ± .017 | 99.902 ± .010 | 3.779 ± .004 |
| .900 | 92.033 ± .005 | 92.088 ± .016 | 91.887 ± .011 | 3.478 ± .004 |

Table 18: Monte Carlo results for various symmetric/asymmetric versions of the CALO2 ES, for matrix elements beyond first order. Z exchange, up-down interference and vacuum polarization are switched ON. The center of mass energy is $\sqrt{s} = 92.3$ GeV. Not available x-sections are set to zero. The wide range is defined by $\theta_{1w} = \theta_{1f} + \delta_{segm}$ and $\theta_{2w} = \theta_{2f} - \delta_{segm}$, and the narrow range by $\theta_{1n} = \theta_{1f} + 2\delta_{segm}$ and $\theta_{2n} = \theta_{2f} - 4\delta_{segm}$; $\delta_{segm} = (\theta_{2f} - \theta_{1f})/16$, $\theta_{1f} = 0.024$ and $\theta_{2f} = 0.058$ rad, respectively.

### 2.7.5 Z and vacuum polarization included

In all the previous comparisons, the small contributions from $s$-channel Z-exchange and $s$-channel photon exchange diagrams were switched off in order to enhance the possibility of seeing more clearly the most important pure photonic higher order corrections. In the following part of numerical comparisons, we restore in the calculations the contributions from these $s$-channel Z-exchange and $s$-channel photon exchange diagrams, together with the effect of vacuum polarization. The comparison of various calculations is done for the semi-realistic ES CALO2 in the versions Wide-Wide, Narrow-Narrow and Narrow-Wide, as defined in Fig. 13. The resulting cross sections are shown for a LEP1 energy in Tab. 18 and Fig. 17. Again, BHLUMI and SABSPV, for values of the energy-cut variable in the experimentally interesting range $0.25 < z_{\rm min} < 0.75$, agree within 0.1% for all the three versions of the ES (WW, NN and WN). BHAGEN95 is also in agreement, in this case, for all the three versions of the ES, due to a slightly bigger correction in these added contributions. We do not expect that switching on the small $s$-channel Z-exchange and $s$-channel photon exchange corrections would change our conclusions for LEP2. Vacuum polarization enters essentially only in the normalization of the SABH cross section, and Z contribution at LEP2 can be safely neglected. We therefore extend the validity of the above exercise to LEP2.

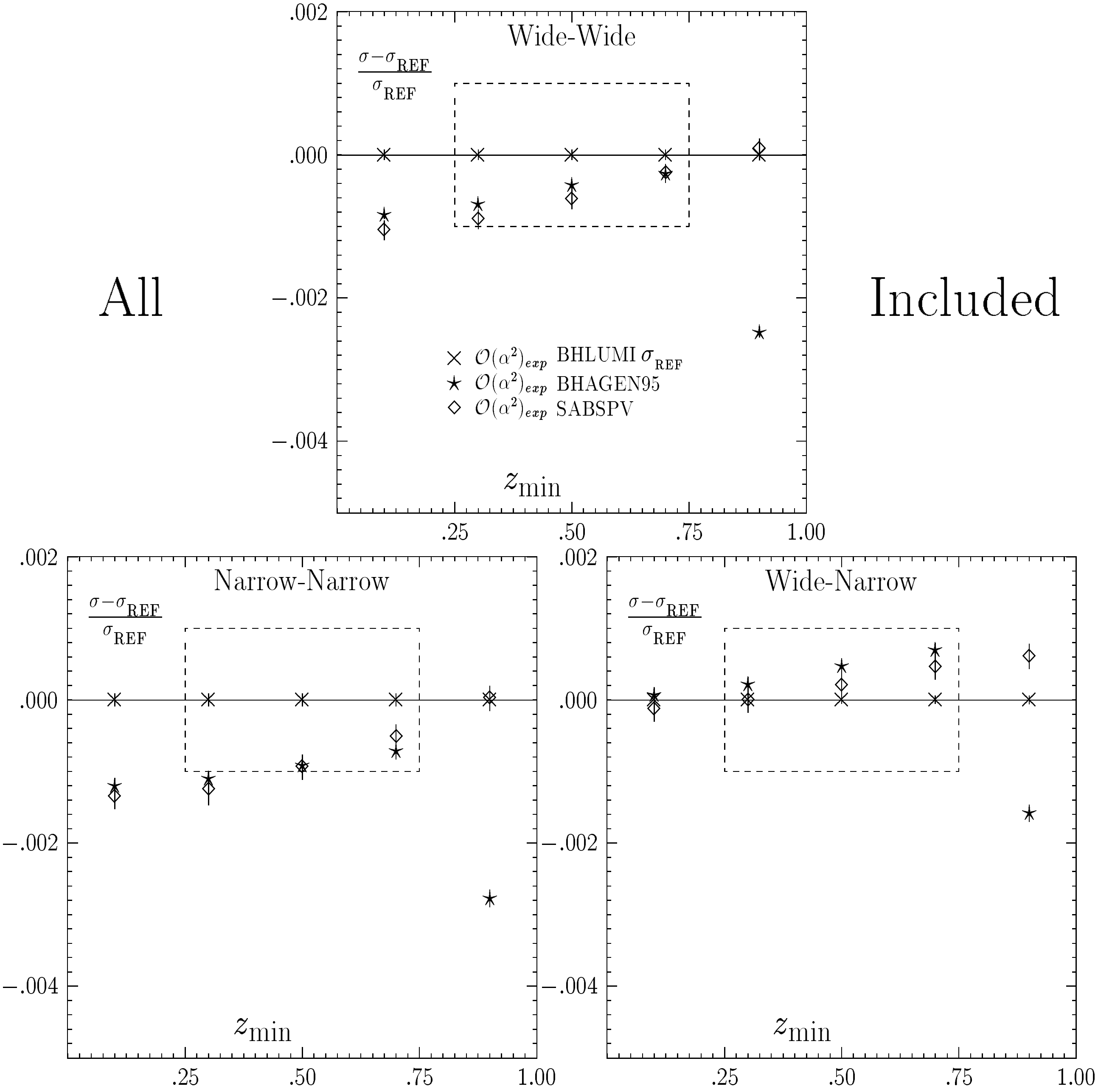

Figure 17: Monte Carlo results for various symmetric/asymmetric versions of the CALO2 ES, for matrix elements beyond first order. Z exchange, up-down interference and vacuum polarization are switched ON. The center of mass energy is $\sqrt{s}$ = 92.3 GeV. Not available x-sections are set to zero. In the plot, the $\mathcal{O}(\alpha^2)^{YFS}_{exp}$ cross section $\sigma_{\rm BHL}$ from BHLUMI 4.x is used as a reference cross section.

## 2.8 The total theoretical error for small-angle Bhabha scattering

In this section we present some supplementary numerical material concerning higher order corrections from MC and non-MC programs, and we summarize on the total theoretical error

| $z_{min}$ | BHLUMI(alf2e) | BHLUMI(alf3e) | BHLUMI(alf2) | NLLBHA(alf2) | NLLBHA(alf3) | NLLBHA(alf3p) |
|---|---|---|---|---|---|---|
| | (a) BARE1 | | | | | |
| .100 | $166.892 \pm .006$ | $-.017 \pm .000$ | $166.988 \pm .021$ | $167.016 \pm .017$ | $166.948 \pm .000$ | $166.966 \pm .000$ |
| .300 | $165.374 \pm .006$ | $-.010 \pm .000$ | $165.471 \pm .021$ | $165.503 \pm .017$ | $165.448 \pm .000$ | $165.421 \pm .000$ |
| .500 | $162.530 \pm .006$ | $-.006 \pm .000$ | $162.594 \pm .021$ | $162.630 \pm .016$ | $162.581 \pm .000$ | $162.527 \pm .000$ |
| .700 | $155.668 \pm .006$ | $-.002 \pm .000$ | $155.620 \pm .020$ | $155.649 \pm .015$ | $155.617 \pm .000$ | $155.528 \pm .000$ |
| .900 | $137.342 \pm .006$ | $.004 \pm .000$ | $137.191 \pm .020$ | $137.205 \pm .014$ | $137.201 \pm .000$ | $137.063 \pm .000$ |
| | (b) SICAL2 | | | | | |
| .000 | $132.816 \pm .006$ | $-.017 \pm .000$ | $132.912 \pm .019$ | $.000 \pm .000$ | $.000 \pm .000$ | $.000 \pm .000$ |
| .200 | $132.553 \pm .006$ | $-.018 \pm .000$ | $132.645 \pm .019$ | $.000 \pm .000$ | $.000 \pm .000$ | $.000 \pm .000$ |
| .400 | $131.985 \pm .006$ | $-.019 \pm .000$ | $132.061 \pm .019$ | $.000 \pm .000$ | $.000 \pm .000$ | $.000 \pm .000$ |
| .600 | $128.672 \pm .006$ | $-.017 \pm .000$ | $128.711 \pm .019$ | $.000 \pm .000$ | $.000 \pm .000$ | $.000 \pm .000$ |
| .800 | $119.013 \pm .006$ | $-.012 \pm .000$ | $119.014 \pm .018$ | $.000 \pm .000$ | $.000 \pm .000$ | $.000 \pm .000$ |

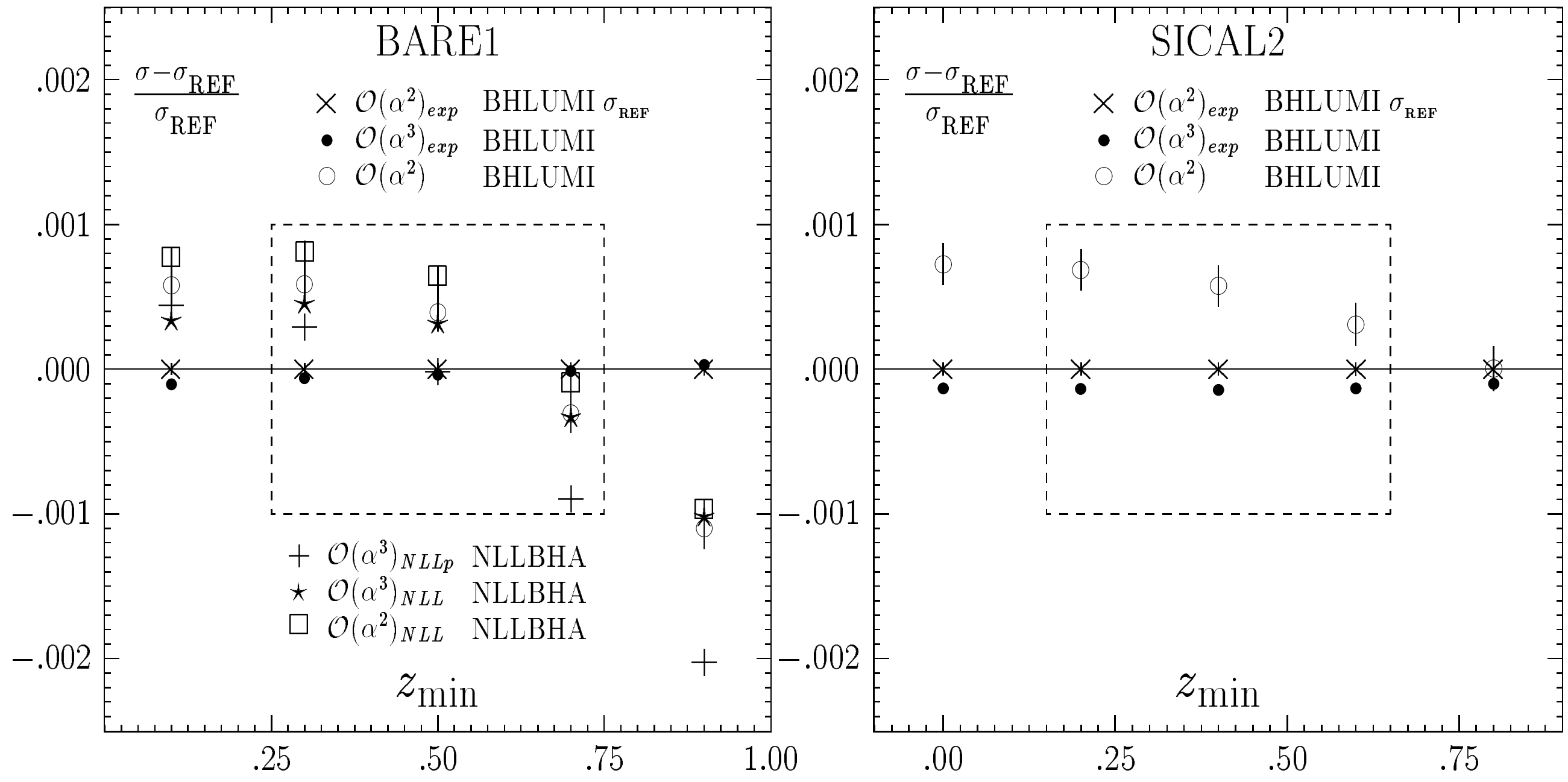

Table 19: In this table/figure we show cross sections for LEP1 center of mass energy, $\sqrt{s} = 92$ GeV. Results from BHLUMI and NLLBHA for the symmetric Wide-Wide ES's BARE1 and SICAL2 are shown. Not available x-sections are set to zero. In the table, column BHLUMI(alf2e) represents $\mathcal{O}(\alpha^2)_{exp}$ BHLUMI 4.02.a, col. BHLUMI(alf2) shows $\mathcal{O}(\alpha^2)$ BHLUMI without exponentiation, col. BHLUMI(alf3e) shows missing $\mathcal{O}(\alpha^3)_{LL}$ in BHLUMI 4.02.a as calulated with the new (unpublished) version of LUMLOG, col. NLLBHA(alf2) shows $\mathcal{O}(\alpha^2)$ result from NLLBHA including NLL corrections, col. NLLBHA(alf3) is the previous plus $\mathcal{O}(\alpha^3)_{LL}$ and col. NLLBHA(alf3p) is the previous plus light pair corrections. In the plot, the $\mathcal{O}(\alpha^2)_{exp}$ cross section $\sigma_{REF}$ from BHLUMI 4.02.a is used as a reference cross section (except for missing $\mathcal{O}(\alpha^3)_{LL}$, for which we show $\sigma/\sigma_{REF}$).

for the SABH process at LEP1 and LEP2.

Let us discuss again the size of the $\mathcal{O}(\alpha^3 L^3)$ and $\mathcal{O}(\alpha^2 L)$ corrections. In the next ta-

| | LEP1 | | LEP2 |
|---|---|---|---|
| Type of correction/error | Ref. [6] | Present | Present |
| (a) Missing photonic $\mathcal{O}(\alpha^2 L)$ | 0.15% | 0.10% | 0.20% |
| (a) Missing photonic $\mathcal{O}(\alpha^3 L^3)$ | 0.008% | 0.015% | 0.03% |
| (c) Vacuum polarization | 0.05% | 0.04% | 0.10% |
| (d) Light pairs | 0.04% | 0.03% | 0.05% |
| (e) Z-exchange | 0.03% | 0.015% | 0.0% |
| Total | 0.16% | 0.11% | 0.25% |

Table 20: Summary of the total (physical+technical) theoretical uncertainty for a typical calorimetric detector. For LEP1, the above estimate is valid for the angular range within $1^\circ - 3^\circ$, and for LEP2 it covers energies up to 176 GeV, and angular range within $1^\circ - 3^\circ$ and $3^\circ - 6^\circ$ (see the text for further comments).

ble/figure 19, we address this question showing once again some results from Tab. 14/Fig. 16, and adding some new numerical results from the BHLUMI event generator and the semianalytical program NLLBHA for the unrealistic ES BARE1 and the realistic ES SICAL2, symmetric WW variants. First, let us recall that in Tab. 14/Fig. 16 the $\mathcal{O}(\alpha^3 L^3)$ effects were included through exponentiation in all calculations, but in most cases they were incomplete. In the case of BHLUMI, the recent version of LUMLOG[6] is able to answer the question: how big is the missing $\mathcal{O}(\alpha^3 L^3)$ in BHLUMI 4.02a. In table/figure 19 we see (black dots) that it is below 0.01% for both BARE1 and SICAL2 ES's. According to our "scaling rules", we conclude that it is below 0.02% at LEP2. Hence, from the *practical* point of view, $\mathcal{O}(\alpha^3 L^3)$ in BHLUMI 4.02a is complete. In table/figure 19 we also include, for the unrealistic BARE1 ES, numerical results from NLLBHA (stars), which includes complete $\mathcal{O}(\alpha^2 L)$ and $\mathcal{O}(\alpha^3 L^3)_{LL}$ corrections. The difference between BHLUMI (crosses) and NLLBHA (stars) should be, in principle, due to $\mathcal{O}(\alpha^2 L)$ (and technical precision), because $\mathcal{O}(\alpha^3 L^3)$ should cancel completely. As we see, the above difference is within the "one per mil box", but for stronger cuts, $z_{min} = 0.9$, it grows slightly beyond 0.1%. Luckily enough, we may push the above exercise in the interesting direction – we have also in table/figure 19 the results from BHLUMI (circles) and NLLBHA (boxes), in which exponentiation and $\mathcal{O}(\alpha^3 L^3)_{LL}$ was removed completely. As we see, these results agree better, even for strong energy cut ($z_{min} = 0.9$). Actually, this result (difference between boxes and circles) represents an interesting quantity: missing $\mathcal{O}(\alpha^2 L)$ in BHLUMI. The above result suggests that it is rather small, below 0.03%. One has to keep in mind that, if the above is true, then the former difference, with $\mathcal{O}(\alpha^3 L^3)_{LL}$ (crosses and stars), is a puzzle and needs to be examined further. In any case, the fact that all the four above results from BHLUMI and NLLBHA are within the "one per mil box" is interesting, encouraging and reinforcing our final conclusion that photonic corrections are under control within 0.1%. For the present time the above interesting comparison is limited to BARE1 ES. For SICAL2 and BARE1 ES's, we see that the difference between BHLUMI with and without exponentiation is quite sizeable, 0.08%, and from that we conclude that the inclusive Yennie-Frautschi-Suura exponentiation in BH-

[6]The new LUMLOG includes final state radiation (in addition to the initial) up to $\mathcal{O}(\alpha^3 L^3)_{LL}$. It was discussed in the Bhabha Working Group and will be included in the next release of BHLUMI.

LUMI is necessary and instrumental for getting good control over the $\mathcal{O}(\alpha^3 L^3)_{LL}$ corrections, even if they are not complete in the matrix element. As a matter of fact, all the other MC codes involved in the present study include exponentiation, and so are on a firm ground from this point of view. In table/figure 19, we show also results from NLLBHA including pair production in addition to the $\mathcal{O}(\alpha^2 L)$ and $\mathcal{O}(\alpha^3 L^3)$ corrections ("plus" marks in the plot). The difference between pluses and stars represents the net effect of the light fermion pair production. For the BARE1 ES, with $z_{min}$ in the experimentally interesting range, it is 0.06% at most. We expect this effect to be about a factor of two smaller for calorimetric ES's.

The total theoretical error for the SABH process at LEP1/LEP2 is summarized in table 20. The errors in the table are understood to be with respect to the cross section calculated for any typical (asymmetric) ES, for the LEP1 experiment in the angular range $1^\circ - 3^\circ$, with respect to the cross section calculated using BHLUMI 4.02a. In the case of LEP2, the estimate extends to the angular range $3^\circ - 6^\circ$, and to the case of the angular range about twice narrower than usual (see the discussion of the numerical results in the previous sections). The entries include combined technical and physical precision. In this table, entry (a) for Missing $\mathcal{O}(\alpha^2 L)$ is based mainly on the agreement between BHLUMI and SABSPV, as seen in tables 14, 16 and 18. It should be stressed that we rely on the agreement between BHLUMI and SABSPV for *all the three types* of ES, Wide-Wide, Wide-Narrow and Narrow Narrow. The agreement between BHLUMI and SABSPV is now better than the one between BHLUMI and OLDBIS+LUMLOG used in the previous best error estimate of Ref. [6]. Noticeably, albeit the agreement between BHLUMI on the one side and BHAGEN95/(OLDBIS + LUMLOG) on the other side is not always below 0.1% for all the ES's considered, it is at least for the experimentally most interesting NW case. This fact is a further reinforcement of the present theoretical error estimate for the SABH process in the NW case, and it is a suggestion for the experimentalists to continue to choose the NW-ES's. The fact that for the unrealistic ES BARE1 the difference between BHLUMI and NLLBHA, see fig. 19, is also within 0.1% confirms this evaluation. Entry (b) is based on table/figure 19. In entry (c), the new improved uncertainty of the vacuum polarization is taken from Tab. 7. We take the biggest of the results from refs. [33, 34]. The light pair production uncertainty, entry (c), is based on new estimates reported during the workshop (see Ref. [7, 8, 12, 15] and Ref. [26, 28]; see also table/figure 19). In tab. 20, we quote for LEP1 the present error due to light fermion pairs contribution to be 0.03%. This is based on all the references quoted above and on the discussion during the WG meetings [29]. The previous estimate in Ref. [6] is therefore confirmed and improved slightly. This is *under the assumption* that the pair effect is corrected for at least in the LL approximation. If the effect is not corrected for[7], then we recommend to use for LEP1 0.04% as an estimate for the missing pair effect (0.06% for LEP2). The material presented at the workshop suggests that the final uncertainty of the light pair contribution will be at the level of 0.015%. In entry (e), the reduced uncertainty of the Z-exchange contribution is based on Ref. [59], work done during this Workshop.

The improvement of the theoretical luminosity error from 0.16% down to 0.11% is basically due to successful comparisons of the programs BHLUMI and SABSPV for a wide range (WW, NN and NW) of experimentally realistic ES's (SICAL2), and also due to an encouraging (although limited to the unrealistic ES BARE1) comparison of un-exponentiated BHLUMI and NLLBHA in table/figure 19. Furthermore, the agreement of BHLUMI, SABSPV, BHAGEN95 and (OLDBIS+LUMLOG) within that same 0.11% error in the NW-ES recommends safely this choice in the experimentally relevant cases. At last, the analysis described in subsection 2.6 shows that the actual Bhabha selections used by the LEP experiments to measure the accelerator luminosity minimize the sensitivity to $O(\alpha^2)$ radiative corrections, thus putting the above conclusions on an even firmer ground. We would like to stress very strongly that the above new estimate 0.11% of the total luminosity error is based on new results which, although pretty stable numerically, are generally still quite fresh and they are *unpublished.* We expect these new results to be published in journals shortly after the workshop, together with the corresponding computer programs.

[7] Production of the light pairs is not included in the standard version of BHLUMI. It is implemented only in the testing unpublished version [29].

The total theoretical error for the SABH process at LEP2 is also summarized in Tab. 20. We assume that the cross section is calculated for any typical (asymmetric) ES for LEP2 experiment, in the typical angular range $1^\circ - 3^\circ$ or $3^\circ - 6^\circ$. The error estimate covers also the "worst case scenario" of the super-narrow angular range (see the example of ES CALO3 in table/figure 17). In entry (a), the estimate of the total photonic uncertainty is based again upon the agreement between BHLUMI and SABSPV on all the variants of ES's considered, and reinforced by the fact that BHAGEN95/OLDBIS+LUMLOG are on the same ground as BHLUMI and SABSPV in the experimentally more interesting NW case (see tables/figures 15 and 17). Note that, sometimes, in the case of other angular range $3^\circ - 6^\circ$ and higher energies, the "scaling laws" from the introduction were used instead of direct calculation to extend the actual numerical results to these situations (see the comments accompanying the relevant tables/plots). We do not see much danger in this because, usually, the large safety margin close to a factor of two was present. Entry (b) is produced out of LEP1 result using the "scaling rule". The vacuum polarization for LEP2 case in the Tab. 20 is taken from Tab. 5 at the $|t| = 36\ \mathrm{GeV}^2$, corresponding to LEP2 energy and the angle of $\theta_{min} = 60$ mrad.

| Type of correction/error | Error estimate |
|---|---|
| (a) Missing $\mathcal{O}(\alpha^2 L)$, $\mathcal{O}(\alpha^3 L^3)$ | $<$ 0.010 % |
| (b) Technical precision (photonic) | 0.040% |
| (c) Vacuum polarization | 0.030% |
| (d) Light fermion pairs | 0.015% |
| (e) Z-contribution | 0.010% |
| Total | 0.053% |

Table 21: *Future* projection of the total (physical+technical) theoretical uncertainty for a typical calorimetric detector, within the $1^\circ - 3^\circ$ angular range at LEP1 energies.

Finally, in view of all the work reviewed during the workshop, we are also able to estimate the precision which will be attained in the next step. It is shown schematically in table 21. At the time when Monte Carlo programs will include the matrix element from $\mathcal{O}(\alpha^2 L)$, the uncertainty due to higher order corrections will be negligible. The dominant contribution will be of *technical* origin and we think that, as we have seen from $\mathcal{O}(\alpha^1)$ comparisons, it can be reduced to 0.04% (provided we can successfully tune two independent Monte Carlo event generators at that precision level, for the same or very similar $\mathcal{O}(\alpha^2)$ matrix elements). The vacuum polarization is now taken according to Ref. [33], and from the discussions during the workshop meetings it was obvious that a further reduction of the uncertainty due to pairs and Z-exchange is also possible. The corresponding work is in progress.

# 3 Large-angle Bhabha scattering

In the present section the LABH process is considered, both at LEP1 and LEP2. The aim of the study, rather than updating the conclusions of Ref. [1] concerning the theoretical accuracy of the LABH process at LEP1, is twofold: on the one hand, the comparison between the semi-analytical benchmarks and the Monte Carlo codes used by the LEP collaborations; on the other one, the study of the LABH process at LEP2, accompanied by the development of dedicated software.

## 3.1 Physics

The main physics interest of Bhabha scattering measurements at large angles (say $\theta > 40°$) around the Z resonance is a precise test of the electroweak sector of the Standard Model. In this angular region more than 80 % of the cross section is due to resonant $s$-channel Z exchange. For $\sqrt{s} = M_Z$ the interference contributions between $s$-channel Z exchange and the other diagrams either vanish or are completely irrelevant, and the $s$-channel photon exchange contribution is small ($\simeq 5 \times 10^{-3}$ of the Z exchange cross section). The only other relevant contribution is $t$-channel photon exchange. For electroweak analyzes, one thus subtracts the $t$-channel and $s - t$ interference contributions from the large-angle experimental data, typically calculated using the ALIBABA [60] semi-analytical (SA) program. After correcting for the effects of real and virtual photon radiation using the analytical programs MIBA [61,62], TOPAZ0 [63,64] or ZFITTER [65,66], the Z exchange cross section $\sigma_Z^0$ may be extracted. For $\sqrt{s} = M_Z$, $\sigma_Z^0 = \sigma_Z^{peak}$ where:

$$\sigma_Z^{peak} = \frac{12\pi\Gamma_e^2}{M_Z^2\Gamma_Z^2} \qquad (2)$$

For the other charged lepton pair decay modes, $\mu^+\mu^-$, $\tau^+\tau^-$, of the Z the quantity $\Gamma_e^2$ in Eqn. (2) is replaced by $\Gamma_e\Gamma_\mu$, $\Gamma_e\Gamma_\tau$, respectively, while for hadronic ($q\bar{q}$) decays it is replaced by $\Gamma_e\Gamma_{had}$. Thus the electronic width of the Z, $\Gamma_e$, which appears in the cross section for all decay modes of the Z, is measured directly and with improved sensitivity (because in this case $\sigma_Z^{peak} \propto \Gamma_e^2$) only in large-angle Bhabha scattering. It is worth noting, however, that in principle the so called $t$-channel subtraction is not unavoidable. Actually, the program TOPAZ0 [63, 64] could be used to fit directly the data for large-angle Bhabha scattering without relying upon $t$-channel subtracted data.

The resulting sensitivity of the backward-forward charge asymmetry in large-angle Bhabha scattering to the important electroweak parameters[8] $\Delta\kappa^{top}$ and $\Delta\kappa^{HIGGS}$

$$\Delta\kappa^{top} = \frac{3\sqrt{2}G_\mu}{16\pi^2}\frac{c_W^2}{s_W^2}m_t^2 \tag{3}$$

$$\Delta\kappa^{HIGGS} = \frac{\sqrt{2}G_\mu M_W^2}{16\pi^2}\left(-\frac{10}{3}\ln\left(\frac{M_H}{M_W}\right)^2-\frac{5}{6}\right) \tag{4}$$

is similar to that of the other dilepton channels $\mu^+\mu^-$, $\tau^+\tau^-$. The above formulae of course indicate only the leading dependence of the one-loop corrections on the masses of the top-quark and the Higgs boson. Actually, at the nowadays precision level a complete electroweak library is mandatory [1].

At the Z peak the purely QED corrections to the large-angle Bhabha cross section are, for typical experimental cuts [67]: $O(\alpha)$, -30 % ; $O(\alpha^2)$, +4 %. These corrections are much larger than those in small-angle Bhabha scattering when typical 'wide'/'narrow' cuts are used [32]: $O(\alpha)$, +5 % ; $O(\alpha^2)$, -1.4 %. Thus theoretical errors on QED radiatively corrected cross sections are expected to be considerably larger in large-angle than in small-angle Bhabha scattering. This is indeed found to be the case in the comparisons between different codes shown below.

In the energy regime of LEP2, the Z-boson effects on the large-angle Bhabha cross section are much smaller than at LEP1. Actually, before entering the details of the comparisons, it is worth noting that large-angle Bhabha scattering shows very different physical features depending on the energy regime at which it is considered. As can be seen from Fig. 18, around the Z peak the cross section is largely dominated by Z-boson annihilation, whereas, already some GeV off resonance, the cross section is largely dominated by $t$-channel photon exchange. From this point of view, large-angle Bhabha scattering at LEP2 is much more similar to small-angle Bhabha scattering than to large-angle Bhabha scattering at LEP1. Hence, at LEP2 the large-angle Bhabha cross section cannot be a useful tool for precise tests of the electroweak sector of the Standard Model, but rather for general QED tests.

The state-of-the-art of large-angle Bhabha scattering up to now can be found in Ref. [1]. In that paper an extensive comparison between two semi-analytical codes, namely ALIBABA [60] and TOPAZ0 [63, 64], is shown. On the other hand, although extensive in the sense that cross sections and asymmetries are considered, that comparison is in some sense limited: actually it involves only semi-analytical codes, on very simple, academic ES's, only at the Z peak.

In view of the above considerations, the tasks of the present Working Group, as far as large-angle Bhabha scattering is concerned, are the following ones:

- involving in the comparisons also the Monte Carlo (MC) codes today available and used by the LEP collaborations;

[8] See [1] for a discussion of pseudo-observables for precision calculations at the Z peak.

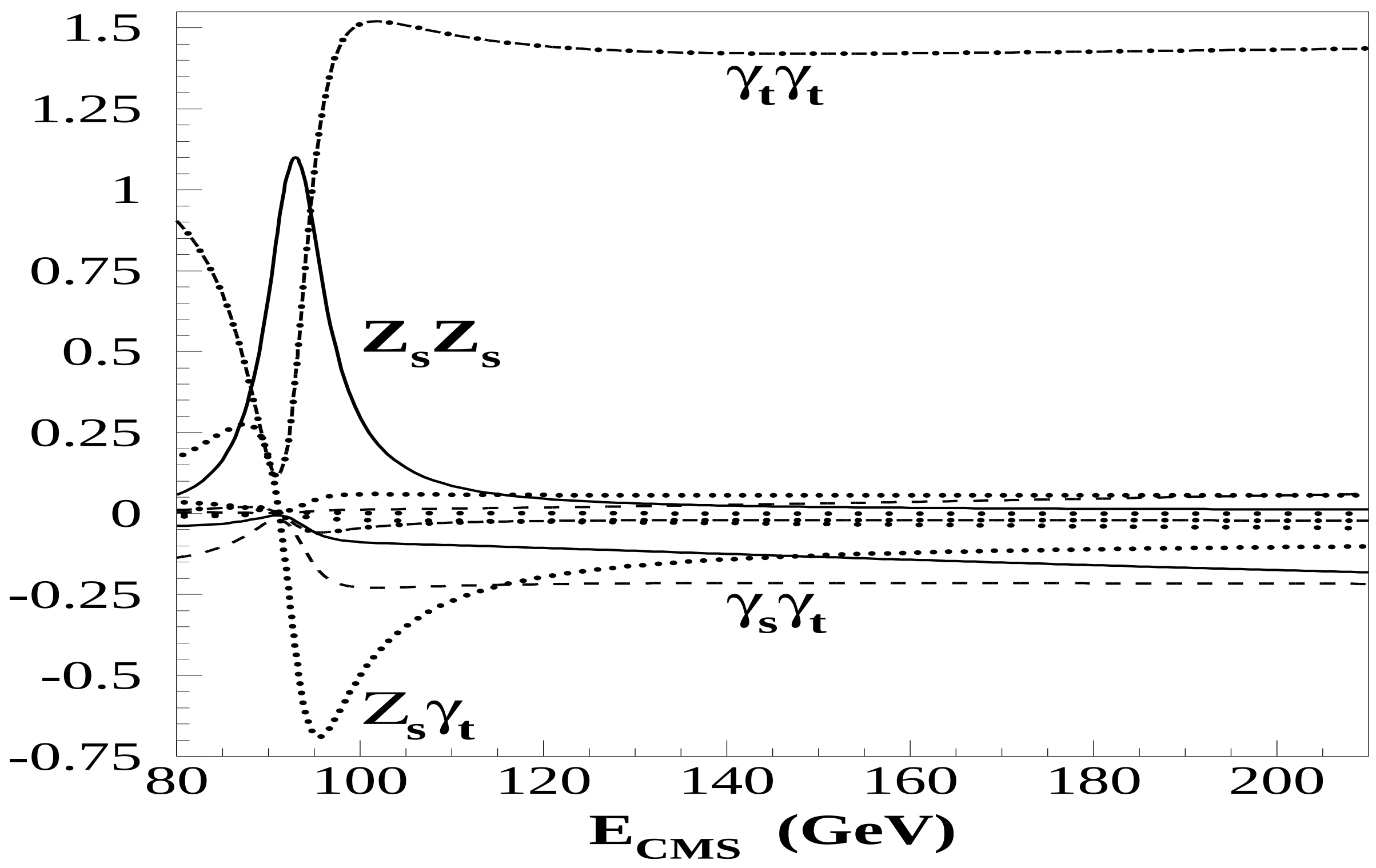

Figure 18: The relative contributions to the integrated cross section at the Born level. The individual contributions are, from top to bottom on the right-hand side of the plot: $\gamma(t)\gamma(t)$, $Z(t)Z(t)$, $\gamma(s)\gamma(s)$, $Z(s)Z(s)$, $Z(s)\gamma(s)$, $Z(s)Z(t)$, $\gamma(s)Z(t)$, $Z(s)\gamma(t)$, $\gamma(t)Z(t)$, and $\gamma(s)\gamma(t)$.

- considering also more realistic, albeit simple, ES's;
- providing results also for the LEP2 energy range, eventually developing dedicated software.

The ES's considered in the present study are the following ones:

- BARE - This ES, for the sake of simplicity, is defined exactly as in [1], namely $40^\circ < \vartheta_- < 140^\circ$, $0^\circ < \vartheta_+ < 180^\circ$, $\vartheta^{max}_{acoll} = 10^\circ, 25^\circ$ and $E_{min} = 1$ GeV for both electron and positron;
- CALO - This ES is defined as above, but with $E_{min} = 20$ GeV for the final fermion energy, which is the electron(positron) energy if there are no photons nearby, whereas it is the electron(positron) plus photon energy if the photon is within a cone of semi-aperture $1^\circ$ from the electron(positron).

For all the cases considered, the input parameters are $M_Z = 91.1887$ GeV, $m_t = 174$ GeV, $m_H = 300$ GeV and $\alpha_s(M_Z) = 0.124$. The predictions by ALIBABA are taken from Ref. [1].

Let us now briefly summarize the features of the codes involved in the study. Here only the general features will be highlighted; for more details the reader is referred to the original literature or to the write-ups presented at the end of this section.

ALIBABA [60] – It is a semi-analytical code, implementing exact $O(\alpha)$ QED and weak corrections. The higher-order QED corrections consist of leading log $O(\alpha^2)$ corrections plus soft-photon exponentiation. Moreover, the weak $O(\alpha)$ corrections are folded with the leading log structure functions. The matching between the exact $O(\alpha)$ QED matrix element and the higher order corrections is performed in additive form. The electroweak library is not up to date. Nonetheless, the code has to be considered as a benchmark.

BHAGEN95 [43] – It is a Monte Carlo integrator for both small and large-angle Bhabha scattering. The value for the cross section is obtained from the event generator BHAGEN94, a structure function based program for all orders resummation, including complete photonic $O(\alpha)$ and leading logarithmic $O(\alpha^2 L^2)$ corrections in all channels, and all relevant electroweak corrections according to BHM/WOH basic formulae from Ref. [1]. The approximations, introduced with the collinear kinematics of initial and final radiation and in its angular distribution, are eliminated for the one hard photon emission by substitution with the exact calculation.

BHAGENE3 [67, 68] – It is a Monte Carlo event generator for large-angle Bhabha scattering and muon pair production. The program includes one-loop and the most important two-loop electroweak as well as QED radiative corrections. The $O(\alpha)$ QED correction uses the exact matrix element. Higher order QED corrections are included in an improved soft photon approximation with exponentiation of initial state radiation. Up to three hard final state photons are generated. Events are generated in the full final state phase space including explicit mass effects in the region of collinear mass singularities. The minimum scattering angle for percent level cross section accuracy is $10^\circ$. Extensive use is made in the program of one and two dimensional look-up tables for fast, flexible and efficient Monte Carlo generation. The program was designed for the Z peak region but may also be used at LEP2 energies.

BHWIDE [69] – It is a new Monte Carlo event generator for large-angle Bhabha scattering at LEP1/SLC and LEP2. It includes multiphoton radiation in the framework of $O(\alpha)$ YFS exponentiation. The $O(\alpha)$ virtual (both weak and QED) corrections are in the current version taken from ALIBABA. The program provides the full event in terms of particle flavors and their four-momenta with an arbitrary number of radiative photons. In many aspects it is similar to the program BHLUMI for small-angle Bhabha scattering and can be considered as its extension to large angles. It has been checked that for the pure QED process BHWIDE at $\mathcal{O}(\alpha)$ (no exponentiation) agrees with the MC program OLDBIS within a statistical accuracy of 0.05%.

SABSPV [46] – It is a new Monte Carlo integrator, originally designed for small-angle Bhabha scattering, but adapted to the treatment of large-angle Bhabha scattering at the LEP2 energy range. It is based on a proper matching of the $O(\alpha)$ corrected cross section for $t$-channel photon exchange and of the leading logarithmic results in the structure function approach. The matching is performed in a factorized form, in order to preserve the classical limit. At present, the effect of up-down interference in the $\gamma(t) - \gamma(t)$ contribution is not taken into account and all the other contributions are corrected at the leading logarithmic level. Due to the present approximations, the theoretical accuracy of the code is of the order of 1%, as far as large-angle Bhabha scattering at LEP2 is concerned.

TOPAZ0 [63, 64] – It is a semi-analytical code, developed for precision physics at LEP1. It includes the state-of-the-art concerning weak and QCD corrections, according to Ref. [1]. As far as QED corrections are concerned, they are exactly treated at $O(\alpha)$ for $s$-channel processes (leptonic and hadronic), at the leading logarithmic level for pure $t$-channel and $s$-$t$ contributions in the Bhabha scattering case. On top of this, higher order QED corrections are taken into account in the structure functions approach, in a factorized form in order to preserve the classical limit. A particular effort has been performed in order to implement as much analytically as possible the experimental cuts typically applied by the LEP collaborations.

UNIBAB [70] – It is a full Monte Carlo event generator that was originally designed for large-angle Bhabha scattering at LEP1 and SLC energies. The QED corrections are implemented in a fully factorized form by assuming $s$-channel dominance and using photon shower algorithms for initial- and final-state radiation, and therefore exponentiation of soft photons and resummation of the logarithms from multiple emission of hard collinear photons is automatic. QED initial-final interference corrections are not yet implemented. The electroweak corrections are based on a library also used by ALIBABA, but updated to include the leading $m_t^4$-dependence and higher order QCD corrections to the Z width.

## 3.2 On Z peak (LEP1)

The situation of the comparisons for LEP1 is summarized in Figs. 19 (BARE) and 20 (CALO) and corresponding tables. Conventionally, the reference cross section with respect to which the relative deviations are computed is taken from TOPAZ0. It has to be stressed that this choice has no particular meaning at all.

Let us begin with commenting the situation of Fig. 19, i.e. for the BARE ES. As far as the comparison between the two semi-analytical codes, ALIBABA and TOPAZ0, is concerned, the agreement is better than 0.1% at the Z peak (energy points n. 4 and 5, corresponding to the smallest experimental error, which is of the order of 0.3% statistical and 0.3% systematic), and deteriorates on the wings, where, on the other hand, the experimental error is larger (for instance, at peak$\pm$2 GeV the experimental error is of the order of 1% statistical and 0.3% systematic). Note that the worst situation is for maximum acollinearity cut of $10^\circ$, above the Z peak, where the codes differ from one another of about 1%: this difference is due to higher order QED effects, as pointed out in Ref. [71] (factorized versus additive formulation). As far as the Monte Carlo codes BHAGENE3 and BHWIDE are concerned, their agreement with the semi-analytical codes at peak is within few per mil, whereas off peak BHWIDE is within 1% and BHAGENE3 can deviate up to 2%.

The situation for the more realistic case, the CALO ES (Fig. 20), is generally better from the point of view of the SA/MC comparisons. Note that ALIBABA is no more involved, since

| No. | $E_{CM}$ | TOPAZ0 | BHWIDE | BHAGENE3 | ALIBABA | BHAGEN95 |
|---|---|---|---|---|---|---|
| | (a) BARE $acol_{max} = 10^o$ | | | | | |
| 1. | 88.45 | $.4579 \pm .0003$ | $.4560 \pm .0004$ | $.4495 \pm .0016$ | $.4575 \pm .0003$ | $.4556 \pm .0002$ |
| 2. | 89.45 | $.6452 \pm .0002$ | $.6429 \pm .0006$ | $.6334 \pm .0023$ | $.6440 \pm .0003$ | $.6403 \pm .0003$ |
| 3. | 90.20 | $.9115 \pm .0002$ | $.9087 \pm .0008$ | $.8997 \pm .0033$ | $.9090 \pm .0004$ | $.9026 \pm .0004$ |
| 4. | 91.19 | $1.1846 \pm .0002$ | $1.1797 \pm .0010$ | $1.1847 \pm .0033$ | $1.1840 \pm .0004$ | $1.1715 \pm .0005$ |
| 5. | 91.30 | $1.1639 \pm .0002$ | $1.1592 \pm .0009$ | $1.1667 \pm .0033$ | $1.1636 \pm .0005$ | $1.1514 \pm .0005$ |
| 6. | 91.95 | $.8738 \pm .0002$ | $.8711 \pm .0007$ | $.8856 \pm .0028$ | $.8769 \pm .0003$ | $.8664 \pm .0003$ |
| 7. | 93.00 | $.4771 \pm .0002$ | $.4761 \pm .0005$ | $.4808 \pm .0019$ | $.4814 \pm .0001$ | $.4756 \pm .0002$ |
| 8. | 93.70 | $.3521 \pm .0002$ | $.3512 \pm .0004$ | $.3521 \pm .0013$ | $.3556 \pm .0001$ | $.3522 \pm .0001$ |
| | (b) BARE $acol_{max} = 25^0$ | | | | | |
| 1. | 88.45 | $.4854 \pm .0003$ | $.4808 \pm .0005$ | $.4699 \pm .0016$ | $.4833 \pm .0003$ | $.4833 \pm .0003$ |
| 2. | 89.45 | $.6746 \pm .0003$ | $.6699 \pm .0006$ | $.6593 \pm .0023$ | $.6727 \pm .0003$ | $.6727 \pm .0003$ |
| 3. | 90.20 | $.9438 \pm .0003$ | $.9387 \pm .0008$ | $.9279 \pm .0033$ | $.9425 \pm .0003$ | $.9425 \pm .0003$ |
| 4. | 91.19 | $1.2198 \pm .0003$ | $1.2130 \pm .0010$ | $1.2169 \pm .0034$ | $1.2187 \pm .0004$ | $1.2187 \pm .0004$ |
| 5. | 91.30 | $1.1989 \pm .0003$ | $1.1924 \pm .0010$ | $1.1995 \pm .0034$ | $1.1982 \pm .0004$ | $1.1982 \pm .0004$ |
| 6. | 91.95 | $.9054 \pm .0002$ | $.9011 \pm .0007$ | $.9124 \pm .0026$ | $.9089 \pm .0003$ | $.9089 \pm .0003$ |
| 7. | 93.00 | $.5040 \pm .0002$ | $.5013 \pm .0005$ | $.4996 \pm .0019$ | $.5054 \pm .0002$ | $.5054 \pm .0002$ |
| 8. | 93.70 | $.3777 \pm .0002$ | $.3749 \pm .0004$ | $.3689 \pm .0013$ | $.3782 \pm .0001$ | $.3782 \pm .0001$ |

BARE $acol_{max} = 10^o$ BARE $acol_{max} = 25^o$

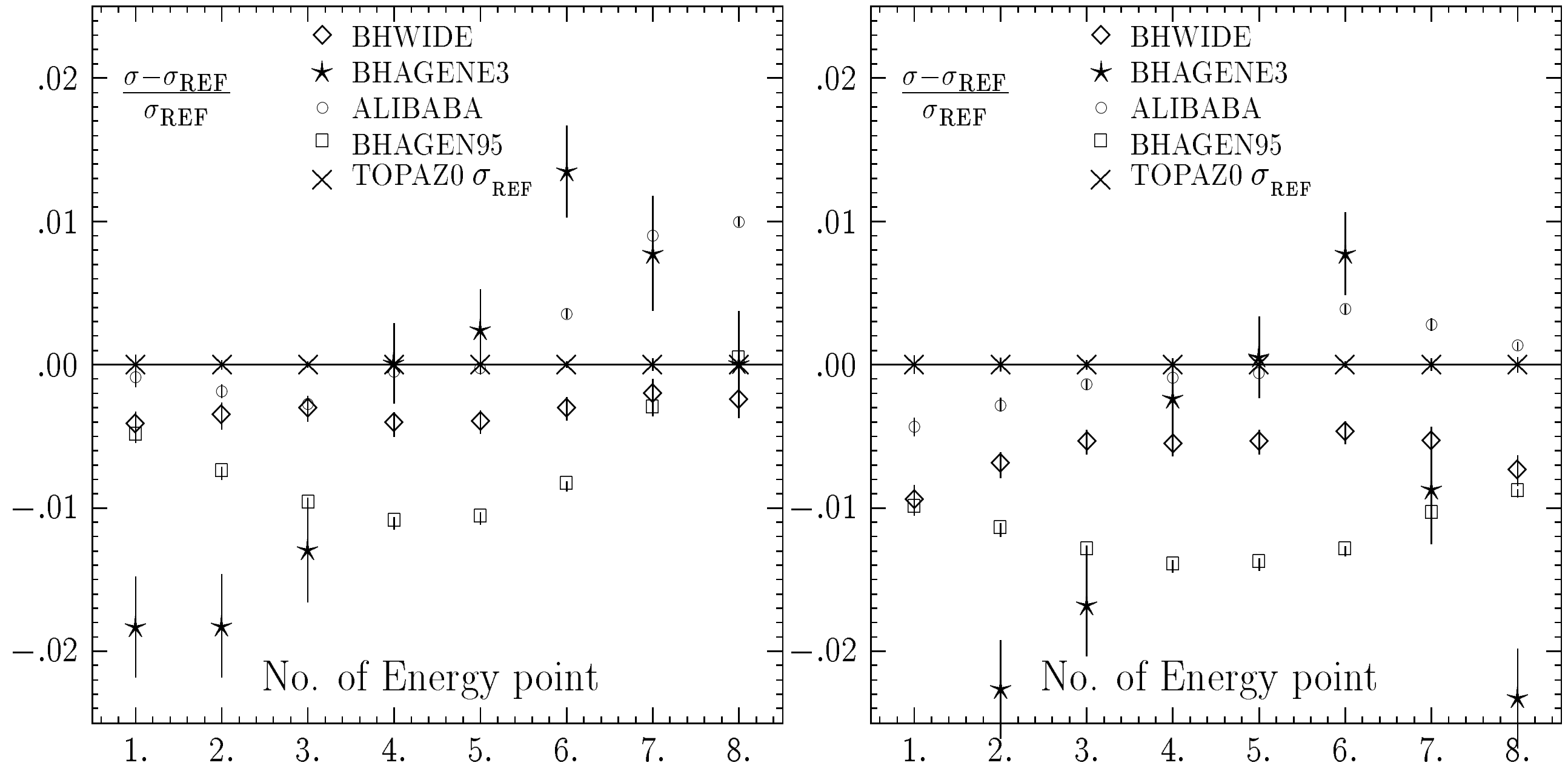

Figure 19: Monte Carlo results for the BARE ES, for two values ($10^o$ and $25^o$) of acollinearity cut. Center of mass energies (in GeV) close to $Z$ peak. In the plots, the cross section $\sigma_{\rm REF}$ from TOPAZ0 is used as a reference cross section. Cross sections in nb.

it cannot manage calorimetric measurements, whereas UNIBAB appears (it is slow for very small minimum fermion energy and therefore it did not contribute to the BARE case). On peak, the agreement between the codes is at the few per mil level; off peak BHWIDE is within 0.5% from TOPAZ0, whereas UNIBAB deviates up to 1% below peak, and BHAGENE3 can differ from TOPAZ0 by about 2%.

| No. | $E_{CM}$ | TOPAZ0 | BHWIDE | BHAGENE3 | UNIBAB | BHAGEN95 |
|---|---|---|---|---|---|---|
| | (a) CALO $acol_{max} = 10^o$ | | | | | |
| 1. | 88.45 | $.4533 \pm .0004$ | $.4523 \pm .0004$ | $.4467 \pm .0008$ | $.4490 \pm .0010$ | $.4501 \pm .0002$ |
| 2. | 89.45 | $.6387 \pm .0004$ | $.6377 \pm .0006$ | $.6302 \pm .0011$ | $.6358 \pm .0012$ | $.6326 \pm .0003$ |
| 3. | 90.20 | $.9023 \pm .0003$ | $.9016 \pm .0008$ | $.8920 \pm .0015$ | $.9021 \pm .0014$ | $.8918 \pm .0004$ |
| 4. | 91.19 | $1.1725 \pm .0001$ | $1.1707 \pm .0010$ | $1.1767 \pm .0021$ | $1.1772 \pm .0016$ | $1.1582 \pm .0005$ |
| 5. | 91.30 | $1.1520 \pm .0001$ | $1.1505 \pm .0009$ | $1.1571 \pm .0020$ | $1.1559 \pm .0016$ | $1.1385 \pm .0005$ |
| 6. | 91.95 | $.8649 \pm .0001$ | $.8646 \pm .0007$ | $.8795 \pm .0015$ | $.8689 \pm .0012$ | $.8579 \pm .0003$ |
| 7. | 93.00 | $.4723 \pm .0001$ | $.4725 \pm .0005$ | $.4796 \pm .0008$ | $.4733 \pm .0008$ | $.4719 \pm .0002$ |
| 8. | 93.70 | $.3486 \pm .0001$ | $.3486 \pm .0004$ | $.3507 \pm .0006$ | $.3486 \pm .0007$ | $.3498 \pm .0001$ |
| | (b) CALO $acol_{max} = 25^o$ | | | | | |
| 1. | 88.45 | $.4769 \pm .0004$ | $.4742 \pm .0004$ | $.4696 \pm .0008$ | $.4733 \pm .0010$ | $.4717 \pm .0002$ |
| 2. | 89.45 | $.6638 \pm .0003$ | $.6615 \pm .0006$ | $.6556 \pm .0011$ | $.6619 \pm .0012$ | $.6554 \pm .0003$ |
| 3. | 90.20 | $.9297 \pm .0003$ | $.9278 \pm .0008$ | $.9207 \pm .0012$ | $.9302 \pm .0014$ | $.9164 \pm .0004$ |
| 4. | 91.19 | $1.2025 \pm .0003$ | $1.1994 \pm .0010$ | $1.2074 \pm .0021$ | $1.2073 \pm .0016$ | $1.1845 \pm .0005$ |
| 5. | 91.30 | $1.1819 \pm .0003$ | $1.1790 \pm .0010$ | $1.1879 \pm .0021$ | $1.1860 \pm .0016$ | $1.1647 \pm .0005$ |
| 6. | 91.95 | $.8924 \pm .0003$ | $.8909 \pm .0007$ | $.9058 \pm .0016$ | $.8965 \pm .0012$ | $.8817 \pm .0003$ |
| 7. | 93.00 | $.4964 \pm .0003$ | $.4954 \pm .0005$ | $.5004 \pm .0009$ | $.4976 \pm .0008$ | $.4929 \pm .0002$ |
| 8. | 93.70 | $.3717 \pm .0003$ | $.3704 \pm .0004$ | $.3690 \pm .0006$ | $.3720 \pm .0007$ | $.3701 \pm .0001$ |

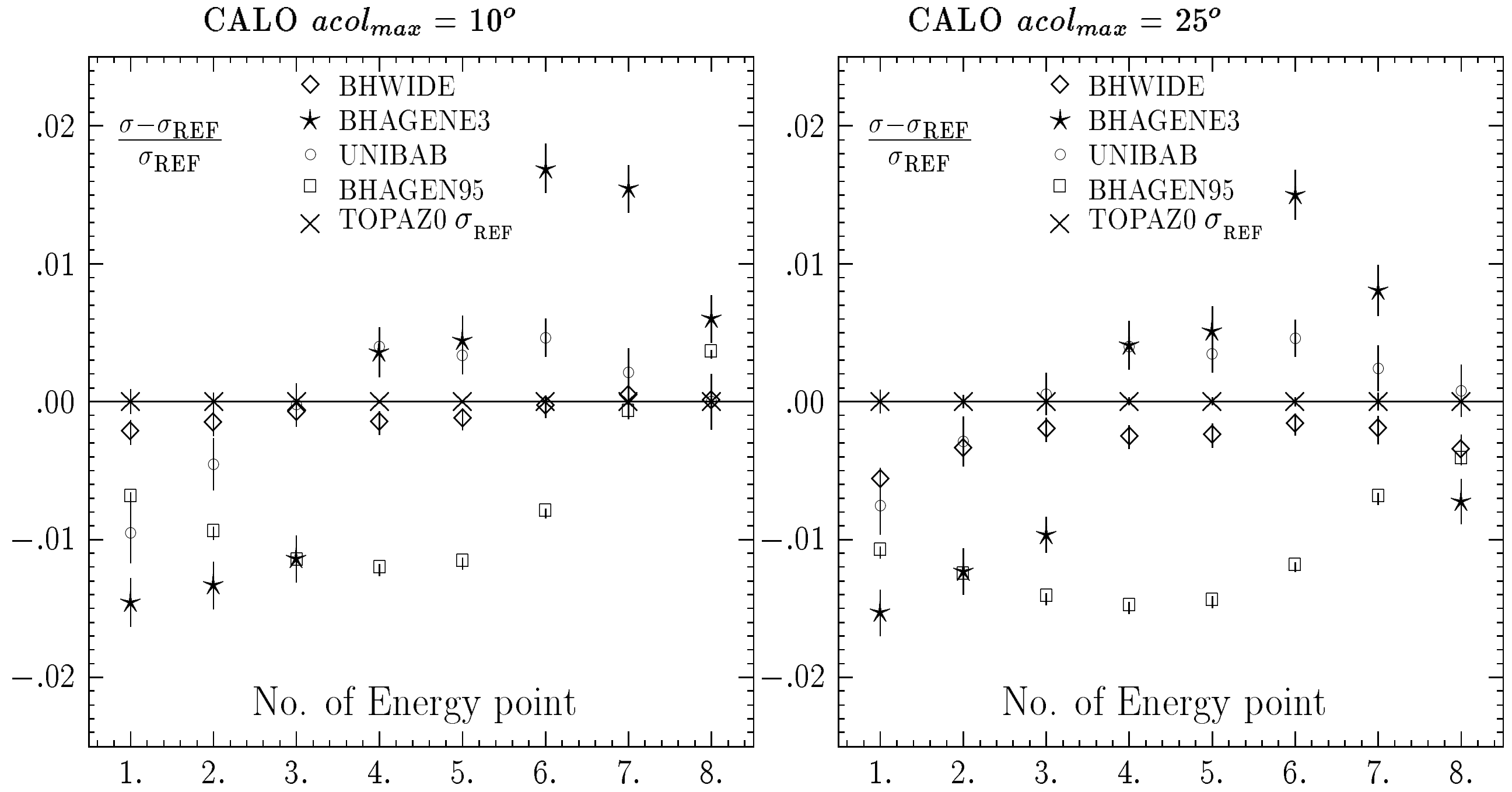

Figure 20: Monte Carlo results for the CALO ES, for two values ($10^o$ and $25^o$) of acollinearity cut. Center of mass energies (in GeV) close to $Z$ peak. In the plots, the cross section $\sigma_{REF}$ from TOPAZ0 is used as a reference cross section. Cross sections in nb.

BHAGEN95 is within 1.5% everywhere for both the BARE and CALO ES's around the $Z$ peak. The agreement is better few GeV above and below the $Z$ resonance. However the implementation of the complete weak and QCD library is very recent and still under tests.

### 3.3 Far off Z peak (LEP2)

| No. | BHWIDE | TOPAZ0 | BHAGENE3 | UNIBAB | SABSPV | BHAGEN95 |
|---|---|---|---|---|---|---|
| | (a) CALO $acol_{max} = 10^o$ | | | | | |
| 1. | $35.257 \pm .040$ | $35.455 \pm .024$ | $34.690 \pm .210$ | $34.498 \pm .157$ | $35.740 \pm .080$ | $35.847 \pm .022$ |
| 2. | $29.899 \pm .034$ | $30.024 \pm .020$ | $28.780 \pm .170$ | $29.189 \pm .134$ | $30.270 \pm .070$ | $30.352 \pm .017$ |
| 3. | $25.593 \pm .029$ | $25.738 \pm .015$ | $24.690 \pm .150$ | $24.976 \pm .115$ | $25.960 \pm .060$ | $26.007 \pm .014$ |
| | (b) CALO $acol_{max} = 25^o$ | | | | | |
| 1. | $39.741 \pm .049$ | $40.487 \pm .025$ | $39.170 \pm .280$ | $39.521 \pm .158$ | $40.240 \pm .100$ | $40.505 \pm .026$ |
| 2. | $33.698 \pm .042$ | $34.336 \pm .017$ | $32.400 \pm .190$ | $33.512 \pm .135$ | $34.100 \pm .080$ | $34.331 \pm .020$ |
| 3. | $28.929 \pm .036$ | $29.460 \pm .013$ | $27.840 \pm .160$ | $28.710 \pm .116$ | $29.280 \pm .070$ | $29.437 \pm .015$ |

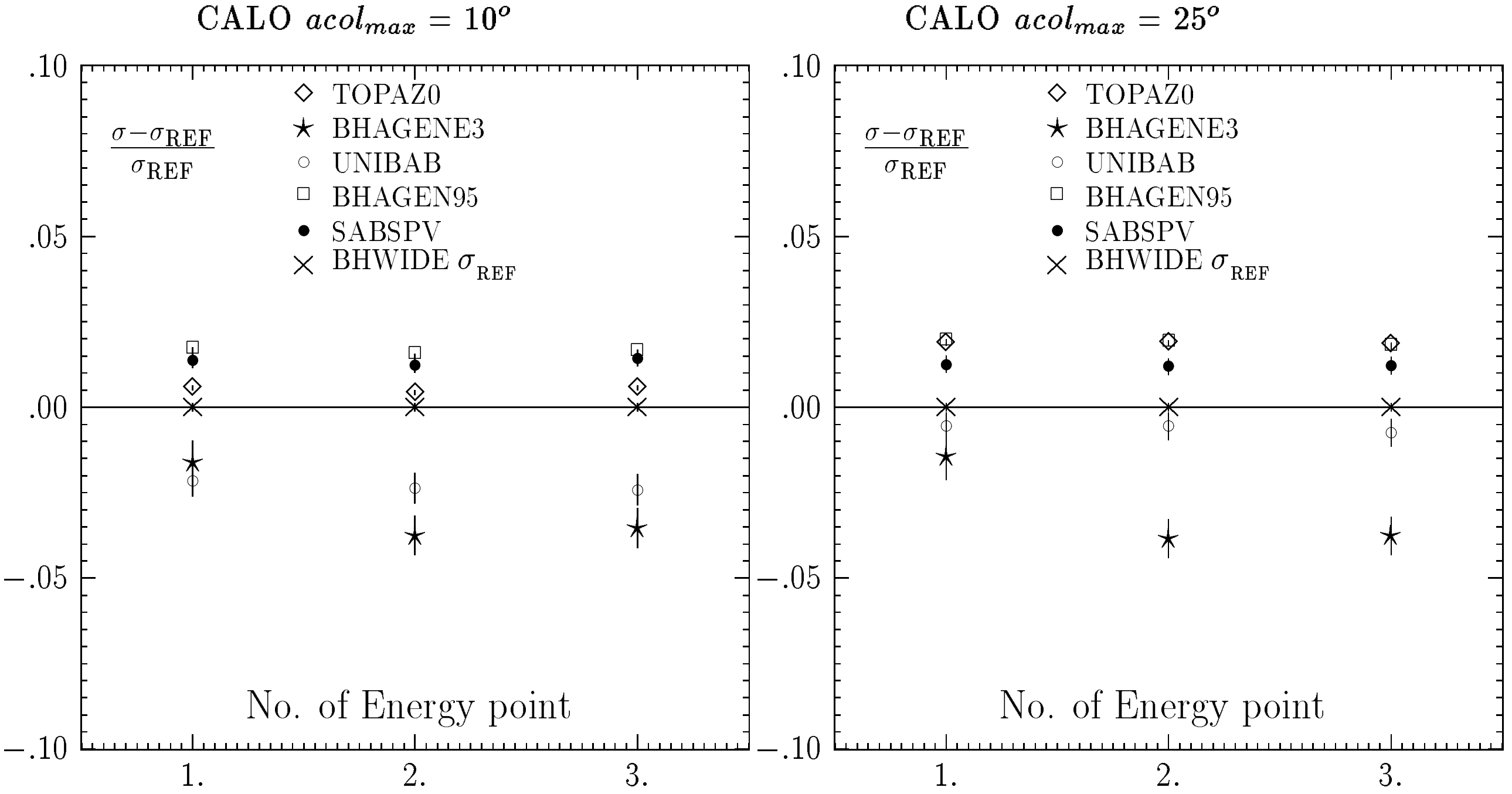

Figure 21: Monte Carlo results for the CALO ES, for two values ($10^o$ and $25^o$) of acollinearity cut. Center of mass energies close to $W$-pair production threshold ($E_{CM}$: 1. 175 GeV, 2. 190 GeV, 3. 205 GeV). In the plots, the cross section $\sigma_{REF}$ from BHWIDE is used as a reference cross section. Cross sections in pb.

The situation of the comparisons for LEP2 is shown in Fig. 21 (CALO) and corresponding table. Conventionally, the reference cross section with respect to which the relative deviations are computed is taken from BHWIDE. It has to be stressed again that this choice has no particular meaning at all. Note that TOPAZ0 has been developed in the Z-dominance approximation, and UNIBAB does not include initial-final interference effects, so that their results are at the leading logarithmic level in the LEP2 energy range. A new entry appears, namely SABSPV, which has been conceived for small-angle Bhabha scattering, and further improved for large-angle Bhabha at LEP2.

BHAGEN95, BHWIDE and SABSPV stay within 2% from one another. More precisely, SABSPV is steadily around 1.5% above BHWIDE and 0.5% below BHAGEN95. BHAGENE3, for both the acollinearity cuts considered, can deviate from the reference cross section up to 5%.

TOPAZ0 and UNIBAB show deviations from the reference cross section (up to 2% above and 3% below, respectively) which depend on the acollinearity cut, and can be presumably traced back to the approximations intrinsic in these Z-peak designed codes. Anyway, the deviations of the two codes from the reference cross section are consistent with what can be expected from leading logarithmic results.

# 4 Short-write-up's of the programs

The aim of the following short-write-up's is to provide quick reference for the reader on basic properties of all event generators used in the numerical comparisons throughout this article. The intention was that details are given only on new and/or unpublished features of the programs (including bugs) while other features are described in general terms with help of references to published works.

## 4.1 BHAGEN95

AUTHORS:

| | |
|---|---|
| **M. Caffo** | INFN and Dipartimento di Fisica dell'Università, Bologna, Italy<br>`caffo@bo.infn.it` |
| **H. Czyż** | University of Silesia, Katowice, Poland, INFN, Università, Bologna, Italy<br>`czyz@bo.infn.it` |
| **E. Remiddi** | INFN and Dipartimento di Fisica dell'Università, Bologna, Italy<br>`remiddi@bo.infn.it` |

GENERAL DESCRIPTION:
BHAGEN95 is a collection of three programs to calculate the cross-section for Bhabha scattering for small and large scattering angles at LEP1 and LEP2 energies. In its present form the integrated cross-section $\sigma$ for a given selection of cuts is calculated as

$$\sigma = \sigma(\text{BHAGEN94}) - \sigma^H(\text{BH94-FO}) + \sigma^H(\text{BHAGEN-1PH}) \ . \qquad (5)$$

$\sigma$(BHAGEN94) is the integrated cross-section obtained with the Monte Carlo event generator

BHAGEN94 [2, 72–76], a structure function based program for all orders resummation, including complete photonic $\mathcal{O}(\alpha)$ and leading logarithmic $\mathcal{O}(\alpha^2 L^2)$ corrections in all channels, and all relevant electroweak corrections according to BHM/WOH basic formulae from [1]. Approximations are introduced with the collinear kinematics of initial and final radiation and in its angular distribution.
$\sigma^H$(BH94-FO) is the integrated cross-section of $\mathcal{O}(\alpha)$ for one hard photon emission obtained with the Monte Carlo event generator BH94-FO [76], the $\mathcal{O}(\alpha)$ expansion of BHAGEN94.
$\sigma^H$(BHAGEN-1PH) is the integrated cross-section obtained with the one hard photon complete matrix element and exact kinematics, implemented in the Monte Carlo event generator BHAGEN-1PH [77].

The subtraction of $\sigma^H$(BH94-FO) and its substitution with $\sigma^H$(BHAGEN-1PH) is to eliminate the error in the contribution coming from the one hard photon emission.

FEATURES OF THE PROGRAM:
The three programs provide cross-sections, which are summed as in Eq. (5) or used to obtain other quantities, such as forward-backward asymmetry. Due to the mentioned substitution procedure, the event generator feature of the constituent programs can not be profited and the use is simply that of a Monte Carlo integrator.
At small-angle we estimate the accuracy in the cross-section evaluation, which comes from the uncontrolled higher orders terms $\mathcal{O}(\alpha^2 L)$ and $\mathcal{O}(\alpha^3 L^3)$ and from the incertitude in $\mathcal{O}(\alpha^2 L^2)$ $s-t$ interference to amount comprehensively to a 0.1%. The error due to approximate two hard photon contribution (strongly dependent on the imposed cuts) is estimated on the basis of the correction required for the one hard photon contribution times $\beta(s) = 0.1$, to account for the increase in perturbative order. All included we estimate at small-angle an accuracy of the order of 0.15% for loose cuts ($z_{min} = 0.3$) and of 0.45% for sharp cuts ($z_{min} = 0.9$) for both LEP1 and LEP2 energies.
At large-angle we estimate the $\mathcal{O}(\alpha^2 L^2)$ $s-t$ interference accuracy up to 1% (depending on cuts) at LEP1 energy, but much smaller at LEP2. The error coming from the approximate treatment of two hard photon emission is estimated as above and is smaller for more stringent acollinearity cut. All included we estimate an accuracy of the order of 1% for both LEP1 and LEP2 energies.

HOW DOES THE CODE WORK:
The three programs run separately. They provide initialization and fiducial volume definition according to input parameters, then starts the generation of events according to some variables which smooth the cross-section behavior. Rejection is performed through the routine `TRIGGER`, where the special cuts can be implemented. The programs stop when the requested number of accepted events is reached or alternatively when the requested accuracy is obtained.

INPUT CARD:
The following data have to be provided in input: mass of the $Z^0$, mass of the top quark, mass of the Higgs, value of $\alpha_S(M_Z^2)$, value of $\Gamma_Z$, the beam energy $E_{beam}$, the minimum energy for leptons $E_{min}$ (larger than 1 GeV), minimum and maximum angle for the scattered electron (positron) with the initial electron (positron) direction, maximum acollinearity allowed between final electron and positron, number of accepted events to be produced, numbers to initialize the random number generator. One may also switch on or off: pairs production, the channels to be considered and the recording of the events. For $\mathcal{O}(\alpha)$ programs one has also to specify the minimum and maximum energy allowed for the photon. For the input of BHAGEN-1PH one has to give also the maximum acoplanarity, and minimum angles of the emitted photon with initial and final fermion directions, if one wants to exclude the contributions with the collinear photons.

DESCRIPTION OF THE OUTPUT:
Each program return the input parameters and the values of the cross-section obtained with weighted and unweighted events, with the relative statistical variance (one standard deviation). Of course due to the efficiency the weighted cross-section is usually much more precise than the unweighted one. The total integrated cross-section is then calculated according to Eq. (5).

AVAILABILITY:
On request to the authors and to be posted on WWW at http://www.bo.infn.it./

## 4.2 BHAGENE3

AUTHORS:

| | |
|---|---|
| **J. Field** | Département de Physique Nucl. et Corpusculaire Univ. de Genève |
| | `jfield@cernvm.cern.ch` |
| **T. Riemann** | DESY, Platanenallee 6, D–15738 Zeuthen |
| | `riemann@ifh.de` |

GENERAL DESCRIPTION:
BHAGENE3 is a Monte Carlo event generator for muon pairs (at all angles) or for Bhabha scattering in the large angle region ($\theta > 10^\circ$). The program, which is intended for use at, or above, the Z peak region contains all tree level diagrams with complete one loop and the leading two loop virtual corrections [78–81] The running $\alpha$ is included with the correct scale in all amplitudes. The O($\alpha$) QED correction uses the exact $ll\gamma$ matrix elements [82,83]. Higher order QED corrections are included in an improved soft photon approximation with exponentiation of initial state radiation. Events with up to three hard photons are generated in the full kinematically allowed phase space including explicit mass effects for near collinear photon radiation. If $n_\gamma^I$, $n_\gamma^F$ are the respective numbers of initial and final state photons, the different final state topologies generated are: $n_\gamma^I/n_\gamma^F = 0/0,\ 1/0,\ 0/1,\ 2/0,\ 1/1,\ 0/2,\ 0/3$ . Initial/Final state interference effects are taken into account only to O($\alpha$). The photon energies are described by scaled variables: $y_i = 2E_\gamma^i/\sqrt{s} < 1$. For $y_i < y_0$ (typically $y_0 = 0.005$) a Born topology (0/0) event is generated. The corresponding cross section contains all virtual (V) corrections and is integrated over the phase space of all soft (S) photons with $y_i < y_0$. Exponentiation of initial state radiation is implemented by modifying the O($\alpha$) partial cross sections and interference terms in such a way that the derivative of the VS cross-section with respect to $y$ is recovered in the $y \to 0$ limit. For example in the $s$-channel photon exchange contribution with initial state radiation:

$$\frac{d\sigma(\gamma_s\gamma_s)_{init}}{d\Omega_l dy} = \frac{\alpha^3}{2\pi s}\frac{1}{y}\left[(\frac{s'}{s})\ln\frac{s}{m_e^2} - 1\right]\left[\frac{u^2+t^2+u'^2+t'^2}{s'^2}\right] \qquad (6)$$

exponentiation is carried out by the replacement $u^2+t^2+u'^2+t'^2 \quad \longrightarrow \quad f(u^2+t^2)+u'^2+t'^2$ where $f = 2\,C_V^i\,y^{\beta_e} - 1$. Events with hard photons are generated according to distributions where the soft photon eikonal factors are corrected by the Gribov-Lipatov [84] kernels. The relative probabilities of different topologies of final state photons are chosen according to the Poisson distribution: $P(n|\bar{N})$ where $n = n_\gamma - 1$ and

$$\bar{N^I} = \beta_e \ln(1/y_0), \quad \bar{N^F} = \beta_f \ln(1/y_0) \qquad (7)$$

A short description of program together with comparisons with other muon pair and wide angle Bhabha codes has been published [67]. A long write-up is also available [68].

| | |
|---|---|
| OIMZ | Z mass (GeV) |
| OIMT | Top quark mass (GeV) |
| OIMH | Higgs boson mass (GeV) |
| OMAS | $\alpha_s(M_Z)$ |
| IOCH | $= 0(\mu^+\mu^-), = 1(e^+e^-)$ |
| IOEXP | = 1 exponentiated , = 0 $O(\alpha)$ |
| OW | collision energy (GeV) |
| OCTC1 | lower $\cos\theta_{l^+}$ in the ODLR frame |
| OCTC2 | lower $\cos\theta_{l^-}$ in the ODLR frame |
| IOXI | initial random number |

Table 22: Variables of the labelled common ICOM. OCTC1,OCTC2 are used in setting up the LUT of the lepton scattering angles. To allow for the effects of the Lorentz boost the angular range should be chosen somewhat wider than that defined by the cuts in the LAB system.

FEATURES OF THE PROGRAM:
The execution of the program has three distinct phases: initialisation, generation and termination. In the initialisation phase all relevant electroweak quantities are calculated from the input parameters $M_Z$, $M_t$ $M_H$ and $\alpha_s$. Also a number of look up tables for quantities such as the lepton scattering angle and photon energies are created for use in the subsequent generation phase. This process is relatively time consuming, so the user should not be surprised if there is some delay between the execution of the program and the start of event generation. In the generation phase events with unit weight are generated by the weight throwing technique. The corresponding 4-vectors are stored in common C4VEC. The user may apply arbitary cuts and produce weighted histograms in subroutine FUSER. Histograms of unit weight events may be produced in subroutine FHIST. In the final, termination, phase the input parameters are printed out together with the exact cross section $\sigma^{CUT}$ and its error, together with all histograms and plots.

HOW TO USE THE PROGRAM:
The program has a very short main program containing definitions of the most important

| | |
|---|---|
| NPAR(1) | <u>1</u>, 0 weak loop corrections ON, OFF |
| NPAR(2) | 2,<u>3</u> parameterisations of had. vac. pol. |
| NPAR(3) | <u>0</u>,1,2 two-loop $\alpha\alpha_s m_t^2$ correction |
| NPAR(4) | <u>1</u>,0 weak box diagrams ON, OFF |
| NPAR(6) | <u>1</u>, 0 two-loop terms $\propto m_t^4$ ON,OFF |
| XPAR(1) | initial lepton charge (-1.D0) |
| XPAR(2) | final lepton charge (-1.D0) |
| XPAR(3) | final lepton colour (1.D0) |
| XPAR(4) | final lepton mass (GeV) |
| XPAR(9) | QCD correction to $\Gamma_q^Z$ (non-$b$ quarks) |
| XPAR(10) | QCD correction to $\Gamma_b^Z$ |
| YMA | maximum value of $\sum E_\gamma/E_{beam}$ (0.99D0) |
| YMI | minimum value of $E_\gamma/E_{beam}$ (0.005D0) |
| WTMAX | maximum value of the event weight (2.0D) |

Table 23: Control parameters defined in SUBROUTINE BHAGENE3. Default values are underlined or given in parentheses.

input parameters, which are stored in the labelled common block ICOM. These variables are described in Table 22 The execution of the program has three distinct phases: (i) Initialisation, (ii) Generation of a single unit weight event, (iii) Termination. Each of these phases is entered via a call to subroutine BHAGENE3 in the main program:
CALL BHAGENE3(MODE,CTP1,CTP2,CTM1,CTM2,CTAC,EP0,EM0)
MODE is set to $-1, 0, 1$ for the initialisation, generation and termination phases respectively. The other parameters of BHAGENE3 define the kinematical cuts to be applied to the generated events:

CTP1 = minimum value of $\cos\theta_{l^+}$
CTP2 = maximum value of $\cos\theta_{l^+}$
CTM1 = minimum value of $\cos\theta_{l^-}$
CTM2 = maximum value of $\cos\theta_{l^-}$
CTAC = maximum value of $\cos\phi_{col}$
EP0 = minimum energy of $l^+$ $(GeV)$
EM0 = minimum energy of $l^-$ $(GeV)$

All these cuts are applied in the laboratory (incoming $e^+, e^-$ centre of mass) system. The angle $\phi_{col}$ is the collinearity angle between the $l^+$ and the $l^-$ (CATC = -1 for a back-to-back configuration). In the calls of BHAGENE3 with MODE = 0,1 only this parameter need be specified. Other initialisation parameters of interest to users are defined in BHAGENE3 itself. A list of the most important of these can be found in Table 23.

AVAILABILITY:
From Compure Physics Communications Program Library, see `http://www.cs.qub.ac.uk/cpc` for more details.

## 4.3 BHLUMI

AUTHORS:

| | |
|---|---|
| **S. Jadach** | Institute of Nuclear Physics, Kraków, ul. Kawiory 26a<br>jadach@cernvm.cern.ch |
| **E. Richter-Wąs** | Institute of Computer Science, Jagellonian University, Kraków<br>erichter@cernvm.cern.ch |
| **B.F.L. Ward** | Department of Physics and Astron., University of Tennessee and SLAC<br>bflw@slac.stanford.edu |
| **Z. Wąs** | CERN and Institute of Nuclear Physics, Kraków, ul. Kawiory 26a<br>wasm@cernvm.cern.ch |

GENERAL DESCRIPTION:
The program is a multiphoton Monte Carlo event generator for low angle Bhabha providing four-momenta of outgoing electron, positron and photons. The first $\mathcal{O}(\alpha^1)_{YFS}$ version was described in ref. [85]. The actual version 4.02.a includes several types of the matrix elements. The most important $\mathcal{O}(\alpha^2)^{prag}_{YFS}$ matrix elements (M.E.) is based on the Yennie-Frautschi-Suura (YFS) exponentiation. This M.E. includes exactly the photonic first order and second order leading-log corrections. In the $\mathcal{O}(\alpha^2)^{prag}_{YFS}$ M.E. the other higher order and subleading contributions are included in the approximate form. The detailed description exists for the version 2.02 in ref. [30]. For the differences between the versions 2.02 and 4.02 the user has to consult ref. [6], the README file in the distribution directory and comments in the main program of the demonstration deck [44]. The only difference between versions 4.02 and 4.02a is correction to an important bug 95a. In order to correct it one has to replace $v^{(2)}_{[1,0]}$ in eq. (3) in Ref. [6] with

$$v^{(2)}_{[1,0]} = (\gamma_p + \gamma_q)\ln\Delta + \frac{3}{2}\gamma - \frac{\alpha}{\pi} - \frac{3}{4}\gamma\ln(1-\tilde{\beta}_1) - \frac{1}{4}\gamma\ln(1-\tilde{\alpha}_1). \quad (8)$$

We also provide patch to correct this in the sorce code of the versions 4.02, see AVAILABILITY below. This correction can affect the result of the program typicaly 0.05%, up to 0.08% for some event selections.

FEATURES OF THE PROGRAM:
BHLUMI consists in fact of the three separate event generators: BHLUM4, OLDBIS and LUMLOG, where OLDBIS is an improved version of the OLDBAB written by Berends and Kleiss at PETRA times, and LUMLOG is an event generator with the inclusive many photons emission strictly collinear to momenta of incomming/outcogoing fermions. M.E. of OLDBIS is limited to $\mathcal{O}(\alpha^1)$ and M.E. of LUMLOG includes exponentiated and non-exponentiated electron structure functions up to $\mathcal{O}(\alpha^3)_{LL}$. BHLUM4 includes four types of the exponentited M.E.: $\mathcal{O}(\alpha^2)^{YFS}_{\mathrm{A,B}}$, $\mathcal{O}(\alpha^1)^{YFS}_{\mathrm{A,B}}$ and four types of the non-exponentited M.E.: $\mathcal{O}(\alpha^2)_{\mathrm{A,B}}$, $\mathcal{O}(\alpha^1)_{\mathrm{A,B}}$ where the cases A and B correspond to two kinds of ansatz employed for modeling the $\mathcal{O}(\alpha^2 L)$, second order NNL, contribution. The BHLUM4 program includes vacuum polarization, $s$-chanel $\gamma$ and $Z$ exchange contributions, see ref. [59] in the approximation suitable for the low angle (below 0.1rad.) scattering. The BHLUM4 does no include so called up-down interferences. However, OLDBIS does include them so it can be used to check how small they are.

HOW DOES THE CODE WORK:

The program ia a full scale Monte Carlo event generator. A single `CALL BHLUMI` produces one event, i.e. the list of the final state four-momenta of electron, positron and photons encoded in the common block. Depending on switch in the input parameters the program provides event with the variable weight WTMOD or with constant weight WTMOD=1. In the constant weight mode the calculation is done for M.E. of the $\mathcal{O}(\alpha^2)^{YFS}_{\rm B}$ type. In the variable weight mode WTMOD corresponds to the above M.E. but the user has also acces to all six types of the M.E. listed above (and even more) and may perform in a single run calculation for various types of the M.E. The choice of one of three sub-generators BHLUM4, OLDBIS or LUMLOG is decided through one of the input parameters. Program requires initialization before producing first MC event. There are many input parameters. The most important ones define the minimum and maximum angle ($t$ chanel transfer). For weighted events it is possible to cover the angular range down to zero angle but the program is realy designed for "double tag" acceptance. It is possible to stop and restart the program from the next event in the series. The distribution directory incudes example demonstrating how to do it.

DESCRIPTION OF THE OUTPUT:

Program prints certain control output. The basic output of the program is the series of the Monte Carlo events and the user decides by himself which events are accepted or rejected according to his favourite selection criteria. The total cross section in nanobarns can be calculated for arbitrary cuts in a standard way

$$\sigma = \sigma_0 \frac{1}{N} \sum_{Accepted\ Events} W_I \tag{9}$$

where the sum of the weights (variable or constant) is over all accepted events, $N$ is total number of generated events and $\sigma_0$ is a reference (normalization) cross section in nanobarns provided by the program at the end of the MC generation. In the analogous standard way one may obtain any arbitrary distribution properly normalized.

AVAILABILITY:

The program is posted on WWW at `http://hpjmiady.ifj.edu.pl` in the form of ".tar.gz" file together with all relevant papers and documentation in postscript. The version 4.02.a which was used to produce all numerical results in this workshop consists of the version 4.02 described in ref. [6] and of the error patch posted in the same location `http://hpjmiady.ifj.edu.pl`. After workshop the equivalent version 4.03 will be released. The new version of BHLUMI will also contain new version of LUMLOG with the final state bremsstrahlung which was used in in the table/figure 19 and improved version of the BHLUMI matrix element without exponentiation which was used in this table/figure.

## 4.4 BHWIDE

AUTHORS:

| | |
|---|---|
| **S. Jadach** | Institute of Nuclear Physics, Kraków, ul. Kawiory 26a<br>`jadach@cernvm.cern.ch` |
| **W. Płaczek** | Dept. of Physics and Astron., Univ. of Tennessee<br>`placzek@hephp02.phys.utk.edu` |
| **B.F.L. Ward** | Dept. of Physics and Astron., Univ. of Tennessee and SLAC<br>`bflw@slac.stanford.edu` |

GENERAL DESCRIPTION:
The program evaluates the large (wide) angle Bhabha cross section at LEP1/SLC and LEP2 energies. The theoretical formulation is based on $\mathcal{O}(\alpha)$ YFS exponentiation, with $\mathcal{O}(\alpha)$ virtual (both weak and QED) corrections taken from ref. [60, 86] as formulated in the program ALIBABA. The YFS exponentiation is realized via Monte Carlo methods based on BHLUMI-type Monte Carlo algorithm, which is explained in refs. [30,85]. Thus, we achieved an event-by-event realization of our calculation in which arbitrary detector cuts are possible and in which infrared singularities are canceled to all orders in $\alpha$. A detailed description of our work can be found in ref. [69].

FEATURES OF THE PROGRAM:
The code is a full-fledged Monte Carlo event generator, so that the final particle four-momenta for the entire $e^+e^- + n\gamma$ final state are available for each event, which may be generated as a weighted or unweighted event, as the user finds more or less convenient accordingly. Thus, it is trivial to impose arbitrary detector cuts on the events. If the user wishes, he/she may also use the original BABAMC [82,83] type of pure weak corrections (there is a simple switch which accomplishes this). The expected accuracy of the program, when all tests are finished, is anticipated at $\sim 0.2\%$ in the $Z$-region and $\sim 0.1\%$ in the LEP2 regime.

HOW DOES THE CODE WORK:
The code works entirely analogous to the MC event generator BHLUMI 2.01 described in ref. [30]. A crude distribution consisting of the primitive Born level distribution and the most dominant part of the YFS form factors, which can be integrated analytically, is used to generate a background population of events. The weight for these events is then computed by the standard rejection techniques involving the ratio of the complete distribution and the crude distribution. As the user wishes, these weights may either be used directly with the events, which have the four-momenta of all final state particles available, or they may be accepted/rejected against a constant maximal weight WTMAX to produce unweighted events via again standard MC methods. Standard final statistics of the run are provided, such as statistical error analysis, total cross section, etc.

DESCRIPTION OF THE OUTPUT:
Program prints certain control output. The basic output of the program is the series of the Monte Carlo events. The total cross section in nanobarns can be calculated for arbitrary cuts in the same standard way as for BHLUMI, i.e. user may imposed arbitrary experimental cuts by rejection.

AVAILABILITY:
The program can be obtained via e-mail from the authors. It will be posted soon on WWW at `http://enigma.phys.utk.edu` as well as on anonymous ftp at `enigma.phys.utk.edu` in the form of ".tar.gz" file together with all relevant papers and documentation in postscript. It will also be available via anonymous ftp at `enigma.phys.utk.edu` in the directory `/pub/BHWIDE`.

## 4.5 NLLBHA

AUTHORS:

| | |
|---|---|
| **A.B. Arbuzov** | Joint Institute for Nuclear Research, Dubna, 141980, Russia<br>arbuzov@thsun1.jinr.dubna.su |
| **E.A. Kuraev** | Joint Institute for Nuclear Research, Dubna, 141980, Russia<br>kuraev@theor.jinrc.dubna.su |
| **L. Trentadue** | CERN TH Division, Universitá di Parma, INFN Sezione di Milano<br>trenta@vxcern.cern.ch |

GENERAL DESCRIPTION:
`NLLBHA` is a semi–analytical program for calculations of radiative QED and electroweak corrections to the small–angle Bhabha scattering at high energies. It takes into account complete (relevant at small angles) first order QED and electroweak corrections, the leading and next–to–leading QED corrections to $\mathcal{O}(\alpha^2)$ and the leading logarithmic contributions to $\mathcal{O}(\alpha^3)$. The corrections due to photon emission as well as the ones due to pair production are included. The theoretical uncertainty of the calculations is less then 0.1%.

FEATURES OF THE PROGRAM:
`NLLBHA` integrates numerically analytical formulae [2, 8, 23]. For the Born cross–section an expansion for small scattering angles is used. The contributions due to real particle emission are integrated over symmetrical detector apertures. For the case of asymmetrical detectors (narrow–wide case) leading logarithmic contributions are calculated (next–to–leading are estimated to be equal or less the ones in the narrow–narrow case). Cuts on the final particles energies are possible. Calorimetric set–up as well as other special experimental conditions are not implemented.

HOW DOES THE CODE WORK:
The code consists of the main part and of a series of subroutines which calculate separately radiative correction (RC) contributions from different Feynman diagrams and configurations. In the main part the flags, the parameters and the constants are defined. Using the flags one can define with their help the event selection (BARE1 symmetric or asymmetric are possible only), the order of corrections, switch on or off different contributions (like Z-boson exchange, vacuum polarization and light pair production). Then the user have to set the parameters: the beam energy, the angular range, the energy cut. The electroweak parameters are calculated with the help of the DIZET package [87].

DESCRIPTION OF THE OUTPUT:
At firdt the code prints the information about the chosen set-up, vacuum polarization (on/off), Z-boson contribution (on/off). Then the code prints the beam energy, the angular range and the electroweak parameters. After calculations it prints, for each value of $x_c$ (energy cut), the Born and the radiatively corrected (to different orders and approximations) cross–sections in [nb]; It also prints a line with the values of the different corrections in percent with respect to the Born cross–section. The normalizations and definitions used do directly correspond to the ones given in [2] where also the origin of all RC contributions can be found.

AVAILABILITY:
The code is available upon request from the authors.

## 4.6 SABSPV

AUTHORS:

| | |
|---|---|
| **M. Cacciari** | DESY, Hamburg, Germany |
| | cacciari@desy.de |
| **G. Montagna** | University of Pavia, Italy |
| | montagna@pavia.pv.infn.it |
| **O. Nicrosini** | CERN - TH Division (Permanent address: INFN Pavia, Italy) |
| | nicrosini@vxcern.cern.ch, nicrosini@pavia.pv.infn.it |
| **F. Piccinini** | INFN Pavia, Italy |
| | piccinini@pavia.pv.infn.it |

GENERAL DESCRIPTION:
SABSPV evaluates small angle Bhabha cross sections, in the angular region used for luminosity measurement at LEP, and large angle Bhabha cross sections at LEP2. The theoretical formulation is based on a suitable matching between an exact fixed order calculation and the resummation of leading log radiative effects provided by the structure function techniques.
The matching between the all-orders leading-log cross section, $\sigma_{LL}^{(\infty)}$, given by the convolution of structure functions with kernel (Born) cross sections, and the $\mathcal{O}(\alpha)$ one is realized according to the following general recipe: the order-$\alpha$ content of the leading-log cross section is extracted by employing the $\mathcal{O}(\alpha)$ expansions of the structure functions, thereby yielding $\sigma_{LL}^{(\alpha)}$. Denoting by $\sigma^{S+V}(k_0)$ the cross section including virtual corrections plus soft photons of energy up to $E_\gamma = k_0 E$, and by $\sigma^H(k_0)$ the radiative $\mathcal{O}(\alpha)$ cross section, the fully corrected cross section can finally be written as

$$\sigma_A = \sigma_{LL}^{(\infty)} - \sigma_{LL}^{(\alpha)} + \sigma^{S+V}(k_0) + \sigma^H(k_0)\ . \qquad (10)$$

Equation (10) is in the additive form. A factorized form can also be supplied. It has the same $\mathcal{O}(\alpha)$ content but also leads to the so-called classical limit, according to which the cross section must vanish in the absence of photonic radiation. It reads

$$\sigma_F = (1 + C^H_{NL})\sigma^{(\infty)}_{LL}, \quad C^H_{NL} \equiv \frac{\sigma^{S+V}(k_0) + \sigma^H(k_0) - \sigma^{(\alpha)}_{LL}}{\sigma} \equiv \frac{\sigma^{(\alpha)}_{NL}}{\sigma}, \tag{11}$$

$\sigma$ being the Born cross section; $C^H_{NL}$ contains the non-log part of the $\mathcal{O}(\alpha)$ cross section, represented by $\sigma^{(\alpha)}_{NL}$.

In order to be flexible with respect to the different kinds of experimental cuts and triggering conditions, it makes use of a multi-dimensional Monte Carlo integration with importance sampling. A detailed description of the formalism adopted and the physical ideas behind it can be found refs. [2,46] and references therein.

FEATURES OF THE PROGRAM:

The code is a Monte Carlo integrator for weighted events. At every step, two kinds of events are fully accessible:

(A) "two-body" events: they include tree-level events and radiative events in the collinear approximation; in this last case, information concerning the equivalent photons lost in the beam pipe is available,

(B) three-body events: they include the radiative events $e^+e^- \to e^+e^-\gamma$ beyond the collinear approximation.

No explicit photons beyond $\mathcal{O}(\alpha)$ are generated; on the generated events, every kind of cuts can be imposed. $\mathcal{O}(\alpha)$ corrections are available for the $\gamma(t)\gamma(t)$ contribution (see for instance [72] for the soft plus virtual corrections and [88, 89] for the hard bremsstrahlung contribution); all the other channels are treated in the leading logarithmic approximation[9]. This theoretical framework does exploit the fact that the $\gamma(t)\gamma(t)$ channel is by far the most dominant one. It is therefore sufficient to evaluate exact order $\alpha$ corrections for this channel only. The other channels, which at the Born level contribute at the level of one per cent in the small angle region and of some per cents in the large angle region at LEP2 energies, can be evaluated in the leading log approximation. Higher order corrections are implemented in the structure function formalism [2]. The overall accuracy of the predictions performed by the code is, generically, of the order of 0.1% in the small angle regime and of the order of 1% in the large angle regime at LEP2 energies.

HOW DOES THE CODE WORK:

The code generates random integration variables within the "fiducial" cuts supplied via the input card (see below). These values are passed to the kinematics subroutines, which construct the full quadrimomenta for electron, positron and photon. The quadrimomenta are then fed to a trigger routine, which either accepts or rejects the event according to the cuts specified in it by the user. The control is then returned to the main integration routine, which generates weighted events, accumulates the cross section result for each single contribution and compose them as described in eqs. (10) and (11). Once in a given number of events the integrations results and the related error estimates are evaluated and written to the output file. The error is also compared to the accuracy limit required, and the run stops when the latter is reached. The program can be restarted from its own output file, by specifying the same "physical" inputs and either a larger number of events or a higher accuracy.

[9] Actually, in the present version of the program the up-down interference contribution is neglected. This is of no practical relevance for the small angle cross section, whereas it introduces an error of the order of some per mil in the large angle cross section at LEP2.

INPUT CARD:
The following data card has to be provided via standard input:

```
46.15D0                                  ! EBEAM
24.D-3  58.D-3  0.D0                     ! T1MIN, T1MAX, E1MIN
24.D-3  58.D-3  0.D0                     ! T2MIN, T2MAX, E2MIN
0.5D0   1.D-2   0.D0  0.D0  0.D0         ! CALOINPUT(1...5)
1                                        ! ISIM
1                                        ! ICALO
1.D5    0.D0    0     'SABSPV.OUT'       ! EVTS, ACCLIM, IRESTART, OUTFILE
```

These parameters have the following meaning:

**1** 46.15D0 - the electron and positron beam energy, EBEAM.

**2** 24.D-3, 58.D-3, 0.D0 - the electron minimum and maximum scattering angle (in radians) and the minimum electron energy (in GeV), T1MIN, T1MAX, E1MIN. These cuts are to be interpreted as "fiducial" cuts within which the events are generated, before going through the triggering routine.

**3** the same for the positron, T2MIN, T2MAX, E2MIN.

**4** 0.5D0, 1.D-2, 0.D0, 0.D0, 0.D0 - inputs that may be required by the cutting routines for the triggers. These values are stored in the vector CALOINPUT(5) via the common block COMMON/CALOS.

**5** 1 - flag for symmetric cuts, ISIM. The user has to specify if the experimental cuts asked for are (1) or not (0) symmetric for electron–positron exchange. If they are, choosing 1 saves computing time.

**6** 1 - flag for choosing the triggering routine, ICALO.

**7** 1.D5, 0.D0, 0, 'SABSPV.OUT' - these are inputs related to the Monte Carlo integration and to the management of the output. Namely, the total number of events, EVTS, the relative accuracy limit aimed at, ACCLIM, the restarting flag, IRESTART (if 1 the program tries to restart execution from the indicated output file, if 0 it reinitializes it), and the output file name, OUTFILE.

DESCRIPTION OF THE OUTPUT:
The output file OUTFILE contains a description of the inputs provided to the code, the results of the Monte Carlo integrations for the various contributions and the final results with their standard statistical error. Moreover informations concerning the random number generator and the cumulants, that can be used to restart the program from where it stopped, are provided.

AVAILABILITY:
The code is available upon request to one of the authors.

## 4.7 UNIBAB

AUTHORS:

**H. Anlauf** TH Darmstadt & Universität Siegen, Germany
anlauf@crunch.ikp.physik.th-darmstadt.de
**T. Ohl** TH Darmstadt, Germany
ohl@crunch.ikp.physik.th-darmstadt.de

GENERAL DESCRIPTION:
UNIBAB is a Monte Carlo event generator designed for large angle Bhabha scattering at LEP and SLC energies. In its original incarnation [90, 91], it was a simple QED dresser describing only multiphoton initial-state radiation, thus focusing on the exponentiation of soft photons and the resummation to all orders of the leading logarithmic corrections of the form $(\alpha/\pi)^n \ln^n(s/m_e^2)$. The first published version, UNIBAB version 2.0 [70] contains improvements in the exclusive photon shower algorithm used for the description of initial-state radiation, and many enhancements, such as final-state radiation using a similar photon shower algorithm. An electroweak library based on ALIBABA [60] was added. Initial and final state corrections are implemented in a fully factorized form. Version 2.1 of the program features the inclusion of longitudinal beam polarization. During this workshop the current version 2.2 was developed, which uses an implementation of the final state photon shower based on the exact lowest order matrix element for the process $Z \to f\bar{f}\gamma$. Also, the electroweak library has been updated slightly to include the leading $m_t^4$-dependence and higher order QCD corrections to the Z width, as discussed in detail in [1].

FEATURES OF THE PROGRAM:
The event generator UNIBAB calculates the QED radiative corrections through a photon shower algorithm. The actual implementation is based on an iterative numerical solution of an Altarelli-Parisi type evolution equation for the electron structure function. The effective matrix element for photon emission from the initial state assumes a factorized form of the radiative matrix element. Therefore it is exact for collinear emission. It also allows to generate finite transverse momenta of the radiated photons. For final state radiation, the algorithm employs an iterated form of the first order matrix element for $Z \to f\bar{f}\gamma$, which gives a reasonable description of exclusive distributions that are sensitive to the details of the approximations used for the multiphoton matrix element, such as acollinearity distributions on the Z peak.
UNIBAB generates only unweighted events. It is implicitly assumed that all scales in the hard subprocess are of the same order of magnitude, and the program does not yet include initial-final interference, thus the program is generally limited to the large angle region. Numerically, the effects from initial-final interference are sufficiently small in the vicinity of the Z peak. For details see the long write-up [70].

HOW DOES THE CODE WORK:
UNIBAB consists of two layers, an external layer with a very simple user interface that allows easy interactive and batch control of the program, and an internal layer with a low level interface to the internal routines. It is however recommended to use the high level interface which automatically takes care of parameter dependencies and properly reinitializes the Monte Carlo when a physics parameter is modified.
In order to run the program, one has to specify several steering parameters that are internally translated into Monte Carlo parameters. The actual physical cuts have to be implemented in an external analyzer. The essential steering parameters are:

- **ctsmin, ctsmax**: cuts on $\cos\theta^*$, where $\theta^*$ is the scattering angle in the boosted subsystem after taking initial state radiation into account.
- **ecut**: minimum energy of the final state fermions.
- **acocut**: maximum acollinearity of the outgoing $e^+e^-$ pair.

An interactive run may look like:

```
set ebeam    45.65         # Beam energy in GeV
set mass1z   91.1888       # Z mass
set mass1t  174            # top quark mass
set mass1h  300            # Higgs mass
set ctsmin -0.8
set ctsmax  0.8
set ecut     20
set acocut   30            # acollinearity cut in degrees
init
generate 100000
close
quit
```

Additional switches control the inclusion or omission of certain contributions like weak box diagrams or $t$-channel diagrams. For more details please consult the manual.

DESCRIPTION OF THE OUTPUT:
**UNIBAB** stores the generated events and all supplementary information for analysis (cross section, Monte Carlo error) in the proposed standard **/hepevt/** common block [72] and must be read from there by a suitable analyzer. A simple yet very flexible tool for implementing a "theorist's detector" is given by **HEPAWK** [92,93], which easily allows to obtain arbitrary distributions from the generated events.

AVAILABILITY:
The current version of **UNIBAB** may be downloaded via anonymous ftp from
**ftp://crunch.ikp.physik.th-darmstadt.de/pub/anlauf/unibab**
along with up-to-date documentation. At the time of this writing (and for historical reasons), the program source and accompanying files are still distributed in the CERN patchy format. Platform-dependent Fortran77 source files will be made available upon request. For the sample test run, **UNIBAB** has also to be linked with the analyzer **HEPAWK** [92,93]. A more modern (auto-configuring) and self-contained version of the Monte Carlo generator will be made available in a future release after the end of the workshop.

# 5 Conclusions and outlook

In this WG, the first systematic comparison of all the existing Monte Carlo event generators for the Bhabha process at LEP1 and LEP2 has been performed. This is one of our main achievements. The other one is that, as a result of these comparisons, the theoretical error of the small-angle Bhabha process is now reduced from 0.16% to 0.11% for typical LEP1 experimental ES's, at the angular range of $1^\circ - 3^\circ$. In parallel, an estimate of the theoretical error of the small-angle Bhabha process at LEP2 has also been fixed at 0.25%, for all possible experimental situations. The theoretical precision of the small-angle Bhabha scattering should be still improved by a factor of two at LEP1, in order to match the experimental precision. From the analysis performed, we conclude that a theoretical error of the order of 0.06% is reasonably feasible at LEP1, and the present study offers a solid ground for the next step in this direction.

As far as the large-angle Bhabha process is concerned, the main result of this WG is that now we have comparisons not only among the semi-analytical benchmarks ALIBABA and TOPAZ0, but also among Monte Carlo event generators and on the Monte Carlo codes versus semianalytical programs. In spite of the fact that the comparisons involving Monte Carlo's do not change the conclusions of the previous LEP1 WG on the theoretical precision of large-angle Bhabha at LEP1 (see [1]), they give information about the performances of the Monte Carlo event generators themselves. In particular (except for some programs which have to be improved, either on the QED libraries or on the pure weak ones), the situation at LEP1 is generally under control with respect to the present experimental accuracy, both on and off Z peak. As far as LEP2 is concerned, a general agreement of the order of 2% has been achieved. There is certainly room for further improvements on this item, but for practical purposes the situation can be considered satisfactory.